\documentclass[longauth]{aa}  
\usepackage[varg]{txfonts}
\usepackage{graphicx}
\usepackage{lscape}
\usepackage{amsmath}
\usepackage{amssymb}
\usepackage{color}
\usepackage{natbib}
\bibpunct{(}{)}{;}{a}{}{,}
\usepackage{url}
\usepackage{multirow}
\usepackage{threeparttable}
\usepackage{booktabs} 
\usepackage{soul}
\usepackage{float}
\usepackage[colorlinks=true]{hyperref}
\usepackage{tablefootnote}
\usepackage{subcaption}
\hypersetup{colorlinks=true,allcolors=blue,citecolor=blue}
\usepackage[font=small,labelfont=bf]{caption}
\usepackage{orcidlink}
\usepackage{siunitx}
\usepackage{lastpage}

\titlerunning{TOI-4552 b: A new ultra-short period rocky world revealed by NIRPS and TESS}
\authorrunning{Srivastava et al.}

\newcommand{\teff}{\ensuremath{T_{\rm eff}}}
\newcommand{\logg}{\mbox{$\log g_*$}\xspace}

\newcommand{\gccc}{\mbox{g\,cm$^{-3}$}\xspace}

\newcommand{\me}{\mbox{$\it{M_{\rm \mathrm{\oplus}}}$}\xspace}
\newcommand{\re}{\mbox{$\it{R_{\rm \mathrm{\oplus}}}$}\xspace}
\newcommand{\mstar}{\mbox{$M_{*}$}\xspace}
\newcommand{\rstar}{\mbox{$R_{*}$}\xspace}

\newcommand{\msol}{\mbox{$\it{M_\mathrm{\odot}}$}\xspace}
\newcommand{\rsol}{\mbox{$\it{R_\mathrm{\odot}}$}\xspace}
\newcommand{\se}{\mbox{$\it{S_{\rm \mathrm{\oplus}}}$}\xspace}

\newcommand{\lstar}{\mbox{$L_{*}$}\xspace}
\newcommand{\lsol}{\mbox{$\it{L_\odot}$}\xspace}

\newcommand{\feh}{\mbox{$[\mathrm{Fe/H}]$}\xspace}
\newcommand{\mh}{\mbox{$[\mathrm{M/H}]$}\xspace}
\newcommand{\alfe}{\mbox{$[\mathrm{\alpha/Fe}]$}\xspace}

\begin{document}

   \title{TOI-4552 b: A new ultra-short period rocky world revealed by NIRPS and TESS}

\author{
Avidaan Srivastava\inst{1,*}\orcidlink{0009-0009-7136-1528},
Ren\'e Doyon\inst{1,2}\orcidlink{0000-0001-5485-4675},
Fran\c{c}ois Bouchy\inst{3}\orcidlink{0000-0002-7613-393X},
\'Etienne Artigau\inst{1,2}\orcidlink{0000-0003-3506-5667},
Charles Cadieux\inst{1}\orcidlink{0000-0001-9291-5555},
Nicole Gromek\inst{4}\orcidlink{0009-0000-1424-7694},
Elisa Delgado-Mena\inst{5,6}\orcidlink{0000-0003-4434-2195},
Yuri S. Messias\inst{1,7}\orcidlink{0000-0002-2425-801X},
Xavier Bonfils\inst{8}\orcidlink{0000-0001-9003-8894},
Roseane de Lima Gomes\inst{1,7}\orcidlink{0000-0002-2023-7641},
Susana C. C. Barros\inst{6,9}\orcidlink{0000-0003-2434-3625},
Bj\"orn Benneke\inst{10,1}\orcidlink{0000-0001-5578-1498},
Marta Bryan\inst{11},
Ryan Cloutier\inst{4}\orcidlink{0000-0001-5383-9393},
Nicolas B. Cowan\inst{12,13}\orcidlink{0000-0001-6129-5699},
Eduardo Cristo\inst{6},
Xavier Delfosse\inst{8}\orcidlink{0000-0001-5099-7978},
Xavier Dumusque\inst{3}\orcidlink{0000-0002-9332-2011},
David Ehrenreich\inst{3,14},
Jonay I. Gonz\'alez Hern\'andez\inst{15,16}\orcidlink{0000-0002-0264-7356},
David Lafreni\`ere\inst{1}\orcidlink{0000-0002-6780-4252},
Izan de Castro Le\~ao\inst{7}\orcidlink{0000-0001-5845-947X},
Christophe Lovis\inst{3}\orcidlink{0000-0001-7120-5837},
Alejandro Su\'arez Mascare\~no\inst{15,16}\orcidlink{0000-0002-3814-5323},
Bruno L. Canto Martins\inst{7}\orcidlink{0000-0001-5578-7400},
Jose Renan De Medeiros\inst{7}\orcidlink{0000-0001-8218-1586},
Lucile Mignon\inst{3,8},
Christoph Mordasini\inst{17}\orcidlink{0000-0002-1013-2811},
Francesco Pepe\inst{3}\orcidlink{0000-0002-9815-773X},
Rafael Rebolo\inst{15,16,18}\orcidlink{0000-0003-3767-7085},
Jason Rowe\inst{19},
Nuno C. Santos\inst{6,9}\orcidlink{0000-0003-4422-2919},
Damien S\'egransan\inst{3},
St\'ephane Udry\inst{3}\orcidlink{0000-0001-7576-6236},
Diana Valencia\inst{11}\orcidlink{0000-0003-3993-4030},
Gregg Wade\inst{20,21},
Jose Manuel Almenara\inst{8}\orcidlink{0000-0003-3208-9815},
Karen A. Collins\inst{22}\orcidlink{0000-0001-6588-9574},
Dennis M. Conti\inst{23}\orcidlink{0000-0003-2239-0567},
George Dransfield\inst{24},
Elsa Ducrot\inst{25,26}\orcidlink{0000-0002-7008-6888},
Zahra Essack\inst{27}\orcidlink{0000-0002-2482-0180},
Dasaev O. Fontinele\inst{7}\orcidlink{0000-0002-3916-6441},
Thierry Forveille\inst{8}\orcidlink{0000-0003-0536-4607},
Marziye Jafariyazani\inst{28}\orcidlink{0000-0001-8019-6661},
Pierrot Lamontagne\inst{1},
Alexandrine L'Heureux\inst{1}\orcidlink{0009-0005-6135-6769},
Khaled Al Moulla\inst{6,3}\orcidlink{0000-0002-3212-5778},
Ares Osborn\inst{8,4,29}\orcidlink{0000-0002-5899-7750},
L\'ena Parc\inst{3}\orcidlink{0000-0002-7382-1913},
David R. Rodriguez\inst{30}\orcidlink{0000-0003-1286-5231},
Richard P. Schwartz\inst{22}\orcidlink{0000-0001-8227-1027},
Madison G. Scott\inst{31}\orcidlink{0009-0006-3846-4558},
Avi Shporer\inst{32}\orcidlink{0000-0002-1839-3120},
Atanas K. Stefanov\inst{15,16}\orcidlink{0000-0002-6059-1178},
Mathilde Timmermans\inst{31,33}\orcidlink{0009-0008-2214-5039},
Amaury H.M.J. Triaud\inst{31}\orcidlink{0000-0002-5510-8751},
Joost P. Wardenier\inst{17,1}\orcidlink{0000-0003-3191-2486},
Drew Weisserman\inst{4}\orcidlink{0000-0002-7992-469X},
Sebasti\'an Z\'u\~niga-Fern\'andez\inst{33}\orcidlink{0000-0002-9350-830X}
}

\institute{
\inst{1}Institut Trottier de recherche sur les exoplan\`etes, D\'epartement de Physique, Universit\'e de Montr\'eal, Montr\'eal, Qu\'ebec, Canada\\
\inst{2}Observatoire du Mont-M\'egantic, Qu\'ebec, Canada\\
\inst{3}Observatoire de Gen\`eve, D\'epartement d’Astronomie, Universit\'e de Gen\`eve, Chemin Pegasi 51, 1290 Versoix, Switzerland\\
\inst{4}Department of Physics \& Astronomy, McMaster University, 1280 Main St W, Hamilton, ON, L8S 4L8, Canada\\
\inst{5}Centro de Astrobiolog\'ia (CAB), CSIC-INTA, Camino Bajo del Castillo s/n, 28692, Villanueva de la Ca\~nada (Madrid), Spain\\
\inst{6}Instituto de Astrof\'isica e Ci\^encias do Espa\c{c}o, Universidade do Porto, CAUP, Rua das Estrelas, 4150-762 Porto, Portugal\\
\inst{7}Departamento de F\'isica Te\'orica e Experimental, Universidade Federal do Rio Grande do Norte, Campus Universit\'ario, Natal, RN, 59072-970, Brazil\\
\inst{8}Univ. Grenoble Alpes, CNRS, IPAG, F-38000 Grenoble, France\\
\inst{9}Departamento de F\'isica e Astronomia, Faculdade de Ci\^encias, Universidade do Porto, Rua do Campo Alegre, 4169-007 Porto, Portugal\\
\inst{10}Department of Earth, Planetary, and Space Sciences, University of California, Los Angeles, CA 90095, USA\\
\inst{11}Department of Physics, University of Toronto, Toronto, ON M5S 3H4, Canada\\
\inst{12}Department of Physics, McGill University, 3600 rue University, Montr\'eal, QC, H3A 2T8, Canada\\
\inst{13}Department of Earth \& Planetary Sciences, McGill University, 3450 rue University, Montr\'eal, QC, H3A 0E8, Canada\\
\inst{14}Centre Vie dans l’Univers, Facult\'e des sciences de l’Universit\'e de Gen\`eve, Quai Ernest-Ansermet 30, 1205 Geneva, Switzerland\\
\inst{15}Instituto de Astrof\'isica de Canarias (IAC), Calle V\'ia L\'actea s/n, 38205 La Laguna, Tenerife, Spain\\
\inst{16}Departamento de Astrof\'isica, Universidad de La Laguna (ULL), 38206 La Laguna, Tenerife, Spain\\
\inst{17}Space Research and Planetary Sciences, Physics Institute, University of Bern, Gesellschaftsstrasse 6, 3012 Bern, Switzerland\\
\inst{18}Consejo Superior de Investigaciones Cient\'ificas (CSIC), E-28006 Madrid, Spain\\
\inst{19}Bishop's University, Dept of Physics and Astronomy, Johnson-104E, 2600 College Street, Sherbrooke, QC, Canada, J1M 1Z7, Canada\\
\inst{20}Department of Physics, Engineering Physics, and Astronomy, Queen’s University, 99 University Avenue, Kingston, ON K7L 3N6, Canada\\
\inst{21}Department of Physics and Space Science, Royal Military College of Canada, 13 General Crerar Cres., Kingston, ON K7P 2M3, Canada\\
\inst{22}Center for astrophysics $\vert$ Harvard \& Smithsonian, 60 Garden Street, Cambridge, MA 02138, USA\\
\inst{23}American Association of Variable Star Observers, Cambridge, USA\\
\inst{24}Department of Physics, University of Oxford, Oxford OX13RH, UK\\
\inst{25}LESIA, Observatoire de Paris, CNRS, Universit\'e Paris Diderot, Universite Pierre et Marie Curie, 5 place Jules Janssen, 92190 Meudon, France\\
\inst{26}AIM, CEA, CNRS, Universit\'e Paris-Saclay, Universite de Paris, F-91191 Gif-sur-Yvette, France\\
\inst{27}Department of Physics and Astronomy, The University of New Mexico, 210 Yale Blvd NE, Albuquerque, NM 87106, USA\\
\inst{28}SETI Institute, Mountain View, CA 94043, USA NASA Ames Research Center, Moffett Field, CA 94035, USA\\
\inst{29}Department of Physics, The University of Warwick, Gibbet Hill Road, Coventry, CV4 7AL, UK\\
\inst{30}Space Telescope Science Institute, 3700 San Martin Drive, Baltimore, MD, 21218, USA\\
\inst{31}School of Physics \& Astronomy, University of Birmingham, Edgbaston, Birmingham B15 2TT, United Kingdom\\
\inst{32}Department of Physics and Kavli Institute for Astrophysics and Space Research, Massachusetts Institute of Technology, Cambridge, MA 02139, USA\\
\inst{33}Astrobiology Research Unit, Universit\'e de Li\`ege, 19C All\'ee du 6 Ao\^ut, 4000 Li\`ege, Belgium\\
\inst{*}\email{avidaan.srivastava@umontreal.ca}
}

   \date{Received December 23, 2025; accepted March 16, 2025}

\abstract
   {A particularly intriguing subclass of rocky exoplanets are the ultra-short period (USP) worlds that orbit their host stars in less than a day. These planets are particularly rare around M dwarf stars, with so far only ten that have a constrained mass and radius.}
   {We present the validation and characterization of the ultra-short period (0.3\,days), Earth-sized planet TOI-4552\,b orbiting a nearby (27.26-pc away) M4.5V dwarf.}
   {Complementing the TESS photometry, ground-based transit observations from LCO, ExTrA and SPECULOOS validated the planetary radius and cleared the field of any contaminants. Speckle imaging with Zorro (Gemini-S) rules out false positive scenarios caused by eclipsing binary sources. Spectroscopic observations with NIRPS and HARPS were used to obtain stellar abundances, constrain the planetary mass, and, in conjunction with the transit observations, estimate the orbital parameters.}
   {TOI-4552 is a quiet star exhibiting no short-term stellar variations seen in photometric or radial velocity data that can be associated to stellar rotation. Long-term photometric data from ASAS-SN also suggests a lack of activity signals. TOI-4552\,b ($M_{\rm p}=1.83\pm0.47\,M_{\oplus}$, $R_{\rm p}=1.11\pm0.04\,R_{\oplus}$) lies between the Earth-like and iron-rich composition tracks on the Mass-Radius diagram. The \texttt{exopie} interior structure model, without constraints from refractory abundance ratio, yields a core mass fraction (CMF) of 0.54$^{+0.17}_{-0.25}$ and a bulk density of 7.74\,g/cm$^3$. Since the CMF spans a wide range due to the large uncertainty on the mass, the definitive interior composition cannot be determined with the current dataset.}
   {TOI-4552\,b hints as being marginally more iron-rich compared to the Earth but confirmation of its status requires additional, precise radial velocity measurements. Combined with its high emission spectroscopic metric (ESM = 19.5), negligible stellar activity and short orbital period, TOI-4552\,b emerges as a compelling target for atmospheric and surface composition studies with JWST.}


\maketitle

\section{Introduction} \label{sec:intro}

The Transiting Exoplanet Survey Satellite (TESS; \citealt{TESS}) mission has been at the forefront of discovering new worlds in the solar neighbourhood via the transit method. As of 2025, more than 7000 planetary candidates have been flagged in TESS datasets around a varied population of stars. These stars include the M dwarfs, cooler (\teff < 4000\,K) and less massive (\mstar = 0.1 -- 0.7\,M$_{\odot}$) than the Sun but more numerous in our galaxy \citep[e.g.,][]{Chabrier2001, Bochanski2010}. Their smaller size facilitates larger signals for transiting planets and their low mass allows for the detection of smaller planets around them. All these factors make exoplanets easier to detect around such stars, which are also known to host a large population of rocky planets \citep{ Bonfils2013,Dressing2015,Gillis2026}. However, it is important to note that many M dwarfs exhibit substantial stellar activity~\citep[e.g.,][] {Galletta2025,Rajpurohit2025,Mignon2023,Roettenbacher2017}, which can significantly hinder planet detection. As a result, while M dwarfs offer clear advantages for exoplanet searches, their intrinsic activity can also present notable observational challenges.

Due to their lower effective temperatures, M-type stars are brighter in the near infrared (NIR) as opposed to the visible (VIS) wavelength band. To extract the maximum amount of spectral information from these stars, a new state-of-the-art instrument, Near Infra-Red Planet Searcher (NIRPS,~\citealt{Bouchy2025, Bouchy2017}), has been developed. NIRPS is installed on the 3.6-m telescope at La Silla Observatory in Chile and works alongside HARPS~\citep{HARPS}. As part of the Guaranteed Time Observations (GTO) sub-programme 2 (SP2) of NIRPS, we follow up TESS transiting candidates to confirm planetary detections and measure their masses. Measurement of both the mass and radius of a planet is crucial to place meaningful constraints on its interior and atmospheric composition through the bulk density.

One particularly interesting subclass of rocky planets are the ultra-short period (USP) worlds, uniquely identified by their orbital period of less than 1 day. While some, such as 55~Cancri\,e \citep{Dawson2010, McArthur2004}, have been studied extensively, their rarity makes it difficult to conduct a wider study. Only 10 USPs orbiting M dwarfs have published masses and radii \citep{Lee2025, toi6255, wolf327, toi1685_1, gj806, Essack2023, gj367, toi1685_2, toi1634, toi1685_3_toi1634, ltt3780, gj1252}. Such a small dataset is not enough to develop a complete understanding about the nature of these worlds. James Webb Space Telescope (JWST; \citealt{jwst, jwst2}) observations have revealed that several USPs are barren worlds lacking substantial atmospheres \citep[e.g.,][]{Kreidberg2019, Luque2025}, likely stripped by the intense stellar irradiation they endure ($>200\,S_{\oplus}$). As a result, interior structure models can be run without atmospheric degeneracies. Since previous studies \citep{Plotnykov2020, Brinkman2024} have concluded that rocky planets exhibit a diverse range of compositions, it is only by expanding the sample size that we can identify any trends and conduct population studies. 

Beyond their compositional significance, USPs offer unique opportunities for surface geological studies with JWST. Their short orbital periods make them amenable to short and repeated observations, and their high irradiation can drive detectable phase-curve signals. Phase-curve and secondary eclipse measurements can constrain albedos, day–night temperature contrasts, and even surface mineralogy in the absence of an atmosphere (e.g., \citealt{Hu2012, Paragas2025}).

We present the discovery and characterization of a USP, TOI-4552\,b, as part of the NIRPS GTO SP2. The article is structured as follows: Section~\ref{sec:obs} discusses the various facilities (TESS, ExTrA, SPECULOOS, LCO, Zorro, HARPS+NIRPS), both space- and ground-based, used to observe the target star and construct our dataset. Section~\ref{sec:stellar_char} covers the various methods used to characterize the star using photometry and HARPS+NIRPS spectroscopy. In Section~\ref{sec:final_analysis} we present our global photometry and RV model and analysis of the results. Section~\ref{sec:discussion} dives deeper into the results and discusses the interior structure modelling and future scientific relevance of TOI-4552\,b. Finally, in Section~\ref{sec:conclusion} we summarise the article and present its conclusions.

\section{Observations} \label{sec:obs}

\subsection{TESS photometry} \label{sec:phot_tess}

\begin{figure}[h]
    \centering
    \includegraphics[width=1\linewidth]{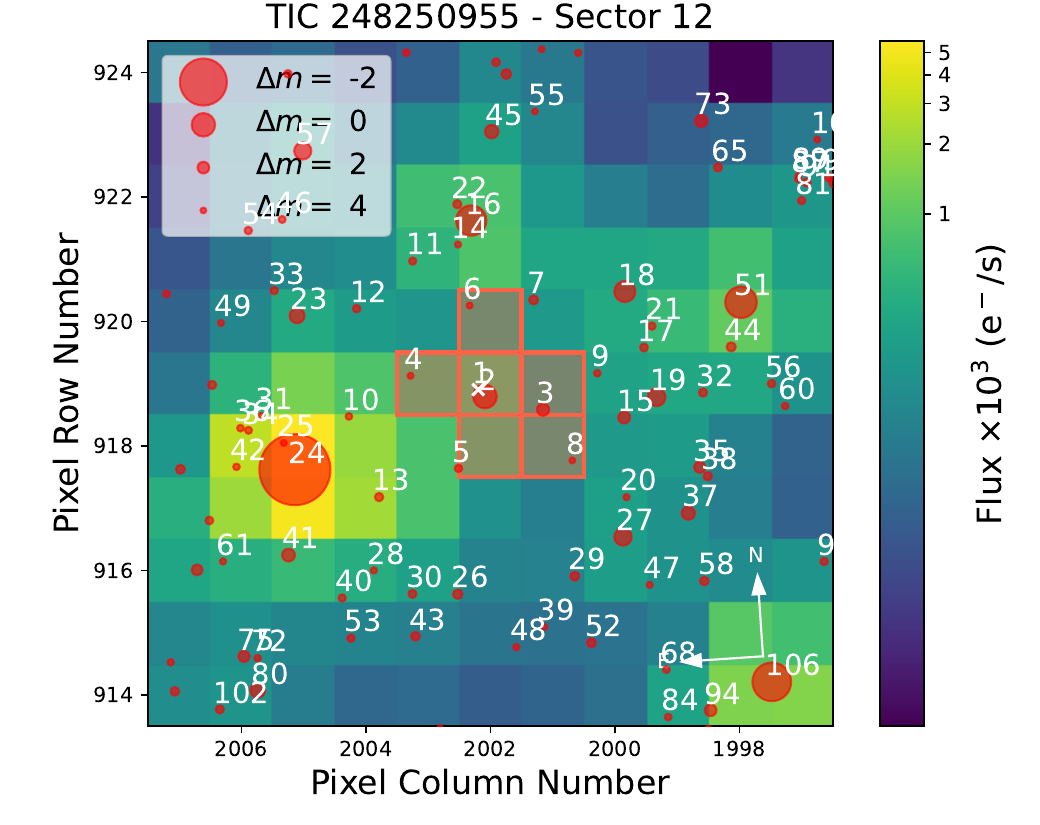}
    \caption{Target pixel file (TPF) of TOI-4552 for TESS sector 12, created using \texttt{tpfplotter} \citep{Aller2020}. The pixels (21\arcsec\ pixel scale) outlined in orange correspond to the aperture mask used to extract the SAP flux for the lightcurve. All Gaia DR3 \citep{GaiaDR3} sources in the field are numbered, with ``1" marking TOI-4552. The size of the red circles is representative of the TESS magnitude of the stars relative to TOI-4552. Since it is a crowded field, TESS SPOC provides lightcurves corrected for the dilution caused by the strongest contaminants. Fig.~\ref{tesstpf_dil} shows the TPF of all three sectors with the primary contaminants.}
    \label{tess_tpf12}
\end{figure}

\begin{figure*}
    \centering
    \includegraphics[width=1\linewidth]{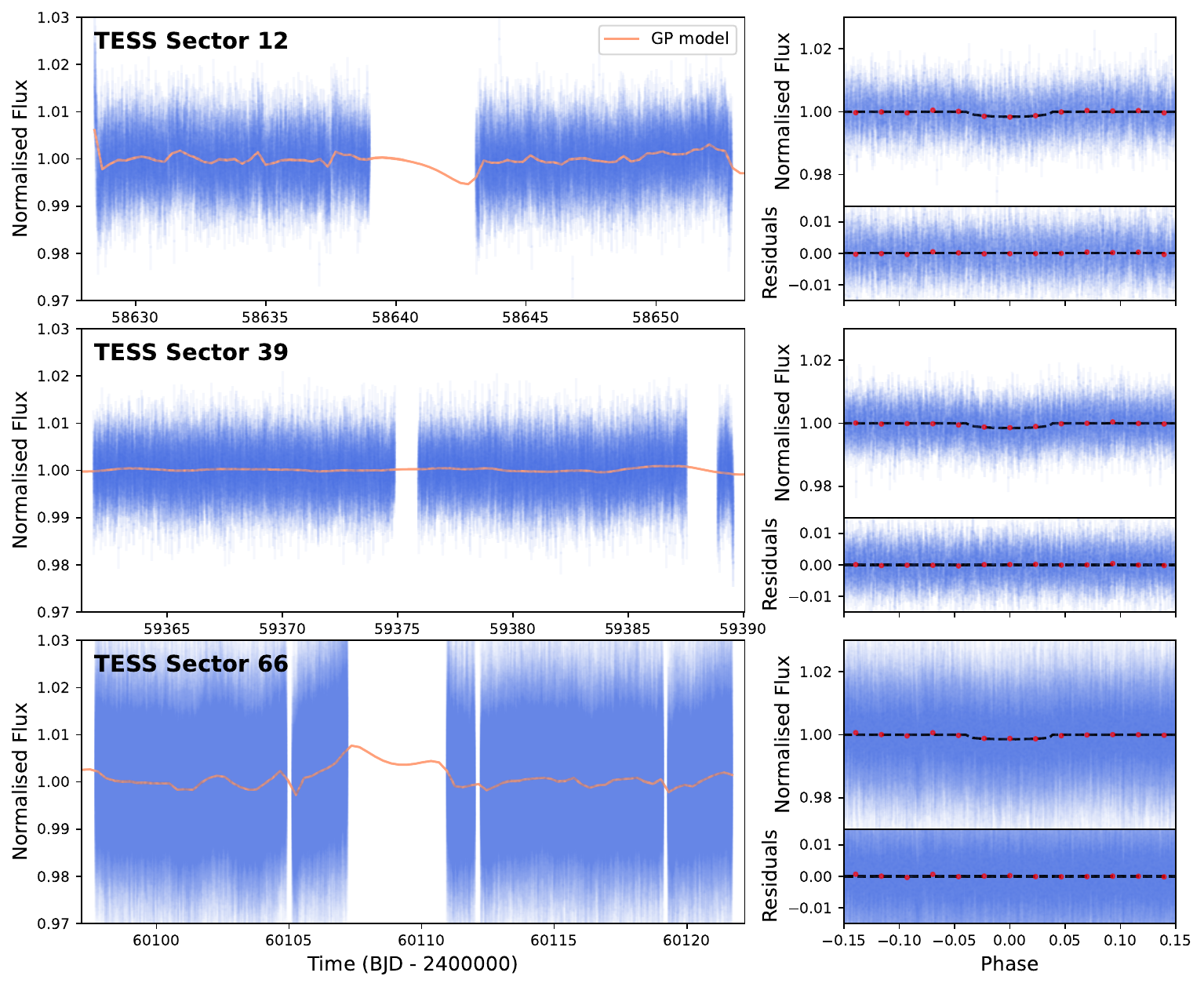}
    \caption{TESS transit lightcurves from sectors 12, 39 and 66. The normalised flux (blue points) as well as the gaussian process (GP) model (orange) used to detrend the lightcurve is plotted on the left. The phase-folded lightcurves centred on the transit of TOI-4552\,b, transit model (black dashed line), binned flux (red points) and the residuals after fitting for the transit are plotted on the right, for each sector. Sectors 12 and 39 were observed by TESS at a 2\,minute cadence, whereas Sector 66 was observed at a 20\,second cadence. Since TOI-4552\,b is a USP, we choose to keep the high cadence of Sector 66 to provide a better sampling at the expense of lower precision.}
    \label{transit_phasefold}
\end{figure*}

TOI-4552\,b (TIC~248250955) was first alerted as a Level 1 planetary candidate \citep{Guerrero2021} on October 28, 2021 by the TESS Science Processing Operations Center (SPOC, \citealt{Jenkins2016}) pipeline at NASA Ames Research Centre on the publicly available TESS data webpage\footnote{\url{https://tev.mit.edu/}}. It has been observed in TESS sectors 12 (year 1), 39 (year 3) and 66 (year 5), each with a 120-second exposure time as well as a 20-second exposure time for sector 66. The target was imaged on CCD 2 of camera 1 for all three sectors. Owing to he short orbital period of 0.3011\,days, each sector contains approximately 70 transits.

Since the TESS pixels are relatively large (21\arcsec\ on sky), the retrieved stellar flux of its targets are often contaminated by a neighbour. \texttt{tpfplotter} \citep{Aller2020} was used to plot the target pixel file (TPF) for Sector 12 (Fig.~\ref{tess_tpf12}) consisting of neighbouring sources from the Gaia DR3 \citep{GaiaDR3}. The TPF highlights the pixels used for the aperture mask to obtain the Simple Aperture Photometry (SAP; \citealt{Twicken2010,Morris2020}) flux in orange. Due to the crowded field, we used \texttt{TESS-cont} \citep{Castro2024} to identify the strongest sources of contamination. These stars in along with their angular separation from TOI-4552 and relative percentage of contamination are: TIC~248250924 (63\arcsec; 17\%), TIC~248250937 (18\arcsec; 7\%), TIC~248251013 (58\arcsec; 2\%), TIC~248250948 (4\arcsec; 1.2\%) and TIC~248250933 (44\arcsec; 1\%). The TPF for all three sectors highlighting these contaminants are presented in Fig.~\ref{tesstpf_dil}.

As per the Data Validation Report (DVR; \citealt{Twicken2018}) produced on October 30, 2023, which presents a joint analysis of the target for all three sectors, the lightcurves were corrected for dilution and crowding effected by a factor in agreement with the aforementioned sources. Additionally, the Centroid Test and Eclipsing Binary Discrimination Test were conducted for validation and ruled in favour of TOI-4552 as the host of the planetary candidate. A Signal-to-Noise ratio (SNR) of 11.8 was reported in the DVR for the fit, denoting that the candidate is likely a planet (SNR\,>\,10). Our analysis uses the normalised Presearch Data Conditioning Simple Aperture Photometry (PDCSAP; \citealt{Smith2012,Stumpe2012,Stumpe2014}) flux which was corrected for dilution in the TESS aperture by known contaminating sources as well as for instrumental systematics. Fig.~\ref{transit_phasefold} showcases the three TESS sectors phase folded at the period and time of conjunction of TOI-4552\,b.  \\

The large 21\arcsec\ pixel scale of TESS can also create some ambiguity regarding which star is hosting the transiting planet candidate if multiple stars are located inside the aperture mask. Since the TESS pixel centred on TOI-4552 shows at least one apparent Gaia-detected contaminant, follow-up with seeing-limited ground-based photometry ($\sim$1–2\arcsec) and diffraction-limited high-angular-resolution imaging is required to attribute the transit event to the correct star and vet false positives from other eclipsing binaries in the aperture mask. Several facilities around the world contribute to this effort under the TESS Follow-up Observing program (TFOP\footnote{\url{https://tess.mit.edu/followup/}}) Working Group Sub-Group 1 (SG1; \citealt{Collins2019}).

\subsection{Ground-based transit confirmations: ExTrA} \label{sec:extra}

ExTrA \citep{Bonfils2015} is a near-infrared (0.85 to 1.55~$\mu$m), multi-object spectrograph fed by three F/8 telescopes each with 60~cm primary mirrors. We used ExTrA to intensely monitor TOI-4552 from 2022 to 2024 over twenty-six individual nights as part of TFOP SG1 using the low resolution (R$\sim$20) mode. On nine of those nights, all three telescopes observed the transit, two telescopes observed the transit on sixteen nights and for one night only one telescope observed the transit totalling 61 full/partial transits used in our analysis (Fig.~\ref{transit-ground}). The data reduction and detrending was performed using the methodology described in \citet{Cointepas2021}. The closest contaminant star, TIC 248250948, lies 3.94\arcsec\ from the target, so we adopted a 3.86\arcsec\ aperture to minimise dilution. The typical seeing during the observations was about 1.3\arcsec, which ensured that most of the stellar flux remained within the chosen aperture. A 1.2\% dilution correction was then applied during the lightcurve extraction and detrending level to remove any residual contamination from nearby sources.

\subsection{Ground-based transit confirmations: LCOGT}\label{sec:lco}

The Las Cumbres Observatory Global Telescope Network (LCOGT; \citealt{Brown2013}) is composed of robotic 1.0~m class telescopes distributed across multiple observatories worldwide and is regularly used for TFOP SG1 follow-up. As part of this program, TOI-4552 was monitored in 2022, yielding four full transits from the Cerro Tololo Inter-American Observatory (CTIO, Chile) and two full transits from the South African Astronomical Observatory (SAAO, South Africa), all in the Sloan-$i'$ photometric bandpass. All data were calibrated using the standard \texttt{BANZAI} pipeline \citep{banzai} and differential photometric data were extracted using {\tt AstroImageJ} \citep{Collins:2017}. The typical photometric aperture radius for all observations was 2.7\arcsec. Fig. \ref{transit-ground} shows the detrended, combined, and phase-folded transit lightcurves used in our joint analysis.

\subsection{Ground-based transit confirmations: SPECULOOS}\label{sec:speculoos}

The Search for habitable Planets EClipsing ULtra-cOOl Stars (SPECULOOS; \citealt{Delrez2018}) southern observatory consists of four robotic 1.0~m telescopes (Europa, Ganymede, Io, Callisto) located at the European Southern Observatory (ESO) Paranal site, suitable for transit observations of late M-type stars such as TOI-4552. Thirteen transits were recorded from 2022 to 2024 with the Ganymede, Europa and Io telescopes in various combination of the Sloan-$g'$, $r'$ and $z'$ filters totalling 20 individual transits used in our analysis. The \texttt{allesfitter} python package \citep{Gunther2021} was used to detrend the transit lightcurves (Fig.~\ref{transit-ground}) following the procedure similar to \citealt{Dransfield2024} and \citealt{Henderson2024}.

\subsection{High resolution imaging: Zorro} \label{sec:gemini}

\begin{figure}[]
    \centering
    \includegraphics[width=1\linewidth]{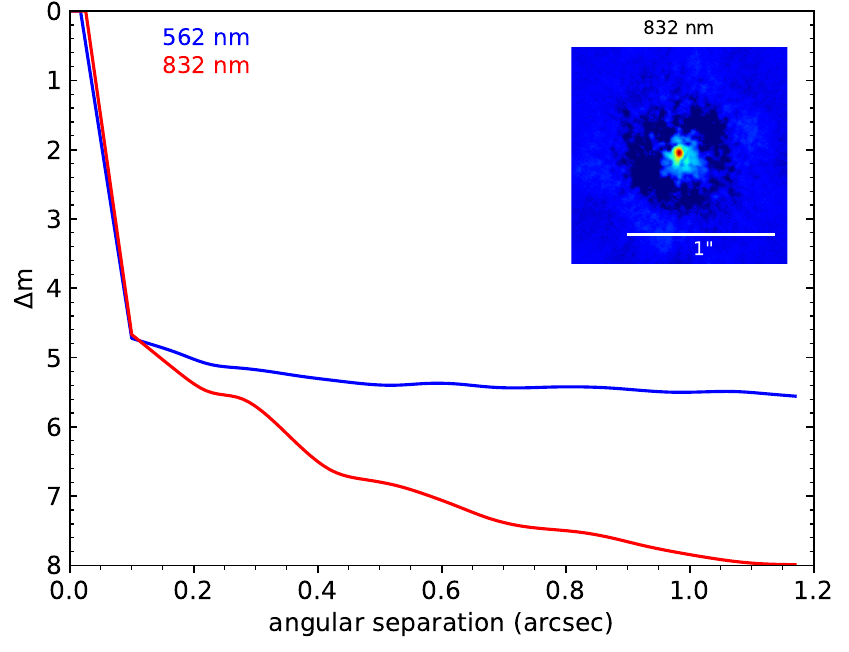}
    \caption{High contrast image of TOI-4552 using Zorro at Gemini-S \citep{Scott2021, Howell2025} in the two simultaneous channels (562-nm and 832-nm) as a function of on-sky separation. Four nights of observations were combined here. The 5-$\sigma$ contrast curve shows no evidence for neighbours or companions contaminating the signal.}
    \label{fig:gemini}
\end{figure}

There may exist additional gravitationally bound or unbound sources, unresolved in Gaia DR3, that may contaminate the derived planetary and stellar parameters. To assess the impact of such sources (if any) and further rule out false positive scenarios, as is the standard procedure for TESS objects of interest (TOIs), we observed TOI-4552 with the Zorro imager \citep{Scott2021} located on the 8.1-m Gemini-South telescope in Chile. Zorro is a fast, dual-channel speckle imager that operates at the diffraction limit for a 6.7\arcsec\ field of view. Observations were made on four individual nights: 2022-05-17, 2022-07-28, 2022-07-30, and 2023-04-08, using the 562~nm and 832~nm bands simultaneously. The techniques and procedures for observation, data reduction and image reconstruction are described in \citep{Howell2025}. Fig.~\ref{fig:gemini} depicts the relative 5-$\sigma$ contrast as a function of on-sky separation; no contaminants were detected within the contrast limit of $\Delta m\lesssim 5$ beyond 0.1\arcsec.

\subsection{NIRPS velocimetry} \label{sec:NIRPS_rv}

TOI-4552 was monitored simultaneously by NIRPS and HARPS from 2023 to 2024 for a total of 60 individual nights. These observations were carried out as part of the NIRPS GTO SP2 corresponding to ground-based RV follow-up of TESS planetary candidates. We used the High Efficiency instrument mode ($\lambda/\Delta \lambda \sim 75,000$) which feeds the spectrograph through a 0.9\arcsec\ fibre centred on the star and a 0.4\arcsec\ fibre observing the simultaneous sky background calibration at a separation of 39\arcsec\ on the sky from the star. The larger science fibre allows for a higher instrument throughput, important for observing faint stars such as TOI-4552. The exposure time per observation was 900 seconds and two consecutive exposures were done every night, except for one, resulting in a total of 119 individual science frames. Because the orbital period (0.3011\,days) is so short, the nightly consecutive observations exhibit a slight but significant phase difference ($\sim7\%$); we therefore choose not to bin the data.

The raw data used in our analysis was processed via the ESO supported NIRPS DRS-3.2.0 \citep{Bouchy2025}. A key challenge in near-infrared velocimetry is contamination from telluric absorption and emission lines introduced by Earth’s atmosphere. Although the NIRPS DRS applies a correction for these features~\citep{Allart2022}, the process is not always perfect. In particular, at times when the systemic velocity of the star ($v_{sys}$, see Table \ref{tab:StellarParameters}) is close to the barycentric Earth radial velocity (BERV), the stellar and telluric lines can overlap, making the correction less effective. If this overlap occurs at multiple epochs, the RV timeseries is offset and imprinted with the harmonics of the Earth's orbital period (365.24\,days). This `BERV overlap' phenomenon and ways to mitigate it are explored in detail in more recent works (\citealt{Frensch2026}; \citealt{Parc2025}; Srivastava et al. in review at A\&A) as well as in Section \ref{sec:rv_analysis}. Alternatively, the NIRPS spectra can be extracted and processed using the \texttt{APERO} framework~\citep{Cook2022} which yielded results consistent with NIRPS DRS-3.2.0, therefore we choose to use the NIRPS DRS-3.2.0 data for our analysis.

The final RV extraction uses the line-by-line (LBL; v0.65.001) framework \citep{Artigau2022} which has been demonstrated to work effectively in the NIR \citep[e.g.,][]{ASM2025, Donati2025}. Since TOI-4552 is faint, creating a high-SNR template for the RV calculation inside the LBL algorithm would be challenging; we therefore instead constructed the template from 111 NIRPS observations of GJ~643, which is much brighter ($V$ = 11.73, $J$ = 7.55) and similar in spectral type (M4.5V) to TOI-4552. The LBL reported median RV uncertainty per-exposure ($\overline{\sigma_{RV}}$) for TOI-4552 is 7.7\,m/s with a root-mean-square (RMS) of 9.2\,m/s. A total of nine exposures of TOI-4552 acquired on 2023-04-15, 2023-04-17, 2024-07-29, 2024-08-23, 2024-09-19, 2024-10-12 and 2024-10-24 were excluded from the final analysis, corresponding to low SNR outliers ($\sigma_{RV}>14$\,m/s) attributed to poor observing conditions.

\subsection{HARPS velocimetry} \label{sec:HARPS_rv}

By design, NIRPS observations are simultaneously accompanied by HARPS, together providing a full coverage of the visible (VIS) and near-infrared (NIR) wavelength ranges. An instrumental problem with the HARPS-EGGS mode fibre shutter forced us to split the 59 total HARPS measurements between its  EGGS (37 observations, $\lambda/\Delta \lambda = 80,000$) and HAM (22 observations, $\lambda/\Delta \lambda = 115,000$) instrument modes. Due to the high magnitude ($V=14.44$) and a late stellar type (M4.5V), HARPS is unable to produce high quality radial-velocity measurements. We used the standard template matching algorithm \texttt{sBART}~\citep{SBART}\footnote{\url{https://github.com/iastro-pt/sBART}} to obtain the RVs, as it has been optimised and tested extensively on HARPS datasets. For EGGS mode observations $\overline{\sigma_{RV}}$ is 9.5\,m/s with a RMS of 27.9\,m/s, while the HAM mode had uncertainties three times larger. Therefore, we only use the HARPS-EGGS observations in our final analysis. The observations from the nights of 2023-04-15 and 2023-04-17 were rejected because the corresponding nights were flagged and removed from the NIRPS timeseries. The final RV timeseries for both NIRPS and HARPS is provided in Table~\ref{tab:RV_timeseries}.

\section{Stellar Characterization of TOI-4552} \label{sec:stellar_char}

\begin{table}[t]
\small
\caption{Stellar parameters for TOI-4552.}
\centering
\renewcommand{\arraystretch}{1.2}
\begin{tabular}{lcc}
\hline\hline
  & \textbf{TOI-4552}  & Source \\
    \hline
    \textbf{Identifiers}& &\\
    TIC ID  & 248250955 & TICv8.2\\
    2MASS ID & J17385163-4738056 &  2MASS\\
    Gaia ID & 5948579462188230144 & Gaia DR3\\
    \hline
    \textbf{Astrometric parameters} & &\\
    Right ascension (J2015.5),  $\alpha$ & 17$^{\mathrm{h}}$ 38$^{\mathrm{m}}$ 51.42$^{\mathrm{s}}$ &Gaia DR3\\
    Declination (J2015.5), $\delta$ & -47$^{\circ}$ 38$^{\prime}$ 13.36$^{\prime \prime}$&Gaia DR3\\
    Parallax (mas) & 36.556 $\pm$ 0.017 &Gaia DR3\\
    Distance (pc) & 27.38$\pm$0.01 &Gaia DR3\\
    $\mu_{\rm{R.A.}}$ (mas yr$^{-1}$) &-134.4654 $\pm$ 0.0185 &Gaia DR3\\
    $\mu_{\rm{Dec}}$ (mas yr$^{-1}$) &-472.5121 $\pm$ 0.0118 &Gaia DR3\\
    $v_{sys}$ (km s$^{-1}$) &-26.2162 $\pm$ 0.9744 &Gaia DR3\\
    $U^{(\theta)}$ (km s$^{-1}$) &-44.918 $\pm$ 0.923 &This work\\
    $V^{(\theta)}$ (km s$^{-1}$) &-50.718 $\pm$ 0.277 &This work\\
    $W^{(\theta)}$ (km s$^{-1}$) &-12.703 $\pm$ 0.146 &This work\\
    \hline
    \textbf{Photometric parameters} & &\\
    TESS (mag) & 11.9251 $\pm$ 0.0073 & TICv8.2\\
    \textit{V} (mag) &14.44 $\pm$ 0.028& TICv8.2\\
    \textit{G} (mag) &13.2136 $\pm$ 0.0004& Gaia DR3\\
    \textit{J} (mag) &10.256 $\pm$ 0.021 & 2MASS\\
    \textit{H} (mag) &9.659 $\pm$ 0.019 & 2MASS\\
    $K_s$ (mag) &9.395 $\pm$ 0.019 & 2MASS\\
    \hline
    \textbf{Bulk parameters} & &\\
    Spectral type & M4.5V & This work \\
    \teff\,(K) &  3258 $\pm$ 115 & This work \\
    $R_\star$  (\rsol)  &  0.2869 $\pm$ 0.0088 & This work \\
    $M_\star$  (\msol)  &  0.2619 $\pm$ 0.0063  & This work \\
    $\rho_\star$  (\gccc)  &  15.803$^{+1.057}_{-1.302}$  & This work \\
    $L_\star$  (\lsol)  &   0.00836 $\pm$ 0.0013  &This work\\
    \feh (dex) &  -0.07 $\pm$ 0.09 &This work\\
    \mh (dex) &  -0.01 $\pm$ 0.07 &This work\\
    \alfe (dex) &  0.06 $\pm$ 0.12 &This work\\
    Fe/Mg (mol. frac.) &  0.91$^{+0.48}_{-0.31}$ &This work\\
    Mg/Si (mol. frac.) &  0.62$^{+0.69}_{-0.33}$ &This work\\
    Si/O (mol. frac.) &  0.10$^{+0.1}_{-0.05}$ &This work\\
    $\log g_*$ (cm\,s$^{-2}$) &  4.940 $\pm$ 0.028  &This work\\
    Macroturbulance (km s$^{-1}$) & 4 $\pm$ 2.25 & This work\\
    Microturbulance (km s$^{-1}$) & 0.75 $\pm$ 0.5 & This work\\
   \hline
\end{tabular}
\\
\begin{tablenotes}
\item Sources: TICv8 \citep{Stassun2019}, 2MASS \citep{2MASS2006}, Gaia DR3 \citep{GaiaDR3}. $^{(\theta)}$ The values for the galactic space velocities (U, V and W) were calculated using the default function of the \texttt{BANYAN $\Sigma$} algorithm~\citep{Gagne2018} using parameters from Gaia DR3. The same algorithm concluded that TOI-4552 is a field star with 99.9\% probability.
\end{tablenotes}
\label{tab:StellarParameters}
\end{table}

\subsection{Empirically derived stellar parameters}

We derived the stellar radius of TOI-4552 using its empirical relation with the $K_s$ absolute magnitude from \citet{Mann2015} for M dwarfs, adopting the $K_s$ magnitudes reported by 2MASS \citep{2MASS2006} and the distance from Gaia DR3 parallax measurements \citep{GaiaDR3}. This yields a radius of $\rstar\,=\,0.2869\pm0.0088\,R_{\odot}$. Similarly, the stellar mass was estimated using the empirical $\mstar - K_s$ relation from \citet{Mann2019}, giving $\mstar\,=\,0.2619\pm0.0063\,M_{\odot}$. From these values, we computed the surface gravity ($\logg \,=\,4.940 \pm 0.028\,$\,cm s$^{-2}$) and stellar luminosity ($\lstar\,=\,0.00836 \pm 0.0013\,L_{\odot}$). The effective temperature ($\teff = 3258 \pm 115$\,K) used in the luminosity calculation was obtained through spectroscopic analysis combined with SED fitting (see Sections~\ref{sec:spec_stellar}, \ref{sec:sed_fit}). All uncertainties were propagated using a Monte Carlo approach, and the final stellar bulk parameters are summarize in Table~\ref{tab:StellarParameters}.

\subsection{Spectral energy distribution fit} \label{sec:sed_fit}

\begin{figure}[]
  \centering
    \includegraphics[width=0.48\textwidth]{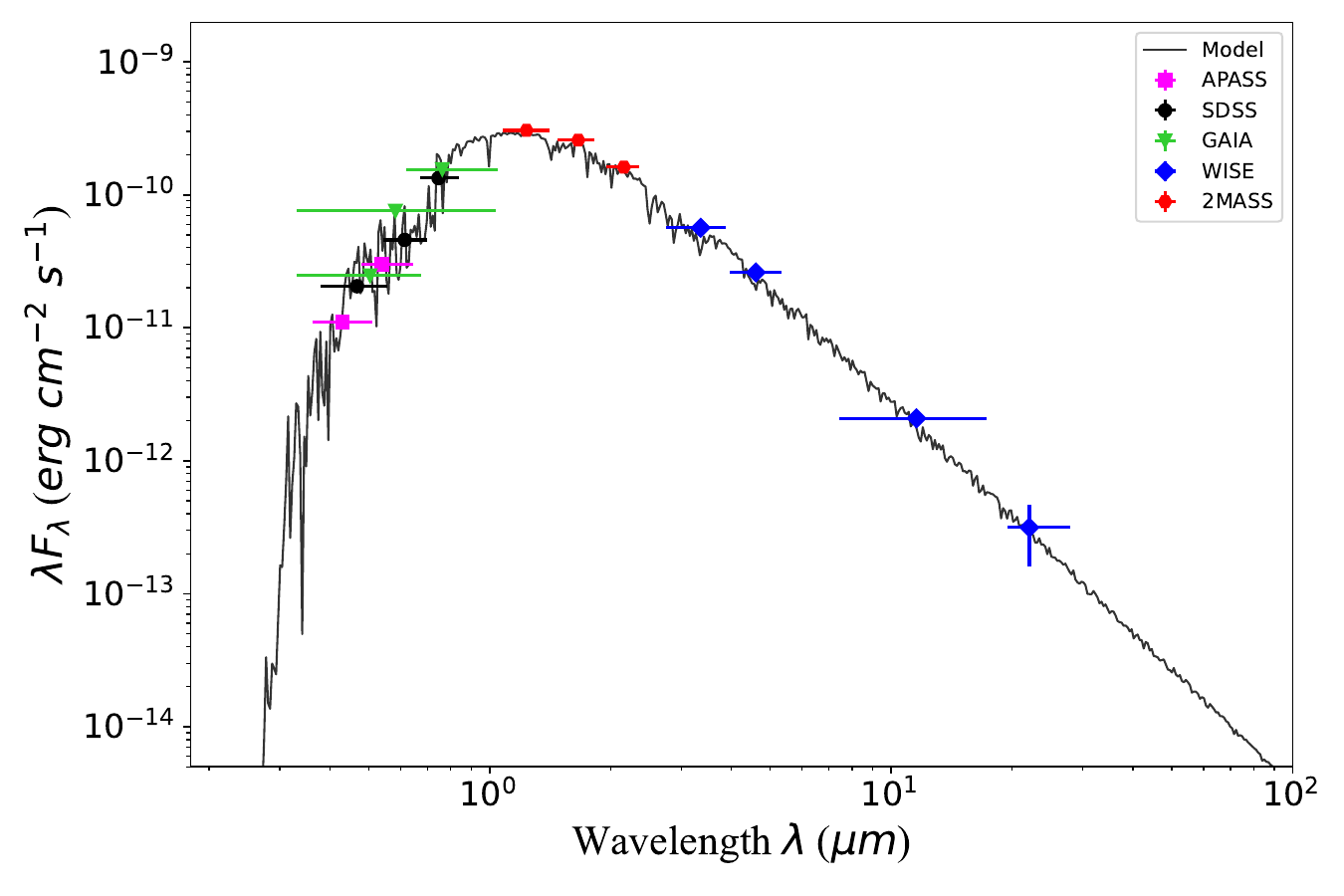}
  \caption{Spectral Energy Distribution (SED) of TOI-4552.
  The best-fitting BT-Settl model (\citealt{Allard2012}) for parameters \teff\,=\,3300\,K, [M/H]\,=\,0\,dex, and $\log\, g_*$\,=\,5.0\,dex is shown by the black curve.
  Photometric data from APASS (magenta), Gaia (green), 2MASS (red), SDSS (black), and WISE (blue) are plotted with horizontal error bars indicating the filter passbands.} 
  \label{fig:SED}  
\end{figure}

Stellar parameters were also derived by fitting a Spectral Energy Distribution (SED) to multi-band photometry using the Virtual Observatory Spectral Analyzer (VOSA) tool \citep{Bayo2008}. The observational data included fluxes from the \textit{GBP}, \textit{G}, and \textit{GRP} bands from the Gaia mission \citep{Gaia2018}, \textit{B} and \textit{V} from APASS \citep{Henden2015}, \textit{g'}, \textit{r'} and \textit{i'} from SDSS Catalogue, release 12 \citep{Shadab2015}, \textit{J}, \textit{H} and \textit{Ks} from 2MASS \citep{2MASS2006} and \textit{W1}, \textit{W2}, \textit{W3}, and \textit{W4} from the WISE mission \citep{WISE2010}. VOSA performs a $\chi^2$ minimization between the observational data and synthetic SEDs from theoretical models, including BT-Settl \citep{Allard2012}, Kurucz \citep{Kurucz1993}, and Castelli \& Kurucz \citep{Castelli2003}. The model that provided the best fit was the BT-Settl with $T_\text{eff}$ = 3300 $\pm$ 50 K, metallicity [M/H] = 0.0 dex, and surface gravity $\log(g)$ = 5.0 dex ($\mathrm{cm}\ \mathrm{s}^{-2}$). The fit accounts for the observed and theoretical fluxes, their errors, the number of data points, and the object distance and radius. The resulting SED is presented in Fig.~\ref{fig:SED}.

Bolometric luminosity was obtained by integrating the SED, giving $\lstar\,=\,0.00770 \pm 0.00003\,L_\odot$. By using the Stefan-Boltzmann law, we obtained the stellar radius $\rstar$\,=\,0.268\,$\pm$\, 0.008\,$R_\odot$. Finally, the stellar mass (\hbox{$\mstar$\,=\,0.259\,$\pm$ 0.012\,$M_\odot$}) was estimated using Equation~6 from \citet{Schweitzer2019}.

A comparison between the empirically computed stellar properties and the ones from the SED fit are showcased in Table~\ref{tab:stellar_param_compare}. Both the mass and the radius of the star are in agreement within 1-$\sigma$ between the two methods. Due to the lower uncertainties, we adopt the empirically derived stellar parameters (Table~\ref{tab:StellarParameters}). 

\begin{table}[t]
    \centering
    \caption{TOI-4552 stellar parameters derived by different methods.}
    \label{tab:stellar_param_compare}
    \small
    \renewcommand{\arraystretch}{1.2}
    \setlength{\tabcolsep}{3.5pt}
    \begin{tabular}{lcc}
        \hline\hline
         & \textbf{Empirical} & \textbf{SED}  \\
        \hline
        $R_*$ ($R_\odot$)  & $0.2869\pm0.0088$ & $0.268\pm0.008$ \\
        $M_*$ ($M_\odot$)  & $0.2619\pm0.0063$ & $0.259\pm0.012$ \\
        $L_*$ ($L_\odot$)  & $0.00836 \pm 0.00130$  & $0.00770\pm0.00003$  \\
        $\log g_*$ (cm s$^{-2}$)  & $4.940\pm0.028$ & $5.0$  \\
        \hline
    \end{tabular}
\end{table}

\subsection{HARPS stellar analysis} \label{sec:spec_stellar}

To derive stellar parameters with the HARPS optical spectrum, we first combined the 33 HARPS-EGGS spectra reduced with the ESPRESSO DRS-3.2.5 pipeline \citep{ESPRESSO} repurposed to work on HARPS data (and corrected by their RV), by using the task {\tt scombine} within IRAF\footnote{IRAF is distributed by National Optical Astronomy Observatories, operated by the Association of Universities for Research in Astronomy, Inc., under contract with the National Science Foundation, USA.} to obtain a high SNR spectrum. Then, we applied the machine learning tool {\tt ODUSSEAS}\footnote{\url{https://github.com/AlexandrosAntoniadis/ODUSSEAS}} \citep{Antoniadis20,Antoniadis24} to derive the effective temperature (\teff) and metallicity (\feh) from the pseudo equivalent widths (EWs) of a set of $\sim$4000 lines in the optical spectra. This tool applies a machine learning model trained with the same lines measured and calibrated in a reference sample of 47 M dwarfs observed with HARPS for which their \feh were obtained from photometric calibrations \citep{Neves12} and their \teff\ from interferometric calibrations \citep{Khata21}. Using this method, we derived a \teff\,=\,3323\,$\pm$\,101\,K and \feh=\,0.00\,$\pm$\,0.12\,dex. The same process was repeated for the same HARPS spectra reduced by the offline DRS-3.5 pipeline as well which yielded \hbox{\teff\,=\,3247 $\pm$ 96\,K} and \hbox{\feh = 0.01 $\pm$ 0.11}.

The HARPS \teff\ values are in agreement with the one obtained from the SED fit as well as the one reported by TESS DVR (\teff\,=\,3265\,$\pm$\,157 K). To accommodate the large uncertainties of each measurement, we combined the four estimates of effective temperature by performing Gaussian resampling by assuming a normal distribution for all four \teff\ centred at the reported value and the uncertainty as the standard deviation. At each step we drew one value from each distribution, averaged them, and repeated this for 10,000 iterations. The median and 1-$\sigma$ scatter of the resulting distribution give our final value of \hbox{\teff\,= 3258 $\pm$ 115\,K}. By doing this we inflate the uncertainty in \teff\ to account for any biases that may affect the uncertainties in the values reported by the four sources.

\begin{table}[t]
    \centering
    \caption{TOI-4552 stellar abundances measured with NIRPS.}
    \label{tab:toi4552_abundances}
    \begin{tabular}{lccc}
        \hline \hline
        \textbf{Element} & \textbf{[X/H]}$^*$  & \textbf{no. of lines} \\
        \hline
        O I  &  -0.02 $\pm$ 0.09  &  31  \\
        Na I  &  0.23 $\pm$ 0.11  &  4  \\
        Mg I  &  -0.12 $\pm$ 0.16  &  1  \\
        Al I  &  -0.32 $\pm$ 0.08  &  5  \\
        Si I  &  0.17 $\pm$ 0.28  &  2  \\
        K I  &  -0.10 $\pm$ 0.31  &  8  \\
        Ca I  &  -0.13 $\pm$ 0.10  &  13  \\
        Ti I &  -0.12 $\pm$ 0.12  &  36  \\
        Cr I &  0.06 $\pm$ 0.09  &  14  \\
        Mn I &  0.28 $\pm$ 0.12  &  5  \\
        Fe I &  -0.07 $\pm$ 0.09  &  25  \\
        \hline
        \multicolumn{3}{l}{$^*$Relative-to-solar abundances}
    \end{tabular}
\end{table}

\subsection{NIRPS stellar analysis}

The stellar elemental abundances were derived from the NIRPS spectra following the methodology of Gromek et al. (in prep), based on the work of \citet{Hejazi2023}, and are summarized in Table~\ref{tab:toi4552_abundances}. We performed a spectral synthesis analysis using the order-merged telluric-corrected NIRPS template spectrum calculated within the LBL framework. Synthetic spectra were generated with MARCS stellar atmosphere models \citep{Gustafsson2008} and the Turbospectrum radiative transfer code \citep{Alvarez1998, Plez2012}, implemented via modified iSpec routines \citep{BC2019, BC2014} using solar abundances from \citet{Asplund2009}. The stellar parameters adopted to generate the model spectra were \teff\,=\,$3258\pm115$\,K, v$_{mac}$\,=\,$4\pm2.25$\,km\,s$^{-1}$, and \hbox{v$_{mic}$\,=\,$0.75\pm0.5$\,km\,s$^{-1}$}. The macroturbulence (v$_{mac}$) and microturbulence (v$_{mic}$) velocities were determined through $\chi^2$-minimization of molecular OH lines, which are especially sensitive to these broadening parameters \citep{Souto2017, Hejazi2023}. Absorption lines were selected from the normalized spectrum, cross-referenced with atomic and molecular features in the VALD line-list \citep{Kupka2011}, and further refined via visual inspection to exclude blended or contaminated features. The oxygen abundance was derived from OH features.

Synthetic spectra were generated for each line by varying the elemental abundance [X/H] between –0.75 and +0.75 dex in 0.25 dex steps which were then interpolated to 0.015 dex resolution. Best-fit abundances were obtained via $\chi^2$ minimization between the model and observed spectrum within a fitting window. For certain lines where the continuum level between the model and the spectra do not match, we fit and apply a uniform flux offset that minimizes the $\chi^2$ between the observed data and the model in the continuum region within 0.5-nm of the spectral line. Final abundances for each element are calculated as weighted averages of the individual abundances, with the weighting equal to the RMSE between the best-fit model and the observed line in each line region, divided by the line depth. Random uncertainties ($\sigma_{ran}$) were calculated as the standard deviation of the line-by-line abundance distribution divided by $\sqrt{N}$, where N is the number of lines used per element. Systematic uncertainties due to \teff, \mh, $\log g_*$, v$_{mac}$, and v$_{mic}$ were estimated by independently resampling each stellar parameter from its Gaussian uncertainty distribution and repeating the analysis over 15 iterations \citep{Hejazi2023}. The total uncertainty is computed by summing the random and systematic components in quadrature. Global metallicity and $\alpha$-enhancement was recomputed following \citet{Hinkel2022}, yielding \mh\,=\,-0.01$\pm$0.07 dex from the summed number ratios of O, Na, Mg, Si, K, Ca, Ti, Cr, Mn and Fe, and \alfe\,=\,0.06$\pm$0.12 using the alpha elements O, Mg, Si, Ca, and Ti. The final stellar abundance measurements and parameters are reported in Tables~\ref{tab:StellarParameters} and \ref{tab:toi4552_abundances}.

It should be noted that the best fit SED model of BT-Settl is in agreement with the spectroscopically derived \teff and \mh values for TOI-4552. We choose to report the values inferred from spectroscopic observations in Table~\ref{tab:StellarParameters}.

\begin{table}[t]
    \centering
    \caption{TOI-4552 b radius derived from transit lightcurves of each individual instrument.}
    \label{tab:toi4552_transit_radii}
    \begin{tabular}{lccc}
        \hline \hline
        \textbf{Instrument} & \textbf{$\delta$ (ppm)} & \textbf{$R_{p}\, (R_{\oplus})$}  & \textbf{no. of transits} \\
        \hline
        TESS & 1195 $\pm$ 89 &  1.083 $\pm$ 0.051  &  224  \\
        ExTrA & 1117 $\pm$ 81 &  1.047 $\pm$ 0.053   &  61   \\
        LCO & 992 $\pm$ 122 &  0.987 $\pm$  0.068   &  6   \\
        SPECULOOS & 1385 $\pm$ 129 &  1.161 $\pm$ 0.064  &  20   \\
        \hline
    \end{tabular}
\end{table}

\subsection{Stellar activity}
\label{sec:stellar_activity}

To constrain any potential stellar activity signals, we analysed the spectroscopic activity indicators in the NIRPS RV computed in the standard LBL reduction framework, namely the differential temperature metric ($\Delta\,T$; \citealt{Artigau2024}) and the second-order derivative of the velocity (D2V; \citealt{Zechmeister2018,Artigau2022}). Fig.~\ref{fig:actvityNIRPS} shows the timeseries for the two indicators and the respective Lomb-Scargle periodogram. No significant peaks are detected above the 99.9\% confidence level on a short time scale apart from aliases of the 1-day RV sampling. For long-term monitoring, we detected no significant trends in the TESS SAP lightcurve. However, we only have three sectors of photometric data, each separated by two years, so the sampling is insufficient to identify any definitive variability. As an alternative, we obtained the publicly available photometric measurements from the All-Sky Automated Survey for Supernovae (ASAS-SN, \citealt{Kochanek2017}) Sky Patrol V2.0\footnote{\url{http://asas-sn.ifa.hawaii.edu/skypatrol/}} \citep{ASASSN1, ASASSN2}. These observations were done using the $V$ and $g'$-band filters and are displayed in Fig. \ref{fig:asassn_phot} along with the periodogram. The highest detected peak is at 765-days (slightly above the 99.99\% confidence level), a possible signature of magnetic activity as can be the case in late M dwarfs \citep{ASM2016}. However, it is of note that is period is very close to the two year harmonic of the Earth's orbital period and is thus likely an artifact of the window function. We conclude that TOI-4552 is a quiet star with no significant short term stellar rotation period detected in RVs and TESS photometry and no obvious magnetic cycle detected in long-term photometry. 

\section{Transit and radial velocity analysis} \label{sec:final_analysis}

\begin{figure*}[h]
    \centering
    \includegraphics[width=1\linewidth]{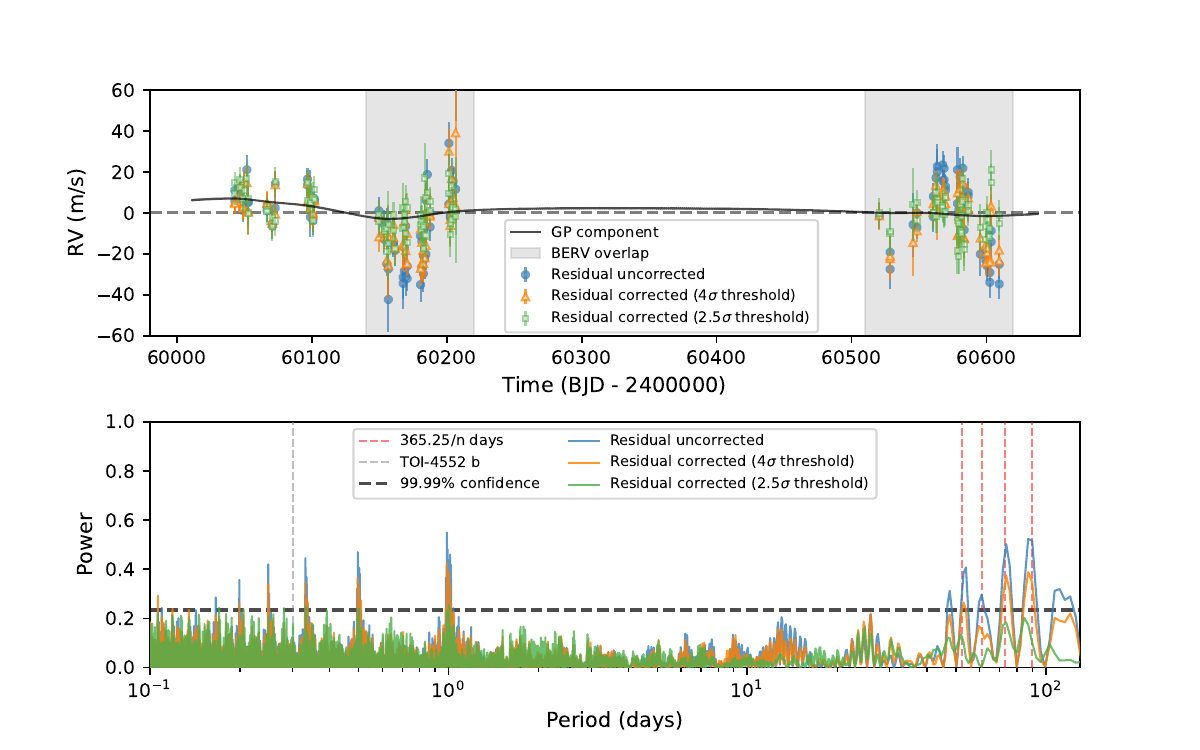}
    \caption{NIRPS RV timeseries of TOI-4552 (top) along with the corresponding periodograms (bottom). Due to the BERV overlap issue discussed in Section~\ref{sec:rv_analysis}, we masked a small number of pixels from each frame that contained stellar lines severely affected by under-corrected tellurics. Outlier pixels were identified using a $\sigma$-clipping threshold, such that any deviations from the median stellar spectrum exceeding the threshold were masked. We present here the unmasked (Residual uncorrected) and masked (Residual corrected) time series at 4$\sigma$ and 2.5$\sigma$ thresholds, along with their corresponding periodograms. After applying the 2.5-$\sigma$ threshold, the peaks associated with harmonics of Earth’s rotation (365.24/n days) fall below the confidence level, indicating that the telluric signals have been effectively suppressed. The telluric residual corrected time series (green) is adopted for our final analysis.}
    \label{fig:bervcrossing}
\end{figure*}

\subsection{Transit lightcurve analysis} \label{sec:transit_analysis}

TOI-4552 was observed in three TESS sectors (12, 39, and 66) and monitored from the ground with ExTrA, LCO and SPECULOOS. No additional transiting candidates or signatures of transit timing variations (TTVs) were observed in our transit sample, making TOI-4552\,b the only currently detected short period transiting planet in the system. We first retrieved the TESS lightcurves using \texttt{lightkurve} \citep{lightkurve}\footnote{\url{https://github.com/lightkurve/lightkurve}}. While these PDCSAP lightcurves were corrected for instrumental systematics, we observed additional residual correlated structures that can impact the retrieved planetary radius. We therefore used a Gaussian Process (GP) with the Mate\'rn-3/2 kernel approximation implemented in \texttt{celerite} \citep{celerite} available natively in \texttt{juliet} to further detrend the lightcurves. The following formalism for the term is used:

\begin{equation}
\label{matern_kernel}
k(\tau) = \alpha^2 \left(1 + \frac{\sqrt{3}\tau}{\beta}\right)\,
\exp\!\left(-\frac{\sqrt{3}\tau}{\beta}\right)
\end{equation}

where $\alpha$ is the amplitude, $\beta$ is the length scale and $\tau$ is the time lag. To prevent the GP from fitting the USP transits, we restrict the lower limit of length scale to be 1-day. The same GP kernel was used inside \texttt{Allesfitter} to detrend the SPECULOOS lightcurves~\citep[e.g.,][]{Dransfield2024, Henderson2024} as well as ExTrA lightcurves~\citep[e.g.,][]{Cointepas2021}. The LCO lightcurves were detrended at the extraction phase via the \texttt{AstroImageJ} software.

Using preliminary period and time of conjugation measurements reported by SPOC ($P_{\rm orb}$\, =\, 0.3011\,days; $t_c$\, =\, 2459361.8822\,BJD), we performed transit fits using \texttt{juliet} \citep{juliet} to ensure that the retrieved radii of the planet are consistent between instruments using stellar parameters from Table~\ref{tab:StellarParameters}. The two coefficient limb darkening formalism from \citet{Kipping2013} was used and both coefficients were fit individually for each instrument. All three TESS sectors (12, 39 and 66) were modelled as independent instruments to incorporate their separate GP fits, lightcurves from LCO-CTIO and LCO-SAAO were modelled separately, all three ExTrA telescopes (ExTrA1, ExTrA2, ExTrA3) were modelled separately, and each of the SPECULOOS telescopes and the various combinations of photometric filters (Europa-$g'$, Europa-$r'$, Ganymede-$r'$, Io-$g'$, Io-$r'$, Io-$z'$) were modelled as different instruments. Table \ref{tab:toi4552_transit_radii} highlights the transit depth ($\delta$) and planetary radius ($R_p$) obtained for each individual instrument as well as the number of total transits observed, which all agree their within 2-$\sigma$ uncertainty. It should be noted that no additional dilution correction was applied to the extracted lightcurves at this point. The same $P_{\rm orb}$ and $t_c$ were later used as priors in the joint fit with RVs (see Section \ref{sec:joint-fit}).

\subsection{Radial velocity analysis: BERV overlap} \label{sec:rv_analysis}

The Earth's atmosphere is not as transparent in NIR compared to VIS wavelength range. Telluric absorption and emission features pollute the entire NIR coverage of NIRPS and while we have methods of correcting them (\citealt{Allart2022, Cook2022}; Srivastava et al. in review), they are not perfect and may leave some residuals that are anchored in the Earth's rest frame. The previously mentioned BERV crossing phenomenon occurs due to these anchored residuals. Following the procedure explored in Srivastava  et al. (in review), we perform a post-processing reduction to mask any pixels, across all observations of TOI-4552, that exhibit variations in flux beyond a 2.5-$\sigma$ threshold relative to the median stellar spectrum. The value for this threshold is partly arbitrary and may not be optimal for other targets. It depends on the SNR of the science observation and the efficiency of the telluric absorption and emission corrections. This aggressive masking of pixels aims to remove telluric residuals at the expense of RV precision. An obvious signature of the dataset being contaminated by telluric residuals is the presence of harmonics of the Earth's orbital period ($365.24/n$ days) in the periodogram. Throughout the development and testing phase of this post-processing tool we analysed the impact of differing thresholds on the RV timeseries (Fig.~\ref{fig:bervcrossing}). The 2.5-$\sigma$ threshold was found to be optimal, as it reduced the power of the peaks associated with harmonics of Earth’s orbital period to below the 0.01\% false-alarm probability (FAP; 99.99\% confidence) in the periodogram.

\subsection{Radial velocity analysis: Gaussian Process detrending} \label{sec:rv_gp}

We also explored alternative methods to correct for the effects of telluric residuals, including the use of a Gaussian Process (GP) to detrend for the systematics. While GP hyperparameters can have a physical interpretation \citep{Stock2023}, we are using it to model systematics and therefore have to carefully choose a kernel and priors for the hyperparameters so as to not overfit the data. Here, we again used the Mate\'rn-3/2 kernel approximation available in \texttt{juliet} (see Section~\ref{sec:transit_analysis} for more details). The GP was applied on two datasets, one not corrected for the telluric residuals (blue points in Fig.~\ref{fig:bervcrossing}) and the other corrected for those features (green points in Fig.~\ref{fig:bervcrossing}). The various models and their results are displayed in Table~\ref{tab:logzCompare}.

Without correcting the BERV overlap systematics, then upon fitting the GP and planet together with wide priors, the detection of the planet is not significant and is statistically similar to the GP-only fit as per the interpretation of the $\Delta\,\mathrm{log}Z$ \citep{Kass1995}. Despite correcting for the BERV overlap to the best of our ability, there may remain some minor correlated noise in the RV timeseries. To account for this we tried to fit the planet with and without a GP, with the scenario of a GP\,+\,Planet fit being more statistically favoured (Table~\ref{tab:logzCompare}), stressing that the identifying the cause and correcting for the systematics is a necessary step. For this fit, the $P_{orb}$, $t_c$ and instrumental white noise jitter ($\sigma_{w}$) were fit as free parameters following their prior distributions used in the joint fit (see Section~\ref{sec:joint-fit}, Table~\ref{tab:posteriors_jointfit}). It should be noted that for the purpose of model comparison, the results of the aforementioned table reflect the various models fit solely on the RV data. Once the best model was identified (GP\,+\,Planet on the telluric residual corrected timeseries), we used it in combination with the transit data to obtain the final parameters detailed in the next section.

\begin{table}[t]
    \centering
    \caption{Comparing the relative Bayesian evidence ($\Delta\mathrm{log}Z$) between various models used to remove systematics from the radial velocity timeseries.}
    \label{tab:logzCompare}
    \small
    \renewcommand{\arraystretch}{1.2}
    \setlength{\tabcolsep}{3.5pt}
    \begin{tabular}{lcc}
        \hline\hline \textbf{Model}
         & \textbf{$K_p$ (m/s)} & \textbf{$\Delta\,\mathrm{log}Z$}  \\
        \hline
        & \textbf{Telluric residual uncorrected data} & \\
        GP only  & -- & 0 \\
        Planet only  & 5.71 $\pm$ 1.78 & -59.56 $\pm$ 0.17\\
        GP + Planet  & 3.04 $\pm$ 1.31  & 2.81 $\pm$ 0.19 \\
        \hline
        & \textbf{Telluric residual corrected data} & \\
        GP only  & -- & 0 \\
        Flat line & -- & -7.57 $\pm$ 0.11 \\
        Planet only  & 4.97 $\pm$ 1.12 & 0.44 $\pm$ 0.13 \\
        GP + Planet  & 4.24 $\pm$ 1.14  & 4.18 $\pm$ 0.16 \\
        \hline
    \end{tabular}
\end{table}

\subsection{Joint fit} \label{sec:joint-fit}

As with the transit photometry analysis, we used \texttt{juliet} for the joint fit. Table \ref{tab:posteriors_jointfit} features all the parameters, priors and posteriors used to do the analysis. The segregation of photometric instruments was identical to the one explained in Section \ref{sec:transit_analysis}. Since the ExTrA lightcurves were ensured to be free of any dilution and contamination effects during the extraction and detrending process, the dilution factor was fit to be unity for this dataset. For all other photometric data, we allowed the dilution factor to vary between 0.5 and 1. While TESS SPOC does correct for dilution as mentioned previously, we decided to allow a variable dilution factor to account for any imperfections. We fit a two coefficient limb darkening model implemented in \texttt{juliet} \citep{Espinoza2016} using the formalism presented in \citet{Kipping2013} for each individual telescope. The posteriors are however not very well constrained, but are consistent within uncertainties across instruments. The stellar density was calculated using the stellar mass and radius from Table \ref{tab:StellarParameters} and we used the parametrisation suggested in \citet{Espinoza2018} for the planetary radius and impact parameter. We additionally fixed the eccentricity to $e\, =\, 0$ and argument of periastron to $\omega\, =\, 90^o$ as the close proximity to the host star almost guarantees a tidally locked, non-eccentric orbit \citep[e.g.,][]{Lyu2024}. To lower the computational load, the fit was performed using detrended ExtrA, LCO and SPECULOOS lightcurves, while the TESS data was detrended simultaneously with the joint fit using the same kernel as described before (Section~\ref{sec:transit_analysis}). As justified in Section~\ref{sec:stellar_activity}, there are no hints of any short-term stellar activity to be fit in our data; therefore, we only use a GP to model any minor systematics as detailed in Section~\ref{sec:rv_gp}. An excess white noise jitter was fit for both NIRPS and HARPS separately as the instrument performances vary significantly.

\begin{figure}
    \centering
    \includegraphics[width=1\linewidth]{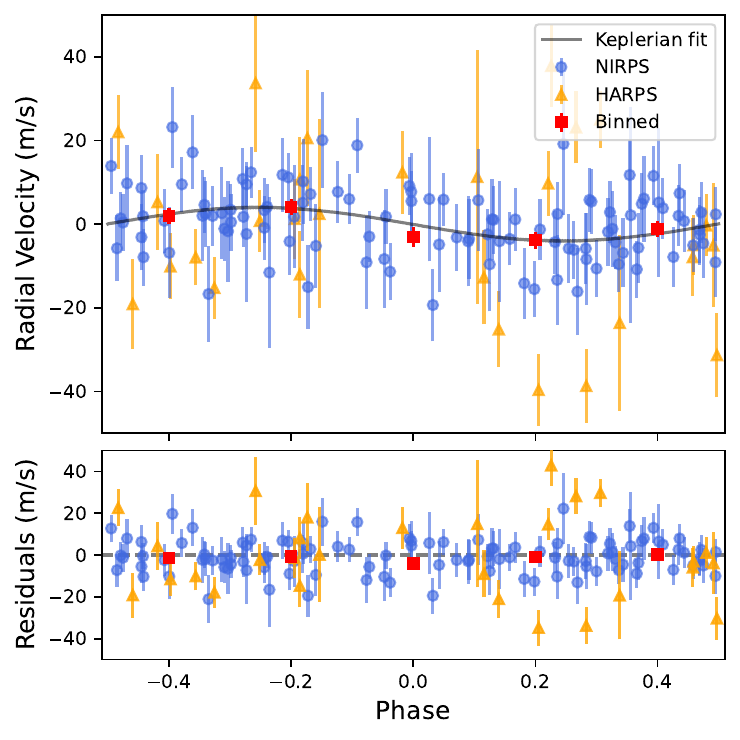}
    \caption{\textit{Top:} NIRPS RV measurements (blue points) and the final fit Keplerian model (grey line), phase-folded according to the $P$ and $t_c$ in Table \ref{tab:planetaryparams}. The red points correspond to the binned RVs. \textit{Bottom:} Residuals of the RVs after subtracting the model.}
    \label{toi4552_rv}
\end{figure}

To run our 1-planet model with 94 parameters, we used the built-in \texttt{dynamic-dynesty} sampler in \texttt{juliet} using 10,000 live points to prevent undersampling. A final planetary radius of $1.11 \pm 0.04\,R_{\oplus}$ was obtained, corresponding to a transit depth of 1258 $\pm$ 55 ppm, consistent with measurements from individual photometric instruments (see Table~\ref{tab:toi4552_transit_radii}). The derived RV semi-amplitude of $4.32 \pm 1.08$ m/s corresponds to a mass of $1.83 \pm 0.47M_{\oplus}$, a mass precision of 25\% (3.9-$\sigma$). The priors and posteriors for the final fit are presented in Table~\ref{tab:posteriors_jointfit}. It should be noted that while we use HARPS in our final analysis, the solution is primarily driven by NIRPS (see Fig.~\ref{toi4552_rv}), thus highlighting the advantage of NIR velocimetry for M dwarf targets.

\section{Discussion} \label{sec:discussion}

\subsection{TOI-4552\,b composition and core mass fraction} \label{sec:composition}

\begin{table}[h]
\tiny
\caption{Planetary parameters for TOI-4552 b.}
\centering
\renewcommand{\arraystretch}{1.5}
\setlength{\tabcolsep}{1pt}
\begin{center}
\begin{tabular}{lc}
\hline\hline
\textbf{Parameter} & \textbf{TOI-4552 b} \\
 \hline
    Orbital period, $\it{P_{\mathrm{orb}}}$ (days)\dotfill  & $ 0.30110032\pm0.00000014$ \\
    Time of conjunction, $t_c$ (BJD)\dotfill & $2459361.88587913\pm0.00024681$ \\
    Planet radius, $\it{R_{\mathrm{p}}}$ (\re)\dotfill  & $1.11 \pm 0.04$ \\
    Planet mass, $\it{M_{\mathrm{p}}}$ (\me)\dotfill  & $ 1.83\pm0.47$ \\
    Planet bulk density, $\rho_\mathrm{p}$ (g~cm$^{-3}$)\dotfill  & $ 7.74 \pm 2.14$ \\
    RV semi-amplitude, $K_{\rm p}$ ($\mathrm{m}\,\mathrm{s}^{-1}$) \dotfill & $4.32^{+1.04}_{-1.08}$ \\
    Orbital inclination, $i$ ($^\circ$)\dotfill & $87.77^{+1.94} _{-1.90}$ \\
    Scaled planetary radius, $R_{\mathrm{p}}$/$R_{*}$ \dotfill &  $0.0352^{+0.0009}_{-0.0009}$ \\
    Impact parameter, $\textit{b}$\dotfill  & $ 0.16^{+0.12} _{-0.11}$ \\
    Semi-major axis, $\it{a}$ (AU)\dotfill  & $0.0056255 \pm 0.0000003$ \\
    Eccentricity, $e$ \dotfill  & $0$ (fixed) \\
    Argument of periastron, $\omega$ (deg) \dotfill  & $90$ (fixed) \\
    Insolation, $\it{S_{\mathrm{p}}}$ (\se)\dotfill  &  $ 263\pm40$ \\
    Equilibrium temperature$^{(\gamma)}$, $T_{\mathrm{eq}}$ (K)\dotfill  & $ 1122\pm10$ \\
    ESM$^{(\gamma)}_{7.5\mu m}$\dotfill  & $ 19.5$ \\
    Dayside temperature$^{(\zeta)}$, $T_{\mathrm{day}}$ (K)\dotfill  & $ 1434\pm10$ \\
    Core mass fraction, CMF \dotfill &  $0.55^{+0.17}_{-0.24}$ \\
 \hline
\end{tabular}
\begin{tablenotes}
\item
\textbf{Notes:} $^{(\gamma)}$ Equilibrium temperature and Emission Spectroscopic Metric (ESM) are calculated as in \citet{Kempton2018}, $^{(\zeta)}$ dayside temperature is computed using Equation 8 in \citet{Morris2022} assuming a bond albedo of A$_\mathrm{B}$ = 0 and heat distribution factor $f\, =\, 2/3$. 
\end{tablenotes}
\label{tab:planetaryparams}
\end{center}
\end{table}

\begin{figure}
    \centering
    \includegraphics[width=\linewidth]{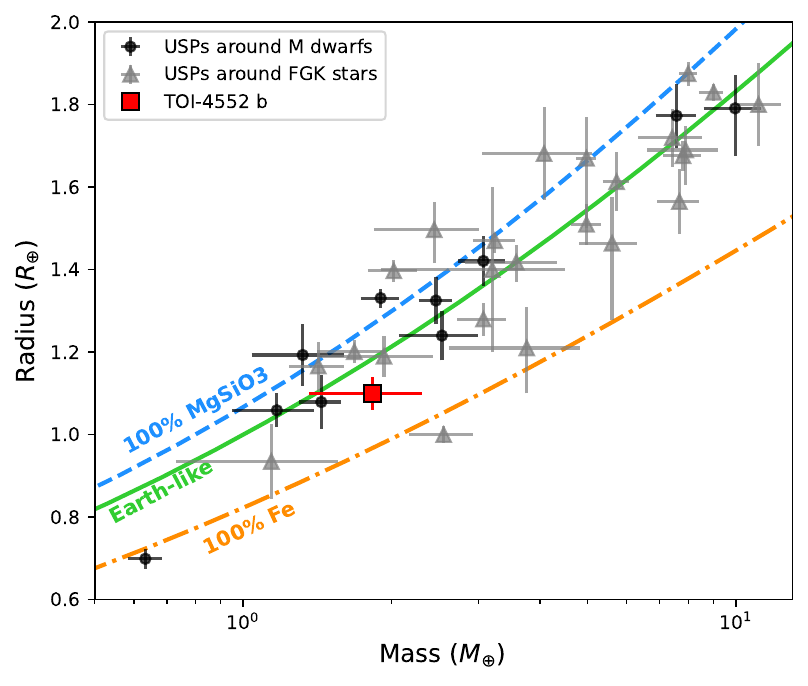}
    \caption{Mass-Radius diagram of all rocky USPs ($R_p\,<\,2R_{\oplus}$, $P\, <\, 1$\,day) around M dwarfs (\teff < 4000\,K) and FGK stars that currently have a measured mass and radius. All planet parameters were obtained through a query search on NASA Exoplanet Archive. Silicate-rich (blue), Earth-like (green) and iron-rich (orange) compositions lines, taken from \citet{Zeng2019}, are overplotted. According to the measured mass and radius, TOI-4552\,b lies in the gap between the Earth-like and pure iron composition lines. A more precise mass measurement is required to obtain better constraints on the composition.}
    \label{fig:usp_mr}
\end{figure}

TOI-4552\,b is the eleventh rocky USP around an M dwarf star to have a constrained mass and radius measurement. The other ten (TOI-6255\,b \citep{toi6255}, GJ~367\,b \citep{gj367}, GJ~1252\,b \citep{gj1252}, Wolf~327\,b \citep{wolf327}, TOI-1685\,b \citep{toi1685_3_toi1634, toi1685_2, toi1685_1}, LTT~3780\,b \citep{ltt3780}, GJ~806\,b \citep{gj806}, TOI-1634\,b \citep{toi1685_3_toi1634}, TOI-1075\,b \citep{Essack2023} and TOI-6324\,b \citep{Lee2025}), along with TOI-4552\,b, are presented in the mass-radius diagram (Fig.~\ref{fig:usp_mr}), with the Earth-like, silicate-rich and pure iron composition lines taken from \citet{Zeng2019}. The masses and radii were taken from the most up to date publication after a query search on the NASA Exoplanet Archive\footnote{\url{https://exoplanetarchive.ipac.caltech.edu/}} \citep{nasa_exoplanet_archive} for planets with an orbital period of $P\,<$ 1\,day, radius of $R\,<\, 2\,R_{\oplus}$, a reported mass and a host start effective temperature of \teff$\, <4000\,K$. Other USPs orbiting FGK stars are also plotted in grey to provide a comparison to the M dwarf sample.

Using the final planetary mass and radius, we derive a bulk density of 7.74$\pm$2.12\,g/cm$^{3}$, higher but within 1-$\sigma$ of that of the Earth ($\rho_{\oplus}$\,=\,5.51\,g/cm$^3$). Most USPs around M dwarfs fall into either Earth-like or silicate-rich composition categories; however, GJ~367\,b (M$_p$\,=\,0.63\,M$_\oplus$; R$_p$\,=\,0.69\,R$_\oplus$) stands out as distinctly iron-rich. In our own Solar System, Mercury is the prime example of a rocky planet with an overabundance of iron compared to Venus, Earth and Mars. Accordingly, planets such as GJ~367\,b are often referred to as ``super-Mercuries''. Per the location in the mass-radius diagram, TOI-4552\,b has a composition partway between the Earth and Mercury, similar to several other USPs around FGK stars.

To better quantify the composition of TOI-4552\,b, we computed an interior structure model using the publicly available \texttt{exopie}\footnote{\url{https://github.com/mplotnyko/exopie}} code \citep{Plotnykov2024}, an updated version of the previous \texttt{SUPEREARTH} code \citep{Plotnykov2020, Valencia2006}. The core mass fraction (CMF) quantifies the proportion of a planet’s total mass contained in its core, and thereby reflects the relative abundances of iron and silicates, which primarily compose the core and mantle, respectively. We first ran \texttt{exopie} assuming a completely rocky planet with no water mass fraction and a
variable (0 - 20\%) amount of iron in the mantle and silicon in the core. The planetary mass and radius were used as priors and no constraints were put on the composition using stellar abundance measurements (Table~\ref{tab:toi4552_abundances}). The results can be found in the corner plot of Fig.~\ref{fig:cmf_no_prior}. The sampler converges on the mass and radius values that are within 1$\sigma$ of our measurements. The resultant CMF of 0.54$^{+0.17}_{-0.24}$ is higher than that of the Earth (CMF$_{\oplus}$\,=\,0.33) and lower than that of Mercury (CMF$_{Mercury}$ = 0.7; \citealt{Szurgot2015}), however consistent with both within the uncertainties. Another relevant metric is the uncompressed density of a planet, which is a measure of the mean density at zero pressure \citep{Faure2007}. Again using \texttt{exopie}, we calculated the uncompressed density of TOI-4552\,b ($\rho_{uncom}$) to be 4.84$\pm$0.70\,g/cm$^3$, lying partway between the Earth's ($\rho_{uncom, Earth}$\,=\,4.4\,g/cm$^3$) and Mercury ($\rho_{uncom, Mercury}$\,=\,5.3\,g/cm$^3$; \citealt{Prentice2005}) but consistent with both.

A common assumption when constraining planetary composition is that, since a planet and its host star form from the same interstellar medium, the planet’s refractory elemental abundances should reflect those of its star \citep{Dorn2015, Hinkel2018}. However, some recent works have questioned the validity of this assumption \citep{Santos2017, Plotnykov2020, Brinkman2024} as the observations suggest that refractory ratios of rocky worlds span a larger range than that of stars. Given the slightly high CMF of TOI-4552\,b, we also ran a second \texttt{exopie} model that did not use the planetary mass and radius as priors. Instead, we adopted the host star’s abundance ratios ([Fe/H], [Mg/H], [Si/H]) to predict the composition of a representative rocky planet formed around this star. Since the [Mg/H] and [Si/H] calculations were done using one and two spectral lines respectively, we instead use [$\alpha$/H] as a proxy for both since they are $\alpha$-elements. The result (Fig.~\ref{fig:cmf_stellar_prior}) suggests a planet with CMF\,=\,0.32$^{+0.09}_{-0.08}$, which is just within 1$\sigma$ of the CMF constrained by the mass and radius. The CMF is not constrained well enough to conclude if TOI-4552\,b is an iron-rich super-Mercury or an Earth-like rocky world as both are statistically plausible from the current dataset.

The formation and evolution pathways of Mercury-like exoplanets remain uncertain. Some hypotheses, such as their being remnants of stripped gas giants \citep{Lin2025} or products of giant impacts \citep{Bonomo2019, Cambioni2024}, are considered unlikely, while other studies \citep{Brinkman2024} find no compelling evidence for the existence of super-Mercuries at all, although GJ~367\,b appears to be a strong candidate, lying close to the iron sequence of the mass-radius diagram (see Fig.~\ref{fig:usp_mr}). Based on planet formation simulations, \citet{Mah2023} propose that super-Mercuries are more likely to form around stars with low Mg/Si ratios (< 1), which is in line, albeit with large uncertainties, with the value derived from the TOI-4552 stellar spectrum analysis (Mg/Si\,=\,0.62$^{+0.67}_{-0.31}$, Table~\ref{tab:StellarParameters}). 

We note that \citet{Brinkman2025} reanalysed the Kepler-100\,b \citep{Weiss2024} and HD~93963A\,b \citep{Serrano2022} systems using additional RV measurements from MAROON-X \citep{MaroonX} and KPF \citep{KPF}. They concluded that the earlier classification of these planets as super-Mercuries was driven by low-precision mass measurements, and that with the higher-precision data the inferred CMFs are instead consistent with Earth-like values. Given the modest 3.9-$\sigma$ mass detection, we cannot exclude that the mass of TOI-4552\,b is slightly overestimated; several planets have had their masses revised downward as more precise RV measurements became available \citep[e.g., LHS~1140\,b][]{Cadieux2024} and the several USPs mentioned prior \citep{Brinkman2025}. We thus conclude that despite the high density, more precise RV measurements are required to confirm or deny TOI-4552\,b's status as a super-Mercury.

Could TOI-4552\,b’s relatively high density simply reflect an underestimated radius? A recent study \citep{Han2025} argues that TESS radii are, on average, underestimated relative to K2/Kepler because TESS’s large pixels can bias the dilution correction. This explanation seems unlikely for TOI-4552. The planetary radii inferred from the dilution-corrected TESS photometry and from ground-based observatories agree within uncertainties (see Table~\ref{tab:toi4552_transit_radii}). Additionally, our final radius of 1.11$\pm$0.04\,R$_\oplus$ was calculated by allowing the dilution factor for the photometric data to be a varying parameter.

\subsection{Potential tidal and rotational deformation}

Since USPs are very close to their host stars, there may be non-insignificant levels of tidal deformation in these rocky plants. TOI-6255\,b~\citep{toi6255} is one such USP on the verge of tidal disruption. Based on the equations presented in \citet{toi6255}, a similar analysis can be performed for TOI-4552\,b. As defined in~\citet{Rappaport2013}, the Roche period ($P_{Roche}$) is the lower limit for the orbital period before the planet is tidally disrupted. The equation:

\begin{equation}
    P_{Roche}\,\approx\,12.6\,\text{h}\,\bigg(\frac{\rho_\text{P}}{1\,g/cm^3}\bigg)^{-\frac{1}{2}}
\end{equation}

approximates the Roche period (in hours), based on the planetary bulk density ($\rho_{\text{P}}$). For TOI-4552\,b the $P_{Roche}$ comes out to be 4.5\,h. Since the orbital period (7.2\,h) is higher, TOI-4552\,b is likely not on the verge of tidal disruption. Furthermore, equations 4, 10 and 11 in~\citet{toi6255} can be used to estimate the fractional change in the radius of TOI-4552\,b due to rotational deformation that can be probed during planetary transit. This change in radius is approximately 2.8\%. With the current precision of the planetary radius being of the order of 4\%, this deformational cannot be accurately measured using the current dataset.

\subsection{Prospects for JWST observations} \label{sec:jwst}

\begin{figure}[]
\includegraphics[width=\columnwidth]{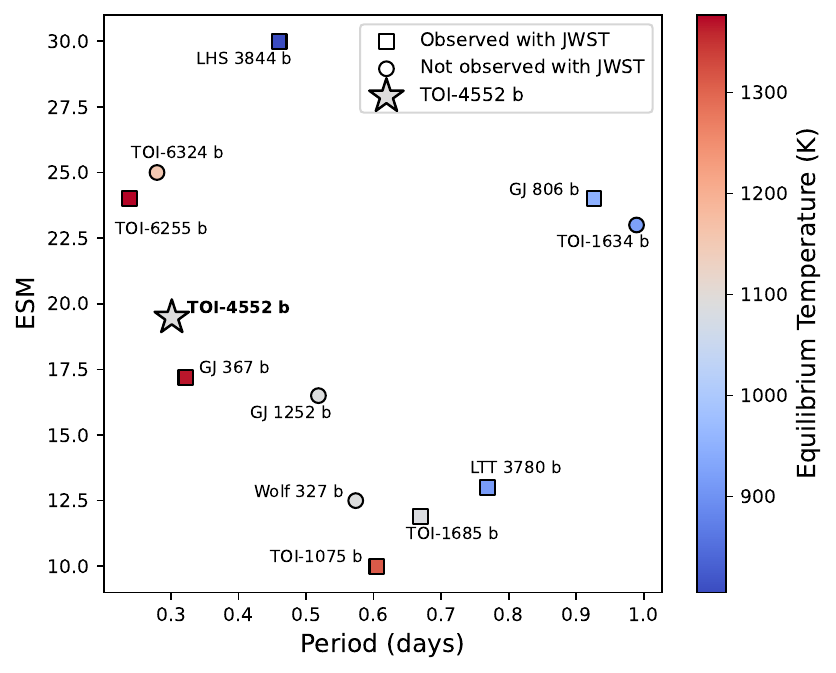}
\centering
\caption[]{Emission spectroscopic metric (ESM) versus period plot of all currently detected transiting rocky USPs with around M dwarfs with a constrained mass and radius. LHS\,3844\,b is additionally plotted, despite not having a published mass, as it is one of the most extensively studied USPs \citep{Kreidberg2019}. The ESMs for all planets was taken from the respective discovery publications. The colour bar indicates the equilibrium temperature and planets that have already been observed with JWST are marked as squares. TOI-4552\,b has an ESM of 19.5 and the third shortest period, making it a high value target for emission spectroscopy.}
\label{usp_esm}
\end{figure}

Characterizing hot rocky planets has become a defining JWST effort, reflected in focused studies \citep[e.g.,][]{DL2021, Luque2025} and the ongoing 500-hour Rocky Worlds survey \citep{Redfield2024}. Due to their short orbital periods of less than one day, USPs are optimal targets to observe and analyse while minimizing the required telescope time. Some USPs orbiting M dwarfs have already been observed  with JWST or have planned future observations: TOI-1685\,b (GO3263, GO4098, GO4195; \citealt{Luque2025}), TOI-6255\,b (GO8864), GJ~367\,b (GO2508; \citealt{Zhang2024}), LHS~3844\,b (GO1846, GO4008, GO7953) and LTT~3780\,b (GO3730). TOI-4552\,b  joins this short list of coveted USPs for follow-up observations. Indeed, TOI-4552\,b features the third shortest orbital period (comparable to those of TOI-6255\,b and GJ~367\,b) and a high Emission Spectroscopic Metric (ESM; \citealt{Kempton2018}) of 19.5 (Table~\ref{tab:planetaryparams}, Fig.~\ref{usp_esm}), making it a favourable USP for emission spectroscopy studies. 

The leading theory of planet formation states that close-in rocky planets had their primordial H$_2$/He atmospheres irradiated away due to the strong X-ray/UV radiation from their M-type host stars, making a CO$_2$/CO dominated secondary atmosphere the likely scenario \citep{Wordsworth2022}. So far there have been no positive atmospheric detections for such worlds (\citealt{Kane2020, Zhang2024, Luque2025}); nevertheless, recent JWST results hint towards a possible CO$_2$/CO atmosphere due to vaporized surface SiO$_2$ around 55~Cnc\,e \citep{Hu2024, Zilinskas2025}.

Formation theories additionally suggest that terrestrial planets start off as molten rocks that cool down on a geological timescale. The time it takes to solidify the molten interior can range from 100~Myr \citep{Hamano2013, Boukare2025} to 10~Gyr \citep{Driscoll2015} depending on the proximity to the host star and the orbital eccentricity of the planet. The day-side temperature of TOI-4552\,b is $T_\mathrm{day}\,=\,1434$\,K, calculated using a heat redistribution factor of $f\,=\,2/3$ following equations from \citet{LM2007, Morris2022}. This temperature is too low to sustain a surface magma ocean \citep{lava_planet2012}. However, the close-in orbit (a\,=\,0.0056\,AU) and any orbital eccentricity, despite the short circularisation timescale of t$_\mathrm{circ}$\,=\,0.25\,Myr \citep{Heller2010, RRM2023}, would subject it to extreme levels of tidal forces resulting in a molten mantle, leading to a higher night-side temperature \citep{Henning2009,Driscoll2015,Herath2024,Boukare2025}. This change in the night-side temperature can be probed through the shape of the phase curve and a non-zero orbital eccentricity could be measured using mid-eclipse timing. Assuming a solid mantle and no internal heating, we predict $T_\mathrm{night}$\,=\,707\,K. Any significant deviations from this value (within uncertainties) could have implications on our understanding of the evolution of such extreme rocky worlds.

In light of recent research related to the characterization of the surface material components of rocky planets with MIRI \citep{Hu2012, Paragas2025}, TOI-4552\,b is a prime candidate for a phase curve study with MIRI following the recommendation by \citet{Hammond2025}. Phase curves are inherently immune to the major drawbacks of transit and secondary eclipse only observations such as the Transit Light Source (TLS;~\citealt{TLS}) effect and missing the transit/eclipse due to unconstrained orbital parameters. In addition, they provide a longer, stable baseline to accurately detect any secondary eclipse or transit features. USPs constitute an attractive subclass of rocky planets which enable full phase curve measurements with a relatively modest observing time investment. Obtaining a phase curve encompassing one transit and two eclipses would constitute an optimal strategy, enabling a stable baseline for the full dataset.

\section{Summary and Conclusion} \label{sec:conclusion}
 
We present the result of the NIRPS-GTO SP2 programme, confirmation of the planetary signal around the mid M dwarf TOI-4552. TOI-4552\,b is an Earth-sized ($R_p\, =\, 1.11 \pm 0.04\,R_{\oplus}$) planet as inferred from photometric lightcurves from TESS, ExTrA, SPECULOOS and LCO observatories. Our radial velocity analysis, which is primarily driven by NIRPS and corroborated by HARPS yields a mass of  $1.83 \pm 0.47\, M_{\oplus}$ corresponding to a bulk density of 
$\rho_\mathrm{p}=7.74\pm2.12$\,g/cm$^{3}$, marginally higher than the Earth's. Both spectroscopy and short and long term photometry show no indication of stellar activity, deeming TOI-4552 as a quiet star.

Based on the planetary mass and radius we ran an interior structure model using \texttt{exopie} which favoured a CMF of 0.54$^{+0.17}_{-0.25}$, partway between an Earth-like and super-Mercury composition. Additional RV measurements with greater precision (e.g., with ESPRESSO) are needed to make a definitive conclusion regarding the composition of TOI-4552\,b.

With recent works focused on investigating exoplanet surface geology using JWST \citep[e.g.,][]{Hu2012, Paragas2025}, TOI-4552\,b emerges as an excellent candidate. Its high emission spectroscopy metric (ESM = 19.5), likely circular orbit, ultra-short period (0.3011 days), and quiet host star make it a particularly favourable target for atmospheric and surface composition characterization with JWST.

\section*{Data availability}

Table~\ref{tab:RV_timeseries} is only available in electronic form at the CDS via anonymous ftp to cdsarc.u-strasbg.fr (130.79.128.5) or \url{http://cdsweb.u-strasbg.fr/cgi-bin/qcat?J/A+A/}.

\begin{acknowledgements}
RD, \'EA, CC, BB, PL, AL \& JPW  acknowledge the financial support of the FRQ-NT through the Centre de recherche en astrophysique du Qu\'ebec as well as the support from the Trottier Family Foundation and the Trottier Institute for Research on Exoplanets.\\
RD \& \'EA  acknowledges support from Canada Foundation for Innovation (CFI) program, the Universit\'e de Montr\'eal and Universit\'e Laval, the Canada Economic Development (CED) program and the Ministere of Economy, Innovation and Energy (MEIE).\\
ED-M, SCB, EC \& NCS  acknowledge the support from FCT - Funda\c{c}\~ao para a Ci\^encia e a Tecnologia through national funds by these grants: UIDB/04434/2020, UIDP/04434/2020.\\
ED-M  further acknowledges the support from FCT through Stimulus FCT contract 2021.01294.CEECIND. ED-M  acknowledges the support by the Ram\'on y Cajal contract RyC2022-035854-I funded by MICIU/AEI/10.13039/501100011033 and by ESF+.\\
The Board of Observational and Instrumental Astronomy (NAOS) at the Federal University of Rio Grande do Norte's research activities are supported by continuous grants from the Brazilian funding agency CNPq. This study was partially funded by the Coordena\c{c}\~ao de Aperfei\c{c}oamento de Pessoal de N\'ivel Superior—Brasil (CAPES) — Finance Code 001 and the CAPES-Print program.\\
0\\
XB, XDe \& TF  acknowledge funding from the French ANR under contract number ANR\-24\-CE49\-3397 (ORVET), and the French National Research Agency in the framework of the Investissements d'Avenir program (ANR-15-IDEX-02), through the funding of the ``Origin of Life" project of the Grenoble-Alpes University.\\
SCB   acknowledges the support from Funda\c{c}\~ao para a Ci\^encia e Tecnologia (FCT) in the form of a work contract through the Scientific Employment Incentive program with reference 2023.06687.CEECIND.\\
NBC  acknowledges support from an NSERC Discovery Grant, a Canada Research Chair, and an Arthur B. McDonald Fellowship, and thanks the Trottier Space Institute for its financial support and dynamic intellectual environment.\\
XDu  acknowledges the support from the European Research Council (ERC) under the European Union’s Horizon 2020 research and innovation programme (grant agreement SCORE No 851555) and from the Swiss National Science Foundation under the grant SPECTRE (No 200021\_215200).\\
This work has been carried out within the framework of the NCCR PlanetS supported by the Swiss National Science Foundation under grants 51NF40\_182901 and 51NF40\_205606.\\
DE  acknowledge support from the Swiss National Science Foundation for project 200021\_200726. The authors acknowledge the financial support of the SNSF.\\
JIGH, ASM, RR \& AKS  acknowledge financial support from the Spanish Ministry of Science, Innovation and Universities (MICIU) projects PID2020-117493GB-I00 and PID2023-149982NB-I00.\\
ICL  acknowledges CNPq research fellowships (Grant No. 313103/2022-4).\\
BLCM  acknowledge CAPES postdoctoral fellowships.\\
BLCM  acknowledges CNPq research fellowships (Grant No. 305804/2022-7).\\
JRM  acknowledges CNPq research fellowships (Grant No. 308928/2019-9).\\
CM  acknowledges the funding from the Swiss National Science Foundation under grant 200021\_204847 “PlanetsInTime”.\\
Co-funded by the European Union (ERC, FIERCE, 101052347). Views and opinions expressed are however those of the author(s) only and do not necessarily reflect those of the European Union or the European Research Council. Neither the European Union nor the granting authority can be held responsible for them.\\
GAW is supported by a Discovery Grant from the Natural Sciences and Engineering Research Council (NSERC) of Canada.\\
We acknowledge funding from the European Research Council under the ERC Grant Agreement n. 337591-ExTrA.\\
AL  acknowledges support from the Fonds de recherche du Qu\'ebec (FRQ) - Secteur Nature et technologies under file \#349961.\\
KAM  acknowledges support from the Swiss National Science Foundation (SNSF) under the Postdoc Mobility grant P500PT\_230225.\\
AKS  acknowledges financial support from La Caixa Foundation (ID 100010434) under the grant LCF/BQ/DI23/11990071.\\
Based on data collected by the SPECULOOS consortium. The ULiege's contribution to SPECULOOS has received funding from the European Research Council under the European Union's Seventh Framework Programme (FP/2007-2013) (grant Agreement n$^\circ$ 336480/SPECULOOS), from the Balzan Prize and Francqui Foundations, from the Belgian Scientific Research Foundation (F.R.S.-FNRS; grant n$^\circ$ T.0109.20), from the University of Liege, and from the ARC grant for Concerted Research Actions financed by the Wallonia-Brussels Federation. This work is supported by a grant from the Simons Foundation (PI Queloz, grant number 327127). This research is in part funded by the European Union's Horizon 2020 research and innovation program (grants agreements n$^{\circ}$ 803193/BEBOP), and from the Science and Technology Facilities Council (STFC; grant n$^\circ$ ST/S00193X/1, and ST/W000385/1)/\\
This research has made use of the NASA Exoplanet Archive, which is operated by the California Institute of Technology, under contract with the National Aeronautics and Space Administration under the Exoplanet Exploration Program.\\
We acknowledge the use of public TESS data from pipelines at the TESS Science Office and at the TESS Science Processing Operations Center. Resources supporting this work were provided by the NASA High-End Computing (HEC) Program through the NASA Advanced Supercomputing (NAS) Division at Ames Research Center for the production of the SPOC data products.\\
\end{acknowledgements}

\bibliography{aa58749-25.bib}{}

@ARTICLE{Han2025,
       author = {{Han}, Te and {Robertson}, Paul and {Brandt}, Timothy D. and {Kanodia}, Shubham and {Ca{\~n}as}, Caleb and {Shporer}, Avi and {Ricker}, George and {Beard}, Corey},
        title = "{Hundreds of TESS Exoplanets Might Be Larger than We Thought}",
      journal = {\apjl},
         year = 2025,
        month = jul,
       volume = {988},
       number = {1},
          eid = {L4},
        pages = {L4},
archivePrefix = {arXiv},
 primaryClass = {astro-ph.EP},
       adsurl = {https://ui.adsabs.harvard.edu/abs/2025ApJ...988L...4H}
}

@ARTICLE{Redfield2024,
       author = {{Redfield}, Seth and {Batalha}, Natasha and {Benneke}, Bj{\"o}rn and {Biller}, Beth and {Espinoza}, Nestor and {France}, Kevin and {Konopacky}, Quinn and {Kreidberg}, Laura and {Rauscher}, Emily and {Sing}, David},
        title = "{Report of the Working Group on Strategic Exoplanet Initiatives with HST and JWST}",
      journal = {arXiv e-prints},
         year = 2024,
        month = apr,
          eid = {arXiv:2404.02932},
        pages = {arXiv:2404.02932},
archivePrefix = {arXiv},
 primaryClass = {astro-ph.IM},
       adsurl = {https://ui.adsabs.harvard.edu/abs/2024arXiv240402932R}
}

@ARTICLE{Stassun2019,
       author = {{Stassun}, Keivan G. and {Oelkers}, Ryan J. and {Paegert}, Martin and {Torres}, Guillermo and {Pepper}, Joshua and {De Lee}, Nathan and {Collins}, Kevin and {Latham}, David W. and {Muirhead}, Philip S. and {Chittidi}, Jay and {Rojas-Ayala}, B{\'a}rbara and {Fleming}, Scott W. and {Rose}, Mark E. and {Tenenbaum}, Peter and {Ting}, Eric B. and {Kane}, Stephen R. and {Barclay}, Thomas and {Bean}, Jacob L. and {Brassuer}, C.~E. and {Charbonneau}, David and {Ge}, Jian and {Lissauer}, Jack J. and {Mann}, Andrew W. and {McLean}, Brian and {Mullally}, Susan and {Narita}, Norio and {Plavchan}, Peter and {Ricker}, George R. and {Sasselov}, Dimitar and {Seager}, S. and {Sharma}, Sanjib and {Shiao}, Bernie and {Sozzetti}, Alessandro and {Stello}, Dennis and {Vanderspek}, Roland and {Wallace}, Geoff and {Winn}, Joshua N.},
        title = "{The Revised TESS Input Catalog and Candidate Target List}",
      journal = {\aj},
         year = 2019,
        month = oct,
       volume = {158},
       number = {4},
          eid = {138},
        pages = {138},
archivePrefix = {arXiv},
 primaryClass = {astro-ph.SR},
       adsurl = {https://ui.adsabs.harvard.edu/abs/2019AJ....158..138S}
}

@ARTICLE{2MASS2006,
       author = {{Skrutskie}, M.~F. and {Cutri}, R.~M. and {Stiening}, R. and {Weinberg}, M.~D. and {Schneider}, S. and {Carpenter}, J.~M. and {Beichman}, C. and {Capps}, R. and {Chester}, T. and {Elias}, J. and {Huchra}, J. and {Liebert}, J. and {Lonsdale}, C. and {Monet}, D.~G. and {Price}, S. and {Seitzer}, P. and {Jarrett}, T. and {Kirkpatrick}, J.~D. and {Gizis}, J.~E. and {Howard}, E. and {Evans}, T. and {Fowler}, J. and {Fullmer}, L. and {Hurt}, R. and {Light}, R. and {Kopan}, E.~L. and {Marsh}, K.~A. and {McCallon}, H.~L. and {Tam}, R. and {Van Dyk}, S. and {Wheelock}, S.},
        title = "{The Two Micron All Sky Survey (2MASS)}",
      journal = {\aj},
         year = 2006,
        month = feb,
       volume = {131},
       number = {2},
        pages = {1163-1183},
       adsurl = {https://ui.adsabs.harvard.edu/abs/2006AJ....131.1163S}
}

@ARTICLE{Gaia2018,
       author = {{Gaia Collaboration} and {Brown}, A.~G.~A. and {Vallenari}, A. and {Prusti}, T. and {de Bruijne}, J.~H.~J. and {Babusiaux}, C. and {Bailer-Jones}, C.~A.~L. and {Biermann}, M. and {Evans}, D.~W. and {Eyer}, L. and {Jansen}, F. and {Jordi}, C. and {Klioner}, S.~A. and {Lammers}, U. and {Lindegren}, L. and {Luri}, X. and {Mignard}, F. and {Panem}, C. and {Pourbaix}, D. and {Randich}, S. and {Sartoretti}, P. and {Siddiqui}, H.~I. and {Soubiran}, C. and {van Leeuwen}, F. and {Walton}, N.~A. and {Arenou}, F. and {Bastian}, U. and {Cropper}, M. and {Drimmel}, R. and {Katz}, D. and {Lattanzi}, M.~G. and {Bakker}, J. and {Cacciari}, C. and {Casta{\~n}eda}, J. and {Chaoul}, L. and {Cheek}, N. and {De Angeli}, F. and {Fabricius}, C. and {Guerra}, R. and {Holl}, B. and {Masana}, E. and {Messineo}, R. and {Mowlavi}, N. and {Nienartowicz}, K. and {Panuzzo}, P. and {Portell}, J. and {Riello}, M. and {Seabroke}, G.~M. and {Tanga}, P. and {Th{\'e}venin}, F. and {Gracia-Abril}, G. and {Comoretto}, G. and {Garcia-Reinaldos}, M. and {Teyssier}, D. and {Altmann}, M. and {Andrae}, R. and {Audard}, M. and {Bellas-Velidis}, I. and {Benson}, K. and {Berthier}, J. and {Blomme}, R. and {Burgess}, P. and {Busso}, G. and {Carry}, B. and {Cellino}, A. and {Clementini}, G. and {Clotet}, M. and {Creevey}, O. and {Davidson}, M. and {De Ridder}, J. and {Delchambre}, L. and {Dell'Oro}, A. and {Ducourant}, C. and {Fern{\'a}ndez-Hern{\'a}ndez}, J. and {Fouesneau}, M. and {Fr{\'e}mat}, Y. and {Galluccio}, L. and {Garc{\'\i}a-Torres}, M. and {Gonz{\'a}lez-N{\'u}{\~n}ez}, J. and {Gonz{\'a}lez-Vidal}, J.~J. and {Gosset}, E. and {Guy}, L.~P. and {Halbwachs}, J. -L. and {Hambly}, N.~C. and {Harrison}, D.~L. and {Hern{\'a}ndez}, J. and {Hestroffer}, D. and {Hodgkin}, S.~T. and {Hutton}, A. and {Jasniewicz}, G. and {Jean-Antoine-Piccolo}, A. and {Jordan}, S. and {Korn}, A.~J. and {Krone-Martins}, A. and {Lanzafame}, A.~C. and {Lebzelter}, T. and {L{\"o}ffler}, W. and {Manteiga}, M. and {Marrese}, P.~M. and {Mart{\'\i}n-Fleitas}, J.~M. and {Moitinho}, A. and {Mora}, A. and {Muinonen}, K. and {Osinde}, J. and {Pancino}, E. and {Pauwels}, T. and {Petit}, J. -M. and {Recio-Blanco}, A. and {Richards}, P.~J. and {Rimoldini}, L. and {Robin}, A.~C. and {Sarro}, L.~M. and {Siopis}, C. and {Smith}, M. and {Sozzetti}, A. and {S{\"u}veges}, M. and {Torra}, J. and {van Reeven}, W. and {Abbas}, U. and {Abreu Aramburu}, A. and {Accart}, S. and {Aerts}, C. and {Altavilla}, G. and {{\'A}lvarez}, M.~A. and {Alvarez}, R. and {Alves}, J. and {Anderson}, R.~I. and {Andrei}, A.~H. and {Anglada Varela}, E. and {Antiche}, E. and {Antoja}, T. and {Arcay}, B. and {Astraatmadja}, T.~L. and {Bach}, N. and {Baker}, S.~G. and {Balaguer-N{\'u}{\~n}ez}, L. and {Balm}, P. and {Barache}, C. and {Barata}, C. and {Barbato}, D. and {Barblan}, F. and {Barklem}, P.~S. and {Barrado}, D. and {Barros}, M. and {Barstow}, M.~A. and {Bartholom{\'e} Mu{\~n}oz}, S. and {Bassilana}, J. -L. and {Becciani}, U. and {Bellazzini}, M. and {Berihuete}, A. and {Bertone}, S. and {Bianchi}, L. and {Bienaym{\'e}}, O. and {Blanco-Cuaresma}, S. and {Boch}, T. and {Boeche}, C. and {Bombrun}, A. and {Borrachero}, R. and {Bossini}, D. and {Bouquillon}, S. and {Bourda}, G. and {Bragaglia}, A. and {Bramante}, L. and {Breddels}, M.~A. and {Bressan}, A. and {Brouillet}, N. and {Br{\"u}semeister}, T. and {Brugaletta}, E. and {Bucciarelli}, B. and {Burlacu}, A. and {Busonero}, D. and {Butkevich}, A.~G. and {Buzzi}, R. and {Caffau}, E. and {Cancelliere}, R. and {Cannizzaro}, G. and {Cantat-Gaudin}, T. and {Carballo}, R. and {Carlucci}, T. and {Carrasco}, J.~M. and {Casamiquela}, L. and {Castellani}, M. and {Castro-Ginard}, A. and {Charlot}, P. and {Chemin}, L. and {Chiavassa}, A. and {Cocozza}, G. and {Costigan}, G. and {Cowell}, S. and {Crifo}, F. and {Crosta}, M. and {Crowley}, C. and {Cuypers}, J. and {Dafonte}, C. and {Damerdji}, Y. and {Dapergolas}, A. and {David}, P. and {David}, M. and {de Laverny}, P. and {De Luise}, F. and {De March}, R. and {de Martino}, D. and {de Souza}, R. and {de Torres}, A. and {Debosscher}, J. and {del Pozo}, E. and {Delbo}, M. and {Delgado}, A. and {Delgado}, H.~E. and {Di Matteo}, P. and {Diakite}, S. and {Diener}, C. and {Distefano}, E. and {Dolding}, C. and {Drazinos}, P. and {Dur{\'a}n}, J. and {Edvardsson}, B. and {Enke}, H. and {Eriksson}, K. and {Esquej}, P. and {Eynard Bontemps}, G. and {Fabre}, C. and {Fabrizio}, M. and {Faigler}, S. and {Falc{\~a}o}, A.~J. and {Farr{\`a}s Casas}, M. and {Federici}, L. and {Fedorets}, G. and {Fernique}, P. and {Figueras}, F. and {Filippi}, F. and {Findeisen}, K. and {Fonti}, A. and {Fraile}, E. and {Fraser}, M. and {Fr{\'e}zouls}, B. and {Gai}, M. and {Galleti}, S. and {Garabato}, D. and {Garc{\'\i}a-Sedano}, F. and {Garofalo}, A. and {Garralda}, N. and {Gavel}, A. and {Gavras}, P. and {Gerssen}, J. and {Geyer}, R. and {Giacobbe}, P. and {Gilmore}, G. and {Girona}, S. and {Giuffrida}, G. and {Glass}, F. and {Gomes}, M. and {Granvik}, M. and {Gueguen}, A. and {Guerrier}, A. and {Guiraud}, J. and {Guti{\'e}rrez-S{\'a}nchez}, R. and {Haigron}, R. and {Hatzidimitriou}, D. and {Hauser}, M. and {Haywood}, M. and {Heiter}, U. and {Helmi}, A. and {Heu}, J. and {Hilger}, T. and {Hobbs}, D. and {Hofmann}, W. and {Holland}, G. and {Huckle}, H.~E. and {Hypki}, A. and {Icardi}, V. and {Jan{\ss}en}, K. and {Jevardat de Fombelle}, G. and {Jonker}, P.~G. and {Juh{\'a}sz}, {\'A}. L. and {Julbe}, F. and {Karampelas}, A. and {Kewley}, A. and {Klar}, J. and {Kochoska}, A. and {Kohley}, R. and {Kolenberg}, K. and {Kontizas}, M. and {Kontizas}, E. and {Koposov}, S.~E. and {Kordopatis}, G. and {Kostrzewa-Rutkowska}, Z. and {Koubsky}, P. and {Lambert}, S. and {Lanza}, A.~F. and {Lasne}, Y. and {Lavigne}, J. -B. and {Le Fustec}, Y. and {Le Poncin-Lafitte}, C. and {Lebreton}, Y. and {Leccia}, S. and {Leclerc}, N. and {Lecoeur-Taibi}, I. and {Lenhardt}, H. and {Leroux}, F. and {Liao}, S. and {Licata}, E. and {Lindstr{\o}m}, H.~E.~P. and {Lister}, T.~A. and {Livanou}, E. and {Lobel}, A. and {L{\'o}pez}, M. and {Managau}, S. and {Mann}, R.~G. and {Mantelet}, G. and {Marchal}, O. and {Marchant}, J.~M. and {Marconi}, M. and {Marinoni}, S. and {Marschalk{\'o}}, G. and {Marshall}, D.~J. and {Martino}, M. and {Marton}, G. and {Mary}, N. and {Massari}, D. and {Matijevi{\v{c}}}, G. and {Mazeh}, T. and {McMillan}, P.~J. and {Messina}, S. and {Michalik}, D. and {Millar}, N.~R. and {Molina}, D. and {Molinaro}, R. and {Moln{\'a}r}, L. and {Montegriffo}, P. and {Mor}, R. and {Morbidelli}, R. and {Morel}, T. and {Morris}, D. and {Mulone}, A.~F. and {Muraveva}, T. and {Musella}, I. and {Nelemans}, G. and {Nicastro}, L. and {Noval}, L. and {O'Mullane}, W. and {Ord{\'e}novic}, C. and {Ord{\'o}{\~n}ez-Blanco}, D. and {Osborne}, P. and {Pagani}, C. and {Pagano}, I. and {Pailler}, F. and {Palacin}, H. and {Palaversa}, L. and {Panahi}, A. and {Pawlak}, M. and {Piersimoni}, A.~M. and {Pineau}, F. -X. and {Plachy}, E. and {Plum}, G. and {Poggio}, E. and {Poujoulet}, E. and {Pr{\v{s}}a}, A. and {Pulone}, L. and {Racero}, E. and {Ragaini}, S. and {Rambaux}, N. and {Ramos-Lerate}, M. and {Regibo}, S. and {Reyl{\'e}}, C. and {Riclet}, F. and {Ripepi}, V. and {Riva}, A. and {Rivard}, A. and {Rixon}, G. and {Roegiers}, T. and {Roelens}, M. and {Romero-G{\'o}mez}, M. and {Rowell}, N. and {Royer}, F. and {Ruiz-Dern}, L. and {Sadowski}, G. and {Sagrist{\`a} Sell{\'e}s}, T. and {Sahlmann}, J. and {Salgado}, J. and {Salguero}, E. and {Sanna}, N. and {Santana-Ros}, T. and {Sarasso}, M. and {Savietto}, H. and {Schultheis}, M. and {Sciacca}, E. and {Segol}, M. and {Segovia}, J.~C. and {S{\'e}gransan}, D. and {Shih}, I. -C. and {Siltala}, L. and {Silva}, A.~F. and {Smart}, R.~L. and {Smith}, K.~W. and {Solano}, E. and {Solitro}, F. and {Sordo}, R. and {Soria Nieto}, S. and {Souchay}, J. and {Spagna}, A. and {Spoto}, F. and {Stampa}, U. and {Steele}, I.~A. and {Steidelm{\"u}ller}, H. and {Stephenson}, C.~A. and {Stoev}, H. and {Suess}, F.~F. and {Surdej}, J. and {Szabados}, L. and {Szegedi-Elek}, E. and {Tapiador}, D. and {Taris}, F. and {Tauran}, G. and {Taylor}, M.~B. and {Teixeira}, R. and {Terrett}, D. and {Teyssandier}, P. and {Thuillot}, W. and {Titarenko}, A. and {Torra Clotet}, F. and {Turon}, C. and {Ulla}, A. and {Utrilla}, E. and {Uzzi}, S. and {Vaillant}, M. and {Valentini}, G. and {Valette}, V. and {van Elteren}, A. and {Van Hemelryck}, E. and {van Leeuwen}, M. and {Vaschetto}, M. and {Vecchiato}, A. and {Veljanoski}, J. and {Viala}, Y. and {Vicente}, D. and {Vogt}, S. and {von Essen}, C. and {Voss}, H. and {Votruba}, V. and {Voutsinas}, S. and {Walmsley}, G. and {Weiler}, M. and {Wertz}, O. and {Wevers}, T. and {Wyrzykowski}, {\L}. and {Yoldas}, A. and {{\v{Z}}erjal}, M. and {Ziaeepour}, H. and {Zorec}, J. and {Zschocke}, S. and {Zucker}, S. and {Zurbach}, C. and {Zwitter}, T.},
        title = "{Gaia Data Release 2. Summary of the contents and survey properties}",
      journal = {\aap},
         year = 2018,
        month = aug,
       volume = {616},
          eid = {A1},
        pages = {A1},
archivePrefix = {arXiv},
 primaryClass = {astro-ph.GA},
       adsurl = {https://ui.adsabs.harvard.edu/abs/2018A&A...616A...1G}
}

@ARTICLE{Kempton2018,
       author = {{Kempton}, Eliza M. -R. and {Bean}, Jacob L. and {Louie}, Dana R. and {Deming}, Drake and {Koll}, Daniel D.~B. and {Mansfield}, Megan and {Christiansen}, Jessie L. and {L{\'o}pez-Morales}, Mercedes and {Swain}, Mark R. and {Zellem}, Robert T. and {Ballard}, Sarah and {Barclay}, Thomas and {Barstow}, Joanna K. and {Batalha}, Natasha E. and {Beatty}, Thomas G. and {Berta-Thompson}, Zach and {Birkby}, Jayne and {Buchhave}, Lars A. and {Charbonneau}, David and {Cowan}, Nicolas B. and {Crossfield}, Ian and {de Val-Borro}, Miguel and {Doyon}, Ren{\'e} and {Dragomir}, Diana and {Gaidos}, Eric and {Heng}, Kevin and {Hu}, Renyu and {Kane}, Stephen R. and {Kreidberg}, Laura and {Mallonn}, Matthias and {Morley}, Caroline V. and {Narita}, Norio and {Nascimbeni}, Valerio and {Pall{\'e}}, Enric and {Quintana}, Elisa V. and {Rauscher}, Emily and {Seager}, Sara and {Shkolnik}, Evgenya L. and {Sing}, David K. and {Sozzetti}, Alessandro and {Stassun}, Keivan G. and {Valenti}, Jeff A. and {von Essen}, Carolina},
        title = "{A Framework for Prioritizing the TESS Planetary Candidates Most Amenable to Atmospheric Characterization}",
      journal = {pasp},
         year = 2018,
        month = nov,
       volume = {130},
       number = {993},
        pages = {114401},
archivePrefix = {arXiv},
 primaryClass = {astro-ph.EP},
       adsurl = {https://ui.adsabs.harvard.edu/abs/2018PASP..130k4401K}
}

@INPROCEEDINGS{Jenkins2016,
       author = {{Jenkins}, Jon M. and {Twicken}, Joseph D. and {McCauliff}, Sean and {Campbell}, Jennifer and {Sanderfer}, Dwight and {Lung}, David and {Mansouri-Samani}, Masoud and {Girouard}, Forrest and {Tenenbaum}, Peter and {Klaus}, Todd and {Smith}, Jeffrey C. and {Caldwell}, Douglas A. and {Chacon}, A.~D. and {Henze}, Christopher and {Heiges}, Cory and {Latham}, David W. and {Morgan}, Edward and {Swade}, Daryl and {Rinehart}, Stephen and {Vanderspek}, Roland},
        title = "{The TESS science processing operations center}",
    booktitle = {Software and Cyberinfrastructure for Astronomy IV},
         year = 2016,
       editor = {{Chiozzi}, Gianluca and {Guzman}, Juan C.},
       series = {Society of Photo-Optical Instrumentation Engineers (SPIE) Conference Series},
       volume = {9913},
        month = aug,
          eid = {99133E},
        pages = {99133E},
       adsurl = {https://ui.adsabs.harvard.edu/abs/2016SPIE.9913E..3EJ}
}

@ARTICLE{Aller2020,
       author = {{Aller}, A. and {Lillo-Box}, J. and {Jones}, D. and {Miranda}, L.~F. and {Barcel{\'o} Forteza}, S.},
        title = "{Planetary nebulae seen with TESS: Discovery of new binary central star candidates from Cycle 1}",
      journal = {\aap},
         year = 2020,
        month = mar,
       volume = {635},
          eid = {A128},
        pages = {A128},
archivePrefix = {arXiv},
 primaryClass = {astro-ph.SR},
       adsurl = {https://ui.adsabs.harvard.edu/abs/2020A&A...635A.128A}
}

@ARTICLE{Guerrero2021,
       author = {{Guerrero}, Natalia M. and {Seager}, S. and {Huang}, Chelsea X. and {Vanderburg}, Andrew and {Garcia Soto}, Aylin and {Mireles}, Ismael and {Hesse}, Katharine and {Fong}, William and {Glidden}, Ana and {Shporer}, Avi and {Latham}, David W. and {Collins}, Karen A. and {Quinn}, Samuel N. and {Burt}, Jennifer and {Dragomir}, Diana and {Crossfield}, Ian and {Vanderspek}, Roland and {Fausnaugh}, Michael and {Burke}, Christopher J. and {Ricker}, George and {Daylan}, Tansu and {Essack}, Zahra and {G{\"u}nther}, Maximilian N. and {Osborn}, Hugh P. and {Pepper}, Joshua and {Rowden}, Pamela and {Sha}, Lizhou and {Villanueva}, Steven, Jr. and {Yahalomi}, Daniel A. and {Yu}, Liang and {Ballard}, Sarah and {Batalha}, Natalie M. and {Berardo}, David and {Chontos}, Ashley and {Dittmann}, Jason A. and {Esquerdo}, Gilbert A. and {Mikal-Evans}, Thomas and {Jayaraman}, Rahul and {Krishnamurthy}, Akshata and {Louie}, Dana R. and {Mehrle}, Nicholas and {Niraula}, Prajwal and {Rackham}, Benjamin V. and {Rodriguez}, Joseph E. and {Rowden}, Stephen J.~L. and {Sousa-Silva}, Clara and {Watanabe}, David and {Wong}, Ian and {Zhan}, Zhuchang and {Zivanovic}, Goran and {Christiansen}, Jessie L. and {Ciardi}, David R. and {Swain}, Melanie A. and {Lund}, Michael B. and {Mullally}, Susan E. and {Fleming}, Scott W. and {Rodriguez}, David R. and {Boyd}, Patricia T. and {Quintana}, Elisa V. and {Barclay}, Thomas and {Col{\'o}n}, Knicole D. and {Rinehart}, S.~A. and {Schlieder}, Joshua E. and {Clampin}, Mark and {Jenkins}, Jon M. and {Twicken}, Joseph D. and {Caldwell}, Douglas A. and {Coughlin}, Jeffrey L. and {Henze}, Chris and {Lissauer}, Jack J. and {Morris}, Robert L. and {Rose}, Mark E. and {Smith}, Jeffrey C. and {Tenenbaum}, Peter and {Ting}, Eric B. and {Wohler}, Bill and {Bakos}, G. {\'A}. and {Bean}, Jacob L. and {Berta-Thompson}, Zachory K. and {Bieryla}, Allyson and {Bouma}, Luke G. and {Buchhave}, Lars A. and {Butler}, Nathaniel and {Charbonneau}, David and {Doty}, John P. and {Ge}, Jian and {Holman}, Matthew J. and {Howard}, Andrew W. and {Kaltenegger}, Lisa and {Kane}, Stephen R. and {Kjeldsen}, Hans and {Kreidberg}, Laura and {Lin}, Douglas N.~C. and {Minsky}, Charlotte and {Narita}, Norio and {Paegert}, Martin and {P{\'a}l}, Andr{\'a}s and {Palle}, Enric and {Sasselov}, Dimitar D. and {Spencer}, Alton and {Sozzetti}, Alessandro and {Stassun}, Keivan G. and {Torres}, Guillermo and {Udry}, Stephane and {Winn}, Joshua N.},
        title = "{The TESS Objects of Interest Catalog from the TESS Prime Mission}",
      journal = {\apjs},
         year = 2021,
        month = jun,
       volume = {254},
       number = {2},
          eid = {39},
        pages = {39},
archivePrefix = {arXiv},
 primaryClass = {astro-ph.EP},
       adsurl = {https://ui.adsabs.harvard.edu/abs/2021ApJS..254...39G}
}

@ARTICLE{Smith2012,
       author = {{Smith}, Jeffrey C. and {Stumpe}, Martin C. and {Van Cleve}, Jeffrey E. and {Jenkins}, Jon M. and {Barclay}, Thomas S. and {Fanelli}, Michael N. and {Girouard}, Forrest R. and {Kolodziejczak}, Jeffery J. and {McCauliff}, Sean D. and {Morris}, Robert L. and {Twicken}, Joseph D.},
        title = "{Kepler Presearch Data Conditioning II - A Bayesian Approach to Systematic Error Correction}",
      journal = {pasp},
         year = 2012,
        month = sep,
       volume = {124},
       number = {919},
        pages = {1000},
archivePrefix = {arXiv},
 primaryClass = {astro-ph.IM},
       adsurl = {https://ui.adsabs.harvard.edu/abs/2012PASP..124.1000S}
}

@ARTICLE{Stumpe2012,
       author = {{Stumpe}, Martin C. and {Smith}, Jeffrey C. and {Van Cleve}, Jeffrey E. and {Twicken}, Joseph D. and {Barclay}, Thomas S. and {Fanelli}, Michael N. and {Girouard}, Forrest R. and {Jenkins}, Jon M. and {Kolodziejczak}, Jeffery J. and {McCauliff}, Sean D. and {Morris}, Robert L.},
        title = "{Kepler Presearch Data Conditioning I{\textemdash}Architecture and Algorithms for Error Correction in Kepler Light Curves}",
      journal = {pasp},
         year = 2012,
        month = sep,
       volume = {124},
       number = {919},
        pages = {985},
archivePrefix = {arXiv},
 primaryClass = {astro-ph.IM},
       adsurl = {https://ui.adsabs.harvard.edu/abs/2012PASP..124..985S}
}

@ARTICLE{Stumpe2014,
       author = {{Stumpe}, Martin C. and {Smith}, Jeffrey C. and {Catanzarite}, Joseph H. and {Van Cleve}, Jeffrey E. and {Jenkins}, Jon M. and {Twicken}, Joseph D. and {Girouard}, Forrest R.},
        title = "{Multiscale Systematic Error Correction via Wavelet-Based Bandsplitting in Kepler Data}",
      journal = {pasp},
         year = 2014,
        month = jan,
       volume = {126},
       number = {935},
        pages = {100},
       adsurl = {https://ui.adsabs.harvard.edu/abs/2014PASP..126..100S}
}

@INPROCEEDINGS{Twicken2010,
       author = {{Twicken}, Joseph D. and {Clarke}, Bruce D. and {Bryson}, Stephen T. and {Tenenbaum}, Peter and {Wu}, Hayley and {Jenkins}, Jon M. and {Girouard}, Forrest and {Klaus}, Todd C.},
        title = "{Photometric analysis in the Kepler Science Operations Center pipeline}",
    booktitle = {Software and Cyberinfrastructure for Astronomy},
         year = 2010,
       editor = {{Radziwill}, Nicole M. and {Bridger}, Alan},
       series = {Society of Photo-Optical Instrumentation Engineers (SPIE) Conference Series},
       volume = {7740},
        month = jul,
          eid = {774023},
        pages = {774023},
       adsurl = {https://ui.adsabs.harvard.edu/abs/2010SPIE.7740E..23T}
}

@MISC{Morris2020,
       author = {{Morris}, Robert L. and {Twicken}, Joseph D. and {Smith}, Jeffrey C. and {Clarke}, Bruce D. and {Jenkins}, Jon M. and {Bryson}, Stephen T. and {Girouard}, Forrest and {Klaus}, Todd C.},
        title = "{Kepler Data Processing Handbook: Photometric Analysis}",
 howpublished = {Kepler Science Document KSCI-19081-003, id. 6. Edited by Jon M. Jenkins.},
         year = 2020,
        month = mar,
          eid = {6},
        pages = {6},
       adsurl = {https://ui.adsabs.harvard.edu/abs/2020ksci.rept....6M}
}

@ARTICLE{Twicken2018,
       author = {{Twicken}, Joseph D. and {Catanzarite}, Joseph H. and {Clarke}, Bruce D. and {Girouard}, Forrest and {Jenkins}, Jon M. and {Klaus}, Todd C. and {Li}, Jie and {McCauliff}, Sean D. and {Seader}, Shawn E. and {Tenenbaum}, Peter and {Wohler}, Bill and {Bryson}, Stephen T. and {Burke}, Christopher J. and {Caldwell}, Douglas A. and {Haas}, Michael R. and {Henze}, Christopher E. and {Sanderfer}, Dwight T.},
        title = "{Kepler Data Validation I{\textemdash}Architecture, Diagnostic Tests, and Data Products for Vetting Transiting Planet Candidates}",
      journal = {pasp},
         year = 2018,
        month = jun,
       volume = {130},
       number = {988},
        pages = {064502},
archivePrefix = {arXiv},
 primaryClass = {astro-ph.EP},
       adsurl = {https://ui.adsabs.harvard.edu/abs/2018PASP..130f4502T}
}

@ARTICLE{Howell2025,
       author = {{Howell}, Steve B. and {Mart{\'\i}nez-V{\'a}zquez}, Clara E. and {Furlan}, Elise and {Scott}, Nicholas J. and {Matson}, Rachel A. and {Littlefield}, Colin and {Clark}, Catherine A. and {Lester}, Kathryn V. and {Hartman}, Zachary D. and {Ciardi}, David R. and {Deveny}, Sarah J.},
        title = "{Nearly a decade of groundbreaking speckle interferometry at the international Gemini observatory}",
      journal = {Frontiers in Astronomy and Space Sciences},
         year = 2025,
        month = jun,
       volume = {12},
          eid = {1608411},
        pages = {1608411},
archivePrefix = {arXiv},
 primaryClass = {astro-ph.IM},
       adsurl = {https://ui.adsabs.harvard.edu/abs/2025FrASS..1208411H}
}

@ARTICLE{GaiaDR3,
       author = {{Gaia Collaboration} and {Vallenari}, A. and {Brown}, A.~G.~A. and {Prusti}, T. and {de Bruijne}, J.~H.~J. and {Arenou}, F. and {Babusiaux}, C. and {Biermann}, M. and {Creevey}, O.~L. and {Ducourant}, C. and {Evans}, D.~W. and {Eyer}, L. and {Guerra}, R. and {Hutton}, A. and {Jordi}, C. and {Klioner}, S.~A. and {Lammers}, U.~L. and {Lindegren}, L. and {Luri}, X. and {Mignard}, F. and {Panem}, C. and {Pourbaix}, D. and {Randich}, S. and {Sartoretti}, P. and {Soubiran}, C. and {Tanga}, P. and {Walton}, N.~A. and {Bailer-Jones}, C.~A.~L. and {Bastian}, U. and {Drimmel}, R. and {Jansen}, F. and {Katz}, D. and {Lattanzi}, M.~G. and {van Leeuwen}, F. and {Bakker}, J. and {Cacciari}, C. and {Casta{\~n}eda}, J. and {De Angeli}, F. and {Fabricius}, C. and {Fouesneau}, M. and {Fr{\'e}mat}, Y. and {Galluccio}, L. and {Guerrier}, A. and {Heiter}, U. and {Masana}, E. and {Messineo}, R. and {Mowlavi}, N. and {Nicolas}, C. and {Nienartowicz}, K. and {Pailler}, F. and {Panuzzo}, P. and {Riclet}, F. and {Roux}, W. and {Seabroke}, G.~M. and {Sordo}, R. and {Th{\'e}venin}, F. and {Gracia-Abril}, G. and {Portell}, J. and {Teyssier}, D. and {Altmann}, M. and {Andrae}, R. and {Audard}, M. and {Bellas-Velidis}, I. and {Benson}, K. and {Berthier}, J. and {Blomme}, R. and {Burgess}, P.~W. and {Busonero}, D. and {Busso}, G. and {C{\'a}novas}, H. and {Carry}, B. and {Cellino}, A. and {Cheek}, N. and {Clementini}, G. and {Damerdji}, Y. and {Davidson}, M. and {de Teodoro}, P. and {Nu{\~n}ez Campos}, M. and {Delchambre}, L. and {Dell'Oro}, A. and {Esquej}, P. and {Fern{\'a}ndez-Hern{\'a}ndez}, J. and {Fraile}, E. and {Garabato}, D. and {Garc{\'\i}a-Lario}, P. and {Gosset}, E. and {Haigron}, R. and {Halbwachs}, J. -L. and {Hambly}, N.~C. and {Harrison}, D.~L. and {Hern{\'a}ndez}, J. and {Hestroffer}, D. and {Hodgkin}, S.~T. and {Holl}, B. and {Jan{\ss}en}, K. and {Jevardat de Fombelle}, G. and {Jordan}, S. and {Krone-Martins}, A. and {Lanzafame}, A.~C. and {L{\"o}ffler}, W. and {Marchal}, O. and {Marrese}, P.~M. and {Moitinho}, A. and {Muinonen}, K. and {Osborne}, P. and {Pancino}, E. and {Pauwels}, T. and {Recio-Blanco}, A. and {Reyl{\'e}}, C. and {Riello}, M. and {Rimoldini}, L. and {Roegiers}, T. and {Rybizki}, J. and {Sarro}, L.~M. and {Siopis}, C. and {Smith}, M. and {Sozzetti}, A. and {Utrilla}, E. and {van Leeuwen}, M. and {Abbas}, U. and {{\'A}brah{\'a}m}, P. and {Abreu Aramburu}, A. and {Aerts}, C. and {Aguado}, J.~J. and {Ajaj}, M. and {Aldea-Montero}, F. and {Altavilla}, G. and {{\'A}lvarez}, M.~A. and {Alves}, J. and {Anders}, F. and {Anderson}, R.~I. and {Anglada Varela}, E. and {Antoja}, T. and {Baines}, D. and {Baker}, S.~G. and {Balaguer-N{\'u}{\~n}ez}, L. and {Balbinot}, E. and {Balog}, Z. and {Barache}, C. and {Barbato}, D. and {Barros}, M. and {Barstow}, M.~A. and {Bartolom{\'e}}, S. and {Bassilana}, J. -L. and {Bauchet}, N. and {Becciani}, U. and {Bellazzini}, M. and {Berihuete}, A. and {Bernet}, M. and {Bertone}, S. and {Bianchi}, L. and {Binnenfeld}, A. and {Blanco-Cuaresma}, S. and {Blazere}, A. and {Boch}, T. and {Bombrun}, A. and {Bossini}, D. and {Bouquillon}, S. and {Bragaglia}, A. and {Bramante}, L. and {Breedt}, E. and {Bressan}, A. and {Brouillet}, N. and {Brugaletta}, E. and {Bucciarelli}, B. and {Burlacu}, A. and {Butkevich}, A.~G. and {Buzzi}, R. and {Caffau}, E. and {Cancelliere}, R. and {Cantat-Gaudin}, T. and {Carballo}, R. and {Carlucci}, T. and {Carnerero}, M.~I. and {Carrasco}, J.~M. and {Casamiquela}, L. and {Castellani}, M. and {Castro-Ginard}, A. and {Chaoul}, L. and {Charlot}, P. and {Chemin}, L. and {Chiaramida}, V. and {Chiavassa}, A. and {Chornay}, N. and {Comoretto}, G. and {Contursi}, G. and {Cooper}, W.~J. and {Cornez}, T. and {Cowell}, S. and {Crifo}, F. and {Cropper}, M. and {Crosta}, M. and {Crowley}, C. and {Dafonte}, C. and {Dapergolas}, A. and {David}, M. and {David}, P. and {de Laverny}, P. and {De Luise}, F. and {De March}, R.},
        title = "{Gaia Data Release 3. Summary of the content and survey properties}",
      journal = {\aap},
         year = 2023,
        month = jun,
       volume = {674},
          eid = {A1},
        pages = {A1},
archivePrefix = {arXiv},
 primaryClass = {astro-ph.GA},
       adsurl = {https://ui.adsabs.harvard.edu/abs/2023A&A...674A...1G}
}

@ARTICLE{Scott2021,
       author = {{Scott}, Nicholas J. and {Howell}, Steve B. and {Gnilka}, Crystal L. and {Stephens}, Andrew W. and {Salinas}, Ricardo and {Matson}, Rachel A. and {Furlan}, Elise and {Horch}, Elliott P. and {Everett}, Mark E. and {Ciardi}, David R. and {Mills}, Dave and {Quigley}, Emmett A.},
        title = "{Twin High-resolution, High-speed Imagers for the Gemini Telescopes: Instrument description and science verification results}",
      journal = {Frontiers in Astronomy and Space Sciences},
         year = 2021,
        month = sep,
       volume = {8},
          eid = {138},
        pages = {138},
       adsurl = {https://ui.adsabs.harvard.edu/abs/2021FrASS...8..138S}
}

@ARTICLE{Collins:2017,
   author = {{Collins}, K.~A. and {Kielkopf}, J.~F. and {Stassun}, K.~G. and
    {Hessman}, F.~V.},
    title = "{AstroImageJ: Image Processing and Photometric Extraction for Ultra-precise Astronomical Light Curves}",
  journal = {\aj},
archivePrefix = "arXiv",
 primaryClass = "astro-ph.IM",
     year = 2017,
    month = feb,
   volume = 153,
      eid = {77},
    pages = {77},
   adsurl = {http://adsabs.harvard.edu/abs/2017AJ....153...77C}
}

@INPROCEEDINGS{collins2019,
       author = {{Collins}, Karen},
        title = "{TESS Follow-up Observing Program Working Group (TFOP WG) Sub Group 1 (SG1): Ground-based Time-series Photometry}",
    booktitle = {American Astronomical Society Meeting Abstracts \#233},
         year = 2019,
       series = {American Astronomical Society Meeting Abstracts},
       volume = {233},
        month = jan,
          eid = {140.05},
        pages = {140.05},
       adsurl = {https://ui.adsabs.harvard.edu/abs/2019AAS...23314005C}
}

@ARTICLE{Brown2013,
       author = {{Brown}, T.~M. and {Baliber}, N. and {Bianco}, F.~B. and {Bowman}, M. and {Burleson}, B. and {Conway}, P. and {Crellin}, M. and {Depagne}, {\'E}. and {De Vera}, J. and {Dilday}, B. and {Dragomir}, D. and {Dubberley}, M. and {Eastman}, J.~D. and {Elphick}, M. and {Falarski}, M. and {Foale}, S. and {Ford}, M. and {Fulton}, B.~J. and {Garza}, J. and {Gomez}, E.~L. and {Graham}, M. and {Greene}, R. and {Haldeman}, B. and {Hawkins}, E. and {Haworth}, B. and {Haynes}, R. and {Hidas}, M. and {Hjelstrom}, A.~E. and {Howell}, D.~A. and {Hygelund}, J. and {Lister}, T.~A. and {Lobdill}, R. and {Martinez}, J. and {Mullins}, D.~S. and {Norbury}, M. and {Parrent}, J. and {Paulson}, R. and {Petry}, D.~L. and {Pickles}, A. and {Posner}, V. and {Rosing}, W.~E. and {Ross}, R. and {Sand}, D.~J. and {Saunders}, E.~S. and {Shobbrook}, J. and {Shporer}, A. and {Street}, R.~A. and {Thomas}, D. and {Tsapras}, Y. and {Tufts}, J.~R. and {Valenti}, S. and {Vander Horst}, K. and {Walker}, Z. and {White}, G. and {Willis}, M.},
        title = "{Las Cumbres Observatory Global Telescope Network}",
      journal = {pasp},
         year = 2013,
        month = sep,
       volume = {125},
       number = {931},
        pages = {1031},
archivePrefix = {arXiv},
 primaryClass = {astro-ph.IM},
       adsurl = {https://ui.adsabs.harvard.edu/abs/2013PASP..125.1031B}
}

@INPROCEEDINGS{banzai,
       author = {{McCully}, Curtis and {Volgenau}, Nikolaus H. and {Harbeck}, Daniel-Rolf and {Lister}, Tim A. and {Saunders}, Eric S. and {Turner}, Monica L. and {Siiverd}, Robert J. and {Bowman}, Mark},
        title = "{Real-time processing of the imaging data from the network of Las Cumbres Observatory Telescopes using BANZAI}",
    booktitle = {Software and Cyberinfrastructure for Astronomy V},
         year = 2018,
       editor = {{Guzman}, Juan C. and {Ibsen}, Jorge},
       series = {Society of Photo-Optical Instrumentation Engineers (SPIE) Conference Series},
       volume = {10707},
        month = jul,
          eid = {107070K},
        pages = {107070K},
archivePrefix = {arXiv},
 primaryClass = {astro-ph.IM},
       adsurl = {https://ui.adsabs.harvard.edu/abs/2018SPIE10707E..0KM}
}

@INPROCEEDINGS{Bonfils2015,
       author = {{Bonfils}, X. and {Almenara}, J.~M. and {Jocou}, L. and {Wunsche}, A. and {Kern}, P. and {Delboulb{\'e}}, A. and {Delfosse}, X. and {Feautrier}, P. and {Forveille}, T. and {Gluck}, L. and {Lafrasse}, S. and {Magnard}, Y. and {Maurel}, D. and {Moulin}, T. and {Murgas}, F. and {Rabou}, P. and {Rochat}, S. and {Roux}, A. and {Stadler}, E.},
        title = "{ExTrA: Exoplanets in transit and their atmospheres}",
    booktitle = {Techniques and Instrumentation for Detection of Exoplanets VII},
         year = 2015,
       editor = {{Shaklan}, Stuart},
       series = {Society of Photo-Optical Instrumentation Engineers (SPIE) Conference Series},
       volume = {9605},
        month = sep,
          eid = {96051L},
        pages = {96051L},
archivePrefix = {arXiv},
 primaryClass = {astro-ph.IM},
       adsurl = {https://ui.adsabs.harvard.edu/abs/2015SPIE.9605E..1LB}
}

@ARTICLE{Cointepas2021,
       author = {{Cointepas}, M. and {Almenara}, J.~M. and {Bonfils}, X. and {Bouchy}, F. and {Astudillo-Defru}, N. and {Murgas}, F. and {Otegi}, J.~F. and {Wyttenbach}, A. and {Anderson}, D.~R. and {Artigau}, {\'E}. and {Canto Martins}, B.~L. and {Charbonneau}, D. and {Collins}, K.~A. and {Collins}, K.~I. and {Correia}, J. -J. and {Curaba}, S. and {Delboulb{\'e}}, A. and {Delfosse}, X. and {D{\'\i}az}, R.~F. and {Dorn}, C. and {Doyon}, R. and {Feautrier}, P. and {Figueira}, P. and {Forveille}, T. and {Gaisne}, G. and {Gan}, T. and {Gluck}, L. and {Helled}, R. and {Hellier}, C. and {Jocou}, L. and {Kern}, P. and {Lafrasse}, S. and {Law}, N. and {Le{\~a}o}, I.~C. and {Lovis}, C. and {Magnard}, Y. and {Mann}, A.~W. and {Maurel}, D. and {de Medeiros}, J.~R. and {Melo}, C. and {Moulin}, T. and {Pepe}, F. and {Rabou}, P. and {Rochat}, S. and {Rodriguez}, D.~R. and {Roux}, A. and {Santos}, N.~C. and {S{\'e}gransan}, D. and {Stadler}, E. and {Ting}, E.~B. and {Twicken}, J.~D. and {Udry}, S. and {Waalkes}, W.~C. and {West}, R.~G. and {W{\"u}nsche}, A. and {Ziegler}, C. and {Ricker}, G. and {Vanderspek}, R. and {Latham}, D.~W. and {Seager}, S. and {Winn}, J. and {Jenkins}, J.~M.},
        title = "{TOI-269 b: an eccentric sub-Neptune transiting a M2 dwarf revisited with ExTrA}",
      journal = {\aap},
         year = 2021,
        month = jun,
       volume = {650},
          eid = {A145},
        pages = {A145},
archivePrefix = {arXiv},
 primaryClass = {astro-ph.EP},
       adsurl = {https://ui.adsabs.harvard.edu/abs/2021A&A...650A.145C}
}

@INPROCEEDINGS{Delrez2018,
       author = {{Delrez}, Laetitia and {Gillon}, Micha{\"e}l. and {Queloz}, Didier and {Demory}, Brice-Olivier and {Almleaky}, Yaseen and {de Wit}, Julien and {Jehin}, Emmanu{\"e}l. and {Triaud}, Amaury H.~M.~J. and {Barkaoui}, Khalid and {Burdanov}, Artem and {Burgasser}, Adam J. and {Ducrot}, Elsa and {McCormac}, James and {Murray}, Catriona and {Silva Fernandes}, Catarina and {Sohy}, Sandrine and {Thompson}, Samantha J. and {Van Grootel}, Val{\'e}rie and {Alonso}, Roi and {Benkhaldoun}, Zouhair and {Rebolo}, Rafael},
        title = "{SPECULOOS: a network of robotic telescopes to hunt for terrestrial planets around the nearest ultracool dwarfs}",
    booktitle = {Ground-based and Airborne Telescopes VII},
         year = 2018,
       editor = {{Marshall}, Heather K. and {Spyromilio}, Jason},
       series = {Society of Photo-Optical Instrumentation Engineers (SPIE) Conference Series},
       volume = {10700},
        month = jul,
          eid = {107001I},
        pages = {107001I},
archivePrefix = {arXiv},
 primaryClass = {astro-ph.IM},
       adsurl = {https://ui.adsabs.harvard.edu/abs/2018SPIE10700E..1ID}
}

@ARTICLE{Gunther2021,
       author = {{G{\"u}nther}, Maximilian N. and {Daylan}, Tansu},
        title = "{Allesfitter: Flexible Star and Exoplanet Inference from Photometry and Radial Velocity}",
      journal = {\apjs},
         year = 2021,
        month = may,
       volume = {254},
       number = {1},
          eid = {13},
        pages = {13},
archivePrefix = {arXiv},
 primaryClass = {astro-ph.EP},
       adsurl = {https://ui.adsabs.harvard.edu/abs/2021ApJS..254...13G}
}

@ARTICLE{Antoniadis24,
       author = {{Antoniadis-Karnavas}, A. and {Sousa}, S.~G. and {Delgado-Mena}, E. and {Santos}, N.~C. and {Andreasen}, D.~T.},
        title = "{ODUSSEAS: Upgraded version with new reference scale and parameter determinations for 82 planet-host M dwarf stars in SWEET-Cat}",
      journal = {\aap},
         year = 2024,
        month = oct,
       volume = {690},
          eid = {A58},
        pages = {A58},
archivePrefix = {arXiv},
 primaryClass = {astro-ph.SR},
       adsurl = {https://ui.adsabs.harvard.edu/abs/2024A&A...690A..58A}
}

@ARTICLE{Antoniadis20,
       author = {{Antoniadis-Karnavas}, A. and {Sousa}, S.~G. and {Delgado-Mena}, E. and {Santos}, N.~C. and {Teixeira}, G.~D.~C. and {Neves}, V.},
        title = "{ODUSSEAS: a machine learning tool to derive effective temperature and metallicity for M dwarf stars}",
      journal = {\aap},
         year = 2020,
        month = apr,
       volume = {636},
          eid = {A9},
        pages = {A9},
archivePrefix = {arXiv},
 primaryClass = {astro-ph.SR},
       adsurl = {https://ui.adsabs.harvard.edu/abs/2020A&A...636A...9A}
}

@ARTICLE{Neves12,
       author = {{Neves}, V. and {Bonfils}, X. and {Santos}, N.~C. and {Delfosse}, X. and {Forveille}, T. and {Allard}, F. and {Nat{\'a}rio}, C. and {Fernandes}, C.~S. and {Udry}, S.},
        title = "{Metallicity of M dwarfs. II. A comparative study of photometric metallicity scales}",
      journal = {\aap},
         year = 2012,
        month = feb,
       volume = {538},
          eid = {A25},
        pages = {A25},
archivePrefix = {arXiv},
 primaryClass = {astro-ph.SR},
       adsurl = {https://ui.adsabs.harvard.edu/abs/2012A&A...538A..25N}
}

@ARTICLE{Khata21,
       author = {{Khata}, Dhrimadri and {Mondal}, Soumen and {Das}, Ramkrishna and {Baug}, Tapas},
        title = "{Estimating T$_{eff}$, radius, and luminosity of M-dwarfs using high-resolution optical and NIR spectral features}",
      journal = {\mnras},
         year = 2021,
        month = oct,
       volume = {507},
       number = {2},
        pages = {1869-1885},
archivePrefix = {arXiv},
 primaryClass = {astro-ph.SR},
       adsurl = {https://ui.adsabs.harvard.edu/abs/2021MNRAS.507.1869K}
}

@ARTICLE{Mann2019,
       author = {{Mann}, Andrew W. and {Dupuy}, Trent and {Kraus}, Adam L. and {Gaidos}, Eric and {Ansdell}, Megan and {Ireland}, Michael and {Rizzuto}, Aaron C. and {Hung}, Chao-Ling and {Dittmann}, Jason and {Factor}, Samuel and {Feiden}, Gregory and {Martinez}, Raquel A. and {Ru{\'\i}z-Rodr{\'\i}guez}, Dary and {Thao}, Pa Chia},
        title = "{How to Constrain Your M Dwarf. II. The Mass-Luminosity-Metallicity Relation from 0.075 to 0.70 Solar Masses}",
      journal = {\apj},
         year = 2019,
        month = jan,
       volume = {871},
       number = {1},
          eid = {63},
        pages = {63},
archivePrefix = {arXiv},
 primaryClass = {astro-ph.SR},
       adsurl = {https://ui.adsabs.harvard.edu/abs/2019ApJ...871...63M}
}

@ARTICLE{Mann2015,
       author = {{Mann}, Andrew W. and {Feiden}, Gregory A. and {Gaidos}, Eric and {Boyajian}, Tabetha and {von Braun}, Kaspar},
        title = "{How to Constrain Your M Dwarf: Measuring Effective Temperature, Bolometric Luminosity, Mass, and Radius}",
      journal = {\apj},
         year = 2015,
        month = may,
       volume = {804},
       number = {1},
          eid = {64},
        pages = {64},
archivePrefix = {arXiv},
 primaryClass = {astro-ph.SR},
       adsurl = {https://ui.adsabs.harvard.edu/abs/2015ApJ...804...64M}
}

@ARTICLE{Artigau2024,
       author = {{Artigau}, {\'E}tienne and {Cadieux}, Charles and {Cook}, Neil J. and {Doyon}, Ren{\'e} and {Dauplaise}, Laurie and {Arnold}, Luc and {Cadieux}, Maya and {Donati}, Jean-Fran{\c{c}}ois and {Cristofari}, Paul and {Delfosse}, Xavier and {Fouqu{\'e}}, Pascal and {Moutou}, Claire and {Larue}, Pierre and {Allart}, Romain},
        title = "{Measuring Sub-Kelvin Variations in Stellar Temperature with High-resolution Spectroscopy}",
      journal = {\aj},
         year = 2024,
        month = dec,
       volume = {168},
       number = {6},
          eid = {252},
        pages = {252},
archivePrefix = {arXiv},
 primaryClass = {astro-ph.SR},
       adsurl = {https://ui.adsabs.harvard.edu/abs/2024AJ....168..252A}
}

@ARTICLE{Artigau2022,
       author = {{Artigau}, {\'E}tienne and {Cadieux}, Charles and {Cook}, Neil J. and {Doyon}, Ren{\'e} and {Vandal}, Thomas and {Donati}, Jean-Fran{\c{c}}ois and {Moutou}, Claire and {Delfosse}, Xavier and {Fouqu{\'e}}, Pascal and {Martioli}, Eder and {Bouchy}, Fran{\c{c}}ois and {Parsons}, Jasmine and {Carmona}, Andres and {Dumusque}, Xavier and {Astudillo-Defru}, Nicola and {Bonfils}, Xavier and {Mignon}, Lucille},
        title = "{Line-by-line Velocity Measurements: an Outlier-resistant Method for Precision Velocimetry}",
      journal = {\aj},
         year = 2022,
        month = sep,
       volume = {164},
       number = {3},
          eid = {84},
        pages = {84},
archivePrefix = {arXiv},
 primaryClass = {astro-ph.IM},
       adsurl = {https://ui.adsabs.harvard.edu/abs/2022AJ....164...84A}
}

@ARTICLE{Kochanek2017,
       author = {{Kochanek}, C.~S. and {Shappee}, B.~J. and {Stanek}, K.~Z. and {Holoien}, T.~W. -S. and {Thompson}, Todd A. and {Prieto}, J.~L. and {Dong}, Subo and {Shields}, J.~V. and {Will}, D. and {Britt}, C. and {Perzanowski}, D. and {Pojma{\'n}ski}, G.},
        title = "{The All-Sky Automated Survey for Supernovae (ASAS-SN) Light Curve Server v1.0}",
      journal = {pasp},
         year = 2017,
        month = oct,
       volume = {129},
       number = {980},
        pages = {104502},
archivePrefix = {arXiv},
 primaryClass = {astro-ph.SR},
       adsurl = {https://ui.adsabs.harvard.edu/abs/2017PASP..129j4502K}
}

@ARTICLE{ASASSN1,
       author = {{Hart}, K. and {Shappee}, B.~J. and {Hey}, D. and {Kochanek}, C.~S. and {Stanek}, K.~Z. and {Lim}, L. and {Dobbs}, S. and {Tucker}, M. and {Jayasinghe}, T. and {Beacom}, J.~F. and {Boright}, T. and {Holoien}, T. and {Ong}, J.~M. Joel and {Prieto}, J.~L. and {Thompson}, T.~A. and {Will}, D.},
        title = "{ASAS-SN Sky Patrol V2.0}",
      journal = {arXiv e-prints},
         year = 2023,
        month = apr,
          eid = {arXiv:2304.03791},
        pages = {arXiv:2304.03791},
archivePrefix = {arXiv},
 primaryClass = {astro-ph.IM},
       adsurl = {https://ui.adsabs.harvard.edu/abs/2023arXiv230403791H}
}

@ARTICLE{ASASSN2,
       author = {{Shappee}, B.~J. and {Prieto}, J.~L. and {Grupe}, D. and {Kochanek}, C.~S. and {Stanek}, K.~Z. and {De Rosa}, G. and {Mathur}, S. and {Zu}, Y. and {Peterson}, B.~M. and {Pogge}, R.~W. and {Komossa}, S. and {Im}, M. and {Jencson}, J. and {Holoien}, T.~W. -S. and {Basu}, U. and {Beacom}, J.~F. and {Szczygie{\l}}, D.~M. and {Brimacombe}, J. and {Adams}, S. and {Campillay}, A. and {Choi}, C. and {Contreras}, C. and {Dietrich}, M. and {Dubberley}, M. and {Elphick}, M. and {Foale}, S. and {Giustini}, M. and {Gonzalez}, C. and {Hawkins}, E. and {Howell}, D.~A. and {Hsiao}, E.~Y. and {Koss}, M. and {Leighly}, K.~M. and {Morrell}, N. and {Mudd}, D. and {Mullins}, D. and {Nugent}, J.~M. and {Parrent}, J. and {Phillips}, M.~M. and {Pojmanski}, G. and {Rosing}, W. and {Ross}, R. and {Sand}, D. and {Terndrup}, D.~M. and {Valenti}, S. and {Walker}, Z. and {Yoon}, Y.},
        title = "{The Man behind the Curtain: X-Rays Drive the UV through NIR Variability in the 2013 Active Galactic Nucleus Outburst in NGC 2617}",
      journal = {\apj},
         year = 2014,
        month = jun,
       volume = {788},
       number = {1},
          eid = {48},
        pages = {48},
archivePrefix = {arXiv},
 primaryClass = {astro-ph.HE},
       adsurl = {https://ui.adsabs.harvard.edu/abs/2014ApJ...788...48S}
}

@ARTICLE{ASM2025,
       author = {{Su{\'a}rez Mascare{\~n}o}, Alejandro and {Artigau}, {\'E}tienne and {Mignon}, Lucile and {Delfosse}, Xavier and {Cook}, Neil J. and {Bouchy}, Fran{\c{c}}ois and {Doyon}, Ren{\'e} and {Gonz{\'a}lez Hern{\'a}ndez}, Jonay I. and {Vandal}, Thomas and {de Castro Le{\~a}o}, Izan and {Stefanov}, Atanas K. and {Faria}, Jo{\~a}o and {Cadieux}, Charles and {Lamontagne}, Pierrot and {Baron}, Fr{\'e}d{\'e}rique and {Barros}, Susana C.~C. and {Benneke}, Bj{\"o}rn and {Bonfils}, Xavier and {Bryan}, Marta and {Martins}, Bruno L. Canto and {Cloutier}, Ryan and {Cowan}, Nicolas B. and {de Freitas}, Daniel Brito and {De Medeiros}, Jose Renan and {Delgado-Mena}, Elisa and {Figueira}, Pedro and {Dumusque}, Xavier and {Ehrenreich}, David and {Lafreni{\`e}re}, David and {Lovis}, Christophe and {Malo}, Lison and {Melo}, Claudio and {Mordasini}, Christoph and {Pepe}, Francesco and {Rebolo}, Rafael and {Rowe}, Jason and {Santos}, Nuno C. and {S{\'e}gransan}, Damien and {Udry}, St{\'e}phane and {Valencia}, Diana and {Wade}, Gregg and {Abreu}, Manuel and {Aguiar}, Jos{\'e} L.~A. and {Al Moulla}, Khaled and {Allain}, Guillaume and {Allart}, Romain and {Arial}, Tomy and {Auger}, Hugues and {Bazinet}, Luc and {Blind}, Nicolas and {Bohlender}, David and {Boisse}, Isabelle and {Boucher}, Anne and {Bourrier}, Vincent and {Bovay}, S{\'e}bastien and {Broeg}, Christopher and {Brousseau}, Denis and {Cabral}, Alexandre and {Carmona}, Andres and {Carteret}, Yann and {Challita}, Zalpha and {Chazelas}, Bruno and {Coelho}, Jo{\~a}o and {Cointepas}, Marion and {Conod}, Uriel and {Cristo}, Eduardo and {Silva}, Ana Rita Costa and {Darveau-Bernier}, Antoine and {Dauplaise}, Laurie and {Delisle}, Jean-Baptiste and {de Lima Gomes}, Roseane and {Forveille}, Thierry and {Frensch}, Yolanda G.~C. and {T{\'e}mich}, F{\'e}lix Gracia and {Fontinele}, Dasaev O. and {Gagn{\'e}}, Jonathan and {Genest}, Fr{\'e}d{\'e}ric and {Genolet}, Ludovic and {Gomes da Silva}, Jo{\~a}o and {Grieves}, Nolan and {Hernandez}, Olivier and {Hobson}, Melissa J. and {Hoeijmakers}, H. Jens and {Hubin}, Norbert and {Jahandar}, Farbod and {Jayawardhana}, Ray and {K{\"a}ufl}, Hans-Ulrich and {Kerley}, Dan and {Kolb}, Johann and {Krishnamurthy}, Vigneshwaran and {Kung}, Benjamin and {L'Heureux}, Alexandrine and {Larue}, Pierre and {Leath}, Henry and {Lim}, Olivia and {Lo Curto}, Gaspare and {Martins}, Allan M. and {Matthews}, Jaymie and {Mayer}, Jean-S{\'e}bastien and {Messias}, Yuri S. and {Metchev}, Stan and {Moranta}, Leslie and {Mounzer}, Dany and {Nari}, Nicola and {Nielsen}, Louise D. and {Osborn}, Ares and {Ouellet}, Mathieu and {Otegi}, Jon and {Parc}, L{\'e}na and {Pasquini}, Luca and {Passegger}, Vera M. and {Pelletier}, Stefan and {Peroux}, C{\'e}line and {Piaulet-Ghorayeb}, Caroline and {Plotnykov}, Mykhaylo and {Pompei}, Emanuela and {Poulin-Girard}, Anne-Sophie and {Rasilla}, Jos{\'e} Luis and {Reshetov}, Vladimir and {Saint-Antoine}, Jonathan and {Sarajlic}, Mirsad and {Saviane}, Ivo and {Schnell}, Robin and {Segovia}, Alex and {Seidel}, Julia and {Silber}, Armin and {Sinclair}, Peter and {Sordet}, Michael and {Sosnowska}, Danuta and {Srivastava}, Avidaan and {Teixeira}, M{\'a}rcio A. and {Thibault}, Simon and {Vall{\'e}e}, Philippe and {Vaulato}, Valentina and {Wardenier}, Joost P. and {Wehbe}, Bachar and {Weisserman}, Drew and {Wevers}, Ivan and {Wildi}, Fran{\c{c}}ois and {Yariv}, Vincent and {Zins}, G{\'e}rard},
        title = "{Diving into the planetary system of Proxima with NIRPS: Breaking the metre per second barrier in the infrared}",
      journal = {\aap},
         year = 2025,
        month = aug,
       volume = {700},
          eid = {A11},
        pages = {A11},
archivePrefix = {arXiv},
 primaryClass = {astro-ph.EP},
       adsurl = {https://ui.adsabs.harvard.edu/abs/2025A&A...700A..11S}
}

@ARTICLE{ASM2016,
       author = {{Su{\'a}rez Mascare{\~n}o}, A. and {Rebolo}, R. and {Gonz{\'a}lez Hern{\'a}ndez}, J.~I.},
        title = "{Magnetic cycles and rotation periods of late-type stars from photometric time series}",
      journal = {\aap},
         year = 2016,
        month = oct,
       volume = {595},
          eid = {A12},
        pages = {A12},
archivePrefix = {arXiv},
 primaryClass = {astro-ph.SR},
       adsurl = {https://ui.adsabs.harvard.edu/abs/2016A&A...595A..12S}
}

@ARTICLE{Bayo2008,
       author = {{Bayo}, A. and {Rodrigo}, C. and {Barrado Y Navascu{\'e}s}, D. and {Solano}, E. and {Guti{\'e}rrez}, R. and {Morales-Calder{\'o}n}, M. and {Allard}, F.},
        title = "{VOSA: virtual observatory SED analyzer. An application to the Collinder 69 open cluster}",
      journal = {\aap},
         year = 2008,
        month = dec,
       volume = {492},
       number = {1},
        pages = {277-287},
archivePrefix = {arXiv},
 primaryClass = {astro-ph},
       adsurl = {https://ui.adsabs.harvard.edu/abs/2008A&A...492..277B}
}

@INPROCEEDINGS{Henden2015,
       author = {{Henden}, Arne A. and {Levine}, Stephen and {Terrell}, Dirk and {Welch}, Douglas L.},
        title = "{APASS - The Latest Data Release}",
    booktitle = {American Astronomical Society Meeting Abstracts \#225},
         year = 2015,
       series = {American Astronomical Society Meeting Abstracts},
       volume = {225},
        month = jan,
          eid = {336.16},
        pages = {336.16},
       adsurl = {https://ui.adsabs.harvard.edu/abs/2015AAS...22533616H}
}

@ARTICLE{Shadab2015,
       author = {{Alam}, Shadab and {Albareti}, Franco D. and {Allende Prieto}, Carlos and {Anders}, F. and {Anderson}, Scott F. and {Anderton}, Timothy and {Andrews}, Brett H. and {Armengaud}, Eric and {Aubourg}, {\'E}ric and {Bailey}, Stephen and {Basu}, Sarbani and {Bautista}, Julian E. and {Beaton}, Rachael L. and {Beers}, Timothy C. and {Bender}, Chad F. and {Berlind}, Andreas A. and {Beutler}, Florian and {Bhardwaj}, Vaishali and {Bird}, Jonathan C. and {Bizyaev}, Dmitry and {Blake}, Cullen H. and {Blanton}, Michael R. and {Blomqvist}, Michael and {Bochanski}, John J. and {Bolton}, Adam S. and {Bovy}, Jo and {Shelden Bradley}, A. and {Brandt}, W.~N. and {Brauer}, D.~E. and {Brinkmann}, J. and {Brown}, Peter J. and {Brownstein}, Joel R. and {Burden}, Angela and {Burtin}, Etienne and {Busca}, Nicol{\'a}s G. and {Cai}, Zheng and {Capozzi}, Diego and {Carnero Rosell}, Aurelio and {Carr}, Michael A. and {Carrera}, Ricardo and {Chambers}, K.~C. and {Chaplin}, William James and {Chen}, Yen-Chi and {Chiappini}, Cristina and {Chojnowski}, S. Drew and {Chuang}, Chia-Hsun and {Clerc}, Nicolas and {Comparat}, Johan and {Covey}, Kevin and {Croft}, Rupert A.~C. and {Cuesta}, Antonio J. and {Cunha}, Katia and {da Costa}, Luiz N. and {Da Rio}, Nicola and {Davenport}, James R.~A. and {Dawson}, Kyle S. and {De Lee}, Nathan and {Delubac}, Timoth{\'e}e and {Deshpande}, Rohit and {Dhital}, Saurav and {Dutra-Ferreira}, Let{\'\i}cia and {Dwelly}, Tom and {Ealet}, Anne and {Ebelke}, Garrett L. and {Edmondson}, Edward M. and {Eisenstein}, Daniel J. and {Ellsworth}, Tristan and {Elsworth}, Yvonne and {Epstein}, Courtney R. and {Eracleous}, Michael and {Escoffier}, Stephanie and {Esposito}, Massimiliano and {Evans}, Michael L. and {Fan}, Xiaohui and {Fern{\'a}ndez-Alvar}, Emma and {Feuillet}, Diane and {Filiz Ak}, Nurten and {Finley}, Hayley and {Finoguenov}, Alexis and {Flaherty}, Kevin and {Fleming}, Scott W. and {Font-Ribera}, Andreu and {Foster}, Jonathan and {Frinchaboy}, Peter M. and {Galbraith-Frew}, J.~G. and {Garc{\'\i}a}, Rafael A. and {Garc{\'\i}a-Hern{\'a}ndez}, D.~A. and {Garc{\'\i}a P{\'e}rez}, Ana E. and {Gaulme}, Patrick and {Ge}, Jian and {G{\'e}nova-Santos}, R. and {Georgakakis}, A. and {Ghezzi}, Luan and {Gillespie}, Bruce A. and {Girardi}, L{\'e}o and {Goddard}, Daniel and {Gontcho}, Satya Gontcho A. and {Gonz{\'a}lez Hern{\'a}ndez}, Jonay I. and {Grebel}, Eva K. and {Green}, Paul J. and {Grieb}, Jan Niklas and {Grieves}, Nolan and {Gunn}, James E. and {Guo}, Hong and {Harding}, Paul and {Hasselquist}, Sten and {Hawley}, Suzanne L. and {Hayden}, Michael and {Hearty}, Fred R. and {Hekker}, Saskia and {Ho}, Shirley and {Hogg}, David W. and {Holley-Bockelmann}, Kelly and {Holtzman}, Jon A. and {Honscheid}, Klaus and {Huber}, Daniel and {Huehnerhoff}, Joseph and {Ivans}, Inese I. and {Jiang}, Linhua and {Johnson}, Jennifer A. and {Kinemuchi}, Karen and {Kirkby}, David and {Kitaura}, Francisco and {Klaene}, Mark A. and {Knapp}, Gillian R. and {Kneib}, Jean-Paul and {Koenig}, Xavier P. and {Lam}, Charles R. and {Lan}, Ting-Wen and {Lang}, Dustin and {Laurent}, Pierre and {Le Goff}, Jean-Marc and {Leauthaud}, Alexie and {Lee}, Khee-Gan and {Lee}, Young Sun and {Licquia}, Timothy C. and {Liu}, Jian and {Long}, Daniel C. and {L{\'o}pez-Corredoira}, Mart{\'\i}n and {Lorenzo-Oliveira}, Diego and {Lucatello}, Sara and {Lundgren}, Britt and {Lupton}, Robert H. and {Mack}, III, Claude E. and {Mahadevan}, Suvrath and {Maia}, Marcio A.~G. and {Majewski}, Steven R. and {Malanushenko}, Elena and {Malanushenko}, Viktor and {Manchado}, A. and {Manera}, Marc and {Mao}, Qingqing and {Maraston}, Claudia and {Marchwinski}, Robert C. and {Margala}, Daniel and {Martell}, Sarah L. and {Martig}, Marie and {Masters}, Karen L. and {Mathur}, Savita and {McBride}, Cameron K. and {McGehee}, Peregrine M. and {McGreer}, Ian D. and {McMahon}, Richard G. and {M{\'e}nard}, Brice and {Menzel}, Marie-Luise and {Merloni}, Andrea and {M{\'e}sz{\'a}ros}, Szabolcs and {Miller}, Adam A. and {Miralda-Escud{\'e}}, Jordi and {Miyatake}, Hironao and {Montero-Dorta}, Antonio D. and {More}, Surhud and {Morganson}, Eric and {Morice-Atkinson}, Xan and {Morrison}, Heather L. and {Mosser}, Ben{\^o}it and {Muna}, Demitri and {Myers}, Adam D. and {Nandra}, Kirpal and {Newman}, Jeffrey A. and {Neyrinck}, Mark and {Nguyen}, Duy Cuong and {Nichol}, Robert C. and {Nidever}, David L. and {Noterdaeme}, Pasquier and {Nuza}, Sebasti{\'a}n E. and {O'Connell}, Julia E. and {O'Connell}, Robert W. and {O'Connell}, Ross and {Ogando}, Ricardo L.~C. and {Olmstead}, Matthew D. and {Oravetz}, Audrey E. and {Oravetz}, Daniel J. and {Osumi}, Keisuke and {Owen}, Russell and {Padgett}, Deborah L. and {Padmanabhan}, Nikhil and {Paegert}, Martin and {Palanque-Delabrouille}, Nathalie and {Pan}, Kaike},
        title = "{The Eleventh and Twelfth Data Releases of the Sloan Digital Sky Survey: Final Data from SDSS-III}",
      journal = {\apjs},
         year = 2015,
        month = jul,
       volume = {219},
       number = {1},
          eid = {12},
        pages = {12},
archivePrefix = {arXiv},
 primaryClass = {astro-ph.IM},
       adsurl = {https://ui.adsabs.harvard.edu/abs/2015ApJS..219...12A}
}

@ARTICLE{WISE2010,
       author = {{Wright}, Edward L. and {Eisenhardt}, Peter R.~M. and {Mainzer}, Amy K. and {Ressler}, Michael E. and {Cutri}, Roc M. and {Jarrett}, Thomas and {Kirkpatrick}, J. Davy and {Padgett}, Deborah and {McMillan}, Robert S. and {Skrutskie}, Michael and {Stanford}, S.~A. and {Cohen}, Martin and {Walker}, Russell G. and {Mather}, John C. and {Leisawitz}, David and {Gautier}, III, Thomas N. and {McLean}, Ian and {Benford}, Dominic and {Lonsdale}, Carol J. and {Blain}, Andrew and {Mendez}, Bryan and {Irace}, William R. and {Duval}, Valerie and {Liu}, Fengchuan and {Royer}, Don and {Heinrichsen}, Ingolf and {Howard}, Joan and {Shannon}, Mark and {Kendall}, Martha and {Walsh}, Amy L. and {Larsen}, Mark and {Cardon}, Joel G. and {Schick}, Scott and {Schwalm}, Mark and {Abid}, Mohamed and {Fabinsky}, Beth and {Naes}, Larry and {Tsai}, Chao-Wei},
        title = "{The Wide-field Infrared Survey Explorer (WISE): Mission Description and Initial On-orbit Performance}",
      journal = {\aj},
         year = 2010,
        month = dec,
       volume = {140},
       number = {6},
        pages = {1868-1881},
archivePrefix = {arXiv},
 primaryClass = {astro-ph.IM},
       adsurl = {https://ui.adsabs.harvard.edu/abs/2010AJ....140.1868W}
}

@ARTICLE{Allard2012,
       author = {{Allard}, F. and {Homeier}, D. and {Freytag}, B.},
        title = "{Models of very-low-mass stars, brown dwarfs and exoplanets}",
      journal = {Philosophical Transactions of the Royal Society of London Series A},
         year = 2012,
        month = jun,
       volume = {370},
       number = {1968},
        pages = {2765-2777},
archivePrefix = {arXiv},
 primaryClass = {astro-ph.SR},
       adsurl = {https://ui.adsabs.harvard.edu/abs/2012RSPTA.370.2765A}
}

@BOOK{Kurucz1993,
       author = {{Kurucz}, Robert L.},
        title = "{SYNTHE spectrum synthesis programs and line data}",
         year = 1993,
       adsurl = {https://ui.adsabs.harvard.edu/abs/1993sssp.book.....K}
}

@INPROCEEDINGS{Castelli2003,
       author = {{Castelli}, F. and {Kurucz}, R.~L.},
        title = "{New Grids of ATLAS9 Model Atmospheres}",
    booktitle = {Modelling of Stellar Atmospheres},
         year = 2003,
       editor = {{Piskunov}, N. and {Weiss}, W.~W. and {Gray}, D.~F.},
       series = {IAU Symposium},
       volume = {210},
        month = jan,
        pages = {A20},
archivePrefix = {arXiv},
 primaryClass = {astro-ph},
       adsurl = {https://ui.adsabs.harvard.edu/abs/2003IAUS..210P.A20C}
}

@ARTICLE{Schweitzer2019,
       author = {{Schweitzer}, A. and {Passegger}, V.~M. and {Cifuentes}, C. and {B{\'e}jar}, V.~J.~S. and {Cort{\'e}s-Contreras}, M. and {Caballero}, J.~A. and {del Burgo}, C. and {Czesla}, S. and {K{\"u}rster}, M. and {Montes}, D. and {Zapatero Osorio}, M.~R. and {Ribas}, I. and {Reiners}, A. and {Quirrenbach}, A. and {Amado}, P.~J. and {Aceituno}, J. and {Anglada-Escud{\'e}}, G. and {Bauer}, F.~F. and {Dreizler}, S. and {Jeffers}, S.~V. and {Guenther}, E.~W. and {Henning}, T. and {Kaminski}, A. and {Lafarga}, M. and {Marfil}, E. and {Morales}, J.~C. and {Schmitt}, J.~H.~M.~M. and {Seifert}, W. and {Solano}, E. and {Tabernero}, H.~M. and {Zechmeister}, M.},
        title = "{The CARMENES search for exoplanets around M dwarfs. Different roads to radii and masses of the target stars}",
      journal = {\aap},
         year = 2019,
        month = may,
       volume = {625},
          eid = {A68},
        pages = {A68},
archivePrefix = {arXiv},
 primaryClass = {astro-ph.SR},
       adsurl = {https://ui.adsabs.harvard.edu/abs/2019A&A...625A..68S}
}

@ARTICLE{juliet,
       author = {{Espinoza}, N{\'e}stor and {Kossakowski}, Diana and {Brahm}, Rafael},
        title = "{juliet: a versatile modelling tool for transiting and non-transiting exoplanetary systems}",
      journal = {\mnras},
         year = 2019,
        month = dec,
       volume = {490},
       number = {2},
        pages = {2262-2283},
archivePrefix = {arXiv},
 primaryClass = {astro-ph.EP},
       adsurl = {https://ui.adsabs.harvard.edu/abs/2019MNRAS.490.2262E}
}

@ARTICLE{Lyu2024,
       author = {{Lyu}, Xintong and {Koll}, Daniel D.~B. and {Cowan}, Nicolas B. and {Hu}, Renyu and {Kreidberg}, Laura and {Rose}, Brian E.~J.},
        title = "{Super-Earth LHS3844b is Tidally Locked}",
      journal = {\apj},
         year = 2024,
        month = apr,
       volume = {964},
       number = {2},
          eid = {152},
        pages = {152},
archivePrefix = {arXiv},
 primaryClass = {astro-ph.EP},
       adsurl = {https://ui.adsabs.harvard.edu/abs/2024ApJ...964..152L}
}

@ARTICLE{Heller2010,
       author = {{Heller}, R. and {Jackson}, B. and {Barnes}, R. and {Greenberg}, R. and {Homeier}, D.},
        title = "{Tidal effects on brown dwarfs: application to the eclipsing binary 2MASS J05352184-0546085. The anomalous temperature reversal in the context of tidal heating}",
      journal = {\aap},
         year = 2010,
        month = may,
       volume = {514},
          eid = {A22},
        pages = {A22},
archivePrefix = {arXiv},
 primaryClass = {astro-ph.SR},
       adsurl = {https://ui.adsabs.harvard.edu/abs/2010A&A...514A..22H}
}

@ARTICLE{Kipping2013,
       author = {{Kipping}, David M.},
        title = "{Efficient, uninformative sampling of limb darkening coefficients for two-parameter laws}",
      journal = {\mnras},
         year = 2013,
        month = nov,
       volume = {435},
       number = {3},
        pages = {2152-2160},
archivePrefix = {arXiv},
 primaryClass = {astro-ph.SR},
       adsurl = {https://ui.adsabs.harvard.edu/abs/2013MNRAS.435.2152K}
}

@ARTICLE{Espinoza2016,
       author = {{Espinoza}, N{\'e}stor and {Jord{\'a}n}, Andr{\'e}s},
        title = "{Limb darkening and exoplanets - II. Choosing the best law for optimal retrieval of transit parameters}",
      journal = {\mnras},
         year = 2016,
        month = apr,
       volume = {457},
       number = {4},
        pages = {3573-3581},
archivePrefix = {arXiv},
 primaryClass = {astro-ph.EP},
       adsurl = {https://ui.adsabs.harvard.edu/abs/2016MNRAS.457.3573E}
}

@ARTICLE{Espinoza2018,
       author = {{Espinoza}, N{\'e}stor},
        title = "{Efficient Joint Sampling of Impact Parameters and Transit Depths in Transiting Exoplanet Light Curves}",
      journal = {Research Notes of the American Astronomical Society},
         year = 2018,
        month = nov,
       volume = {2},
       number = {4},
          eid = {209},
        pages = {209},
archivePrefix = {arXiv},
 primaryClass = {astro-ph.EP},
       adsurl = {https://ui.adsabs.harvard.edu/abs/2018RNAAS...2..209E}
}

@ARTICLE{Bouchy2025,
       author = {{Bouchy}, Fran{\c{c}}ois and {Doyon}, Ren{\'e} and {Pepe}, Francesco and {Melo}, Claudio and {Artigau}, {\'E}tienne and {Malo}, Lison and {Wildi}, Fran{\c{c}}ois and {Baron}, Fr{\'e}d{\'e}rique and {Delfosse}, Xavier and {De Medeiros}, Jose Renan and {Rebolo}, Rafael and {Santos}, Nuno C. and {Wade}, Gregg and {Allart}, Romain and {Al Moulla}, Khaled and {Blind}, Nicolas and {Cadieux}, Charles and {Canto Martins}, Bruno L. and {Cook}, Neil J. and {Dumusque}, Xavier and {Frensch}, Yolanda and {Genest}, Fr{\'e}d{\'e}ric and {Gonz{\'a}lez Hern{\'a}ndez}, Jonay I. and {Grieves}, Nolan and {Lo Curto}, Gaspare and {Lovis}, Christophe and {Mignon}, Lucile and {Nielsen}, Louise D. and {Poulin-Girard}, Anne-Sophie and {Rasilla}, Jos{\'e} Luis and {Reshetov}, Vladimir and {Sosnowska}, Danuta and {Sordet}, Michael and {Saint-Antoine}, Jonathan and {Su{\'a}rez Mascare{\~n}o}, Alejandro and {Thibault}, Simon and {Vall{\'e}e}, Philippe and {Vandal}, Thomas and {Abreu}, Manuel and {Aguiar}, Jos{\'e} L.~A. and {Allain}, Guillaume and {Arial}, Tomy and {Auger}, Hugues and {Barros}, Susana C.~C. and {Bazinet}, Luc and {Benneke}, Bj{\"o}rn and {Bonfils}, Xavier and {Boucher}, Anne and {Bourrier}, Vincent and {Bovay}, S{\'e}bastien and {Broeg}, Christopher and {Brousseau}, Denis and {Bruniquel}, Vincent and {Bryan}, Marta and {Cabral}, Alexandre and {Carmona}, Andres and {Carteret}, Yann and {Challita}, Zalpha and {Chazelas}, Bruno and {Cloutier}, Ryan and {Coelho}, Jo{\~a}o and {Cointepas}, Marion and {Conod}, Uriel and {Cowan}, Nicolas B. and {Cristo}, Eduardo and {Gomes da Silva}, Jo{\~a}o and {Dauplaise}, Laurie and {Darveau-Bernier}, Antoine and {de Lima Gomes}, Roseane and {de Freitas}, Daniel Brito and {Delgado-Mena}, Elisa and {Delisle}, Jean-Baptiste and {Ehrenreich}, David and {Faria}, Jo{\~a}o and {Figueira}, Pedro and {Fontinele}, Dasaev O. and {Forveille}, Thierry and {Gagn{\'e}}, Jonathan and {Genolet}, Ludovic and {T{\'e}mich}, F{\'e}lix Gracia and {Hernandez}, Olivier and {Hobson}, Melissa J. and {Hoeijmakers}, Jens and {Hubin}, Norbert and {Jahandar}, Farbod and {Jayawardhana}, Ray and {K{\"a}ufl}, Hans-Ulrich and {Kerley}, Dan and {Kolb}, Johann and {Krishnamurthy}, Vigneshwaran and {Lafreni{\`e}re}, David and {Lamontagne}, Pierrot and {Larue}, Pierre and {Leath}, Henry and {L'Heureux}, Alexandrine and {de Castro Le{\~a}o}, Izan and {Lim}, Olivia and {Martins}, Allan M. and {Matthews}, Jaymie and {Mayer}, Jean-S{\'e}bastien and {Messias}, Yuri S. and {Metchev}, Stan and {Moranta}, Leslie and {Mordasini}, Christoph and {Mounzer}, Dany and {Nari}, Nicola and {Osborn}, Ares and {Ouellet}, Mathieu and {Otegi}, Jon and {Parc}, L{\'e}na and {Pasquini}, Luca and {Passegger}, Vera M. and {Pelletier}, Stefan and {Peroux}, C{\'e}line and {Piaulet-Ghorayeb}, Caroline and {Plotnykov}, Mykhaylo and {Pompei}, Emanuela and {Rowe}, Jason and {Sarajlic}, Mirsad and {Segovia}, Alex and {Seidel}, Julia and {S{\'e}gransan}, Damien and {Schnell}, Robin and {Costa Silva}, Ana Rita and {Srivastava}, Avidaan and {Stefanov}, Atanas K. and {Teixeira}, M{\'a}rcio A. and {Udry}, St{\'e}phane and {Valencia}, Diana and {Vaulato}, Valentina and {Wardenier}, Joost P. and {Wehbe}, Bachar and {Weisserman}, Drew and {Wevers}, Ivan and {Yariv}, Vincent and {Zins}, G{\'e}rard},
        title = "{NIRPS joining HARPS at ESO 3.6 m: On-sky performance and science objectives}",
      journal = {\aap},
         year = 2025,
        month = aug,
       volume = {700},
          eid = {A10},
        pages = {A10},
archivePrefix = {arXiv},
 primaryClass = {astro-ph.IM},
       adsurl = {https://ui.adsabs.harvard.edu/abs/2025A&A...700A..10B}
}

@ARTICLE{Bouchy2017,
       author = {{Bouchy}, F. and {Doyon}, R. and {Artigau}, {\'E}. and {Melo}, C. and {Hernandez}, O. and {Wildi}, F. and {Delfosse}, X. and {Lovis}, C. and {Figueira}, P. and {Canto Martins}, B.~L. . and {Gonz{\'a}lez Hern{\'a}ndez}, J.~I. . and {Thibault}, S. and {Reshetov}, V. and {Pepe}, F. and {Santos}, N.~C. and {de Medeiros}, J.~R. . and {Rebolo}, R. and {Abreu}, M. and {Adibekyan}, V.~Z. and {Bandy}, T. and {Benz}, W. and {Blind}, N. and {Bohlender}, D. and {Boisse}, I. and {Bovay}, S. and {Broeg}, C. and {Brousseau}, D. and {Cabral}, A. and {Chazelas}, B. and {Cloutier}, R. and {Coelho}, J. and {Conod}, U. and {Cumming}, A. and {Delabre}, B. and {Genolet}, L. and {Hagelberg}, J. and {Jayawardhana}, R. and {K{\"a}ufl}, H. -U. and {Lafreni{\`e}re}, D. and {de Castro Le{\~a}o}, I. . and {Malo}, L. and {de Medeiros Martins}, A. . and {Matthews}, J.~M. and {Metchev}, S. and {Oshagh}, M. and {Ouellet}, M. and {Parro}, V.~C. and {Rasilla Pi{\~n}eiro}, J.~L. . and {Santos}, P. and {Sarajlic}, M. and {Segovia}, A. and {Sordet}, M. and {Udry}, S. and {Valencia}, D. and {Vall{\'e}e}, P. and {Venn}, K. and {Wade}, G.~A. and {Saddlemyer}, L.},
        title = "{Near-InfraRed Planet Searcher to Join HARPS on the ESO 3.6-metre Telescope}",
      journal = {The Messenger},
         year = 2017,
        month = sep,
       volume = {169},
        pages = {21-27},
       adsurl = {https://ui.adsabs.harvard.edu/abs/2017Msngr.169...21B}
}

@ARTICLE{Allart2022,
       author = {{Allart}, R. and {Lovis}, C. and {Faria}, J. and {Dumusque}, X. and {Sosnowska}, D. and {Figueira}, P. and {Silva}, A.~M. and {Mehner}, A. and {Pepe}, F. and {Cristiani}, S. and {Rebolo}, R. and {Santos}, N.~C. and {Adibekyan}, V. and {Cupani}, G. and {Di Marcantonio}, P. and {D'Odorico}, V. and {Gonz{\'a}lez Hern{\'a}ndez}, J.~I. and {Martins}, C.~J.~A.~P. and {Milakovi{\'c}}, D. and {Nunes}, N.~J. and {Sozzetti}, A. and {Su{\'a}rez Mascare{\~n}o}, A. and {Tabernero}, H. and {Zapatero Osorio}, M.~R.},
        title = "{Automatic model-based telluric correction for the ESPRESSO data reduction software. Model description and application to radial velocity computation}",
      journal = {\aap},
         year = 2022,
        month = oct,
       volume = {666},
          eid = {A196},
        pages = {A196},
archivePrefix = {arXiv},
 primaryClass = {astro-ph.EP},
       adsurl = {https://ui.adsabs.harvard.edu/abs/2022A&A...666A.196A}
}

@ARTICLE{Donati2025,
       author = {{Donati}, J. -F. and {Cristofari}, P.~I. and {Moutou}, C. and {L'Heureux}, A. and {Cook}, N.~J. and {Artigau}, E. and {Alencar}, S.~H.~P. and {Gaidos}, E. and {Vidotto}, A. and {Petit}, P. and {Carmona}, A. and {Ray}, T.},
        title = "{Six-year SPIRou monitoring of the young planet-host AU Mic}",
      journal = {\aap},
         year = 2025,
        month = aug,
       volume = {700},
          eid = {A227},
        pages = {A227},
archivePrefix = {arXiv},
 primaryClass = {astro-ph.SR},
       adsurl = {https://ui.adsabs.harvard.edu/abs/2025A&A...700A.227D}
}

@ARTICLE{Stock2023,
       author = {{Stock}, Stephan and {Kemmer}, Jonas and {Kossakowski}, Diana and {Sabotta}, Silvia and {Reffert}, Sabine and {Quirrenbach}, Andreas},
        title = "{Gaussian processes for radial velocity modeling. Better rotation periods and planetary parameters with the quasi-periodic kernel and constrained priors}",
      journal = {\aap},
         year = 2023,
        month = jun,
       volume = {674},
          eid = {A108},
        pages = {A108},
archivePrefix = {arXiv},
 primaryClass = {astro-ph.EP},
       adsurl = {https://ui.adsabs.harvard.edu/abs/2023A&A...674A.108S}
}

@ARTICLE{Wordsworth2022,
       author = {{Wordsworth}, Robin and {Kreidberg}, Laura},
        title = "{Atmospheres of Rocky Exoplanets}",
      journal = {\araa},
         year = 2022,
        month = aug,
       volume = {60},
        pages = {159-201},
archivePrefix = {arXiv},
 primaryClass = {astro-ph.EP},
       adsurl = {https://ui.adsabs.harvard.edu/abs/2022ARA&A..60..159W}
}

@ARTICLE{Luque2025,
       author = {{Luque}, Rafael and {Coy}, Brandon Park and {Xue}, Qiao and {Feinstein}, Adina D. and {Ahrer}, Eva-Maria and {Changeat}, Quentin and {Zhang}, Michael and {Moran}, Sarah E. and {Bean}, Jacob L. and {Kite}, Edwin and {Weiner Mansfield}, Megan and {Pall{\'e}}, Enric},
        title = "{A Dark, Bare Rock for TOI-1685 b from a JWST NIRSpec G395H Phase Curve}",
      journal = {\aj},
         year = 2025,
        month = jul,
       volume = {170},
       number = {1},
          eid = {49},
        pages = {49},
archivePrefix = {arXiv},
 primaryClass = {astro-ph.EP},
       adsurl = {https://ui.adsabs.harvard.edu/abs/2025AJ....170...49L}
}

@ARTICLE{DL2021,
       author = {{Diamond-Lowe}, Hannah and {Youngblood}, Allison and {Charbonneau}, David and {King}, George and {Teal}, D.~J. and {Bastelberger}, Sandra and {Corrales}, Lia and {Kempton}, Eliza M. -R.},
        title = "{The High-energy Spectrum of the Nearby Planet-hosting Inactive Mid-M Dwarf LHS 3844}",
      journal = {\aj},
         year = 2021,
        month = jul,
       volume = {162},
       number = {1},
          eid = {10},
        pages = {10},
archivePrefix = {arXiv},
 primaryClass = {astro-ph.SR},
       adsurl = {https://ui.adsabs.harvard.edu/abs/2021AJ....162...10D}
}

@ARTICLE{Zilinskas2025,
       author = {{Zilinskas}, M. and {van Buchem}, C.~P.~A. and {Zieba}, S. and {Miguel}, Y. and {Sandford}, E. and {Hu}, R. and {Patel}, J.~A. and {Bello-Arufe}, A. and {Janssen}, L.~J. and {Tsai}, S. -M. and {Dragomir}, D. and {Zhang}, M.},
        title = "{Characterising the atmosphere of 55 Cancri e: 1D forward model grid for current and future JWST observations}",
      journal = {\aap},
         year = 2025,
        month = may,
       volume = {697},
          eid = {A34},
        pages = {A34},
archivePrefix = {arXiv},
 primaryClass = {astro-ph.EP},
       adsurl = {https://ui.adsabs.harvard.edu/abs/2025A&A...697A..34Z}
}

@ARTICLE{Hu2024,
       author = {{Hu}, Renyu and {Bello-Arufe}, Aaron and {Zhang}, Michael and {Paragas}, Kimberly and {Zilinskas}, Mantas and {van Buchem}, Christiaan and {Bess}, Michael and {Patel}, Jayshil and {Ito}, Yuichi and {Damiano}, Mario and {Scheucher}, Markus and {Oza}, Apurva V. and {Knutson}, Heather A. and {Miguel}, Yamila and {Dragomir}, Diana and {Brandeker}, Alexis and {Demory}, Brice-Olivier},
        title = "{A secondary atmosphere on the rocky exoplanet 55 Cancri e}",
      journal = {\nat},
         year = 2024,
        month = jun,
       volume = {630},
       number = {8017},
        pages = {609-612},
archivePrefix = {arXiv},
 primaryClass = {astro-ph.EP},
       adsurl = {https://ui.adsabs.harvard.edu/abs/2024Natur.630..609H}
}

@ARTICLE{Morris2022,
       author = {{Morris}, Brett M. and {Heng}, Kevin and {Jones}, Kathryn and {Piaulet}, Caroline and {Demory}, Brice-Olivier and {Kitzmann}, Daniel and {Jens Hoeijmakers}, H.},
        title = "{Physically-motivated basis functions for temperature maps of exoplanets}",
      journal = {\aap},
         year = 2022,
        month = apr,
       volume = {660},
          eid = {A123},
        pages = {A123},
archivePrefix = {arXiv},
 primaryClass = {astro-ph.EP},
       adsurl = {https://ui.adsabs.harvard.edu/abs/2022A&A...660A.123M}
}

@ARTICLE{Kane2020,
       author = {{Kane}, Stephen R. and {Roettenbacher}, Rachael M. and {Unterborn}, Cayman T. and {Foley}, Bradford J. and {Hill}, Michelle L.},
        title = "{A Volatile-poor Formation of LHS 3844b Based on Its Lack of Significant Atmosphere}",
      journal = {psj},
         year = 2020,
        month = sep,
       volume = {1},
       number = {2},
          eid = {36},
        pages = {36},
archivePrefix = {arXiv},
 primaryClass = {astro-ph.EP},
       adsurl = {https://ui.adsabs.harvard.edu/abs/2020PSJ.....1...36K}
}

@ARTICLE{Paragas2025,
       author = {{Paragas}, Kimberly and {Knutson}, Heather A. and {Hu}, Renyu and {Ehlmann}, Bethany L. and {Alemanno}, Giulia and {Helbert}, J{\"o}rn and {Maturilli}, Alessandro and {Zhang}, Michael and {Iyer}, Aishwarya and {Rossman}, George},
        title = "{A New Spectral Library for Modeling the Surfaces of Hot, Rocky Exoplanets}",
      journal = {\apj},
         year = 2025,
        month = mar,
       volume = {981},
       number = {2},
          eid = {130},
        pages = {130},
archivePrefix = {arXiv},
 primaryClass = {astro-ph.EP},
       adsurl = {https://ui.adsabs.harvard.edu/abs/2025ApJ...981..130P}
}

@ARTICLE{Hu2012,
       author = {{Hu}, Renyu and {Ehlmann}, Bethany L. and {Seager}, Sara},
        title = "{Theoretical Spectra of Terrestrial Exoplanet Surfaces}",
      journal = {\apj},
         year = 2012,
        month = jun,
       volume = {752},
       number = {1},
          eid = {7},
        pages = {7},
archivePrefix = {arXiv},
 primaryClass = {astro-ph.EP},
       adsurl = {https://ui.adsabs.harvard.edu/abs/2012ApJ...752....7H}
}

@ARTICLE{Henning2009,
       author = {{Henning}, Wade G. and {O'Connell}, Richard J. and {Sasselov}, Dimitar D.},
        title = "{Tidally Heated Terrestrial Exoplanets: Viscoelastic Response Models}",
      journal = {\apj},
         year = 2009,
        month = dec,
       volume = {707},
       number = {2},
        pages = {1000-1015},
archivePrefix = {arXiv},
 primaryClass = {astro-ph.EP},
       adsurl = {https://ui.adsabs.harvard.edu/abs/2009ApJ...707.1000H}
}

@ARTICLE{TLS,
       author = {{Rackham}, Benjamin V. and {Apai}, D{\'a}niel and {Giampapa}, Mark S.},
        title = "{The Transit Light Source Effect: False Spectral Features and Incorrect Densities for M-dwarf Transiting Planets}",
      journal = {\apj},
         year = 2018,
        month = feb,
       volume = {853},
       number = {2},
          eid = {122},
        pages = {122},
archivePrefix = {arXiv},
 primaryClass = {astro-ph.EP},
       adsurl = {https://ui.adsabs.harvard.edu/abs/2018ApJ...853..122R}
}

@ARTICLE{toi6255,
       author = {{Dai}, Fei and {Howard}, Andrew W. and {Halverson}, Samuel and {Orell-Miquel}, Jaume and {Pall{\'e}}, Enric and {Isaacson}, Howard and {Fulton}, Benjamin and {Price}, Ellen M. and {Plotnykov}, Mykhaylo and {Rogers}, Leslie A. and {Valencia}, Diana and {Paragas}, Kimberly and {Greklek-McKeon}, Michael and {Gomez Barrientos}, Jonathan and {Knutson}, Heather A. and {Petigura}, Erik A. and {Weiss}, Lauren M. and {Lee}, Rena and {Brinkman}, Casey L. and {Huber}, Daniel and {Stef{\'a}nsson}, Gumundur and {Masuda}, Kento and {Giacalone}, Steven and {Lu}, Cicero X. and {Kite}, Edwin S. and {Hu}, Renyu and {Gaidos}, Eric and {Zhang}, Michael and {Rubenzahl}, Ryan A. and {Winn}, Joshua N. and {Han}, Te and {Beard}, Corey and {Holcomb}, Rae and {Householder}, Aaron and {Gilbert}, Gregory J. and {Lubin}, Jack and {Ong}, J.~M. Joel and {Polanski}, Alex S. and {Saunders}, Nicholas and {Van Zandt}, Judah and {Yee}, Samuel W. and {Zhang}, Jingwen and {Zink}, Jon and {Holden}, Bradford and {Baker}, Ashley and {Brodheim}, Max and {Crossfield}, Ian J.~M. and {Deich}, William and {Edelstein}, Jerry and {Gibson}, Steven R. and {Hill}, Grant M. and {Jelinsky}, Sharon R. and {Kassis}, Marc and {Laher}, Russ R. and {Lanclos}, Kyle and {Lilley}, Scott and {Payne}, Joel N. and {Rider}, Kodi and {Robertson}, Paul and {Roy}, Arpita and {Schwab}, Christian and {Shaum}, Abby P. and {Sirk}, Martin M. and {Smith}, Chris and {Vandenberg}, Adam and {Walawender}, Josh and {Wang}, Sharon X. and {Wang}, Shin-Ywan (Cindy) and {Wishnow}, Edward and {Wright}, Jason T. and {Yeh}, Sherry and {Caballero}, Jos{\'e} A. and {Morales}, Juan C. and {Murgas}, Felipe and {Nagel}, Evangelos and {Reiners}, Ansgar and {Schweitzer}, Andreas and {Tabernero}, Hugo M. and {Zechmeister}, Mathias and {Spencer}, Alton and {Ciardi}, David R. and {Clark}, Catherine A. and {Lund}, Michael B. and {Caldwell}, Douglas A. and {Collins}, Karen A. and {Schwarz}, Richard P. and {Barkaoui}, Khalid and {Watkins}, Cristilyn and {Shporer}, Avi and {Narita}, Norio and {Fukui}, Akihiko and {Srdoc}, Gregor and {Latham}, David W. and {Jenkins}, Jon M. and {Ricker}, George R. and {Seager}, Sara and {Vanderspek}, Roland},
        title = "{An Earth-sized Planet on the Verge of Tidal Disruption}",
      journal = {\aj},
         year = 2024,
        month = sep,
       volume = {168},
       number = {3},
          eid = {101},
        pages = {101},
archivePrefix = {arXiv},
 primaryClass = {astro-ph.EP},
       adsurl = {https://ui.adsabs.harvard.edu/abs/2024AJ....168..101D}
}

@ARTICLE{gj367,
       author = {{Lam}, Kristine W.~F. and {Csizmadia}, Szil{\'a}rd and {Astudillo-Defru}, Nicola and {Bonfils}, Xavier and {Gandolfi}, Davide and {Padovan}, Sebastiano and {Esposito}, Massimiliano and {Hellier}, Coel and {Hirano}, Teruyuki and {Livingston}, John and {Murgas}, Felipe and {Smith}, Alexis M.~S. and {Collins}, Karen A. and {Mathur}, Savita and {Garcia}, Rafael A. and {Howell}, Steve B. and {Santos}, Nuno C. and {Dai}, Fei and {Ricker}, George R. and {Vanderspek}, Roland and {Latham}, David W. and {Seager}, Sara and {Winn}, Joshua N. and {Jenkins}, Jon M. and {Albrecht}, Simon and {Almenara}, Jose M. and {Artigau}, Etienne and {Barrag{\'a}n}, Oscar and {Bouchy}, Fran{\c{c}}ois and {Cabrera}, Juan and {Charbonneau}, David and {Chaturvedi}, Priyanka and {Chaushev}, Alexander and {Christiansen}, Jessie L. and {Cochran}, William D. and {De Meideiros}, Jos{\'e} R. and {Delfosse}, Xavier and {D{\'\i}az}, Rodrigo F. and {Doyon}, Ren{\'e} and {Eigm{\"u}ller}, Philipp and {Figueira}, Pedro and {Forveille}, Thierry and {Fridlund}, Malcolm and {Gaisn{\'e}}, Guillaume and {Goffo}, Elisa and {Georgieva}, Iskra and {Grziwa}, Sascha and {Guenther}, Eike and {Hatzes}, Artie P. and {Johnson}, Marshall C. and {Kab{\'a}th}, Petr and {Knudstrup}, Emil and {Korth}, Judith and {Lewin}, Pablo and {Lissauer}, Jack J. and {Lovis}, Christophe and {Luque}, Rafael and {Melo}, Claudio and {Morgan}, Edward H. and {Morris}, Robert and {Mayor}, Michel and {Narita}, Norio and {Osborne}, Hannah L.~M. and {Palle}, Enric and {Pepe}, Francesco and {Persson}, Carina M. and {Quinn}, Samuel N. and {Rauer}, Heike and {Redfield}, Seth and {Schlieder}, Joshua E. and {S{\'e}gransan}, Damien and {Serrano}, Luisa M. and {Smith}, Jeffrey C. and {{\v{S}}ubjak}, J{\'a}n and {Twicken}, Joseph D. and {Udry}, St{\'e}phane and {Van Eylen}, Vincent and {Vezie}, Michael},
        title = "{GJ 367b: A dense, ultrashort-period sub-Earth planet transiting a nearby red dwarf star}",
      journal = {Science},
         year = 2021,
        month = dec,
       volume = {374},
       number = {6572},
        pages = {1271-1275},
archivePrefix = {arXiv},
 primaryClass = {astro-ph.EP},
       adsurl = {https://ui.adsabs.harvard.edu/abs/2021Sci...374.1271L}
}

@ARTICLE{gj1252,
       author = {{Shporer}, Avi and {Collins}, Karen A. and {Astudillo-Defru}, Nicola and {Irwin}, Jonathan and {Bonfils}, Xavier and {Collins}, Kevin I. and {Matthews}, Elisabeth and {Winters}, Jennifer G. and {Anderson}, David R. and {Armstrong}, James D. and {Charbonneau}, David and {Cloutier}, Ryan and {Daylan}, Tansu and {Gan}, Tianjun and {G{\"u}nther}, Maximilian N. and {Hellier}, Coel and {Horne}, Keith and {Huang}, Chelsea X. and {Jensen}, Eric L.~N. and {Kielkopf}, John and {Palle}, Enric and {Sefako}, Ramotholo and {Stassun}, Keivan G. and {Tan}, Thiam-Guan and {Vanderburg}, Andrew and {Ricker}, George R. and {Latham}, David W. and {Vanderspek}, Roland and {Seager}, Sara and {Winn}, Joshua N. and {Jenkins}, Jon M. and {Colon}, Knicole and {Dressing}, Courtney D. and {L{\'e}epine}, S{\'e}bastien and {Muirhead}, Philip S. and {Rose}, Mark E. and {Twicken}, Joseph D. and {Villasenor}, Jesus Noel},
        title = "{GJ 1252 b: A 1.2 R$_{{\ensuremath{\oplus}}}$ Planet Transiting an M3 Dwarf at 20.4 pc}",
      journal = {\apjl},
         year = 2020,
        month = feb,
       volume = {890},
       number = {1},
          eid = {L7},
        pages = {L7},
archivePrefix = {arXiv},
 primaryClass = {astro-ph.EP},
       adsurl = {https://ui.adsabs.harvard.edu/abs/2020ApJ...890L...7S}
}

@ARTICLE{wolf327,
       author = {{Murgas}, F. and {Pall{\'e}}, E. and {Orell-Miquel}, J. and {Carleo}, I. and {Pe{\~n}a-Mo{\~n}ino}, L. and {P{\'e}rez-Torres}, M. and {Watkins}, C.~N. and {Jeffers}, S.~V. and {Azzaro}, M. and {Barkaoui}, K. and {Belinski}, A.~A. and {Caballero}, J.~A. and {Charbonneau}, D. and {Cheryasov}, D.~V. and {Ciardi}, D.~R. and {Collins}, K.~A. and {Cort{\'e}s-Contreras}, M. and {de Leon}, J. and {Duque-Arribas}, C. and {Enoc}, G. and {Esparza-Borges}, E. and {Fukui}, A. and {Gerald{\'\i}a-Gonz{\'a}lez}, S. and {Gilbert}, E.~A. and {Hatzes}, A.~P. and {Hayashi}, Y. and {Henning}, Th. and {Herrero}, E. and {Jenkins}, J.~M. and {Lillo-Box}, J. and {Lodieu}, N. and {Lund}, M.~B. and {Luque}, R. and {Montes}, D. and {Nagel}, E. and {Narita}, N. and {Parviainen}, H. and {Polanski}, A.~S. and {Reffert}, S. and {Schlecker}, M. and {Sch{\"o}fer}, P. and {Schwarz}, R.~P. and {Schweitzer}, A. and {Seager}, S. and {Stassun}, K.~G. and {Tabernero}, H.~M. and {Terada}, Y. and {Twicken}, J.~D. and {Vanaverbeke}, S. and {Winn}, J.~N. and {Zambelli}, R. and {Amado}, P.~J. and {Quirrenbach}, A. and {Reiners}, A. and {Ribas}, I.},
        title = "{Wolf 327b: A new member of the pack of ultra-short-period super-Earths around M dwarfs}",
      journal = {\aap},
         year = 2024,
        month = apr,
       volume = {684},
          eid = {A83},
        pages = {A83},
archivePrefix = {arXiv},
 primaryClass = {astro-ph.EP},
       adsurl = {https://ui.adsabs.harvard.edu/abs/2024A&A...684A..83M}
}

@ARTICLE{toi1685_1,
       author = {{Burt}, Jennifer A. and {Hooton}, Matthew J. and {Mamajek}, Eric E. and {Barrag{\'a}n}, Oscar and {Millholland}, Sarah C. and {Fairnington}, Tyler R. and {Fisher}, Chloe and {Halverson}, Samuel P. and {Huang}, Chelsea X. and {Brady}, Madison and {Seifahrt}, Andreas and {Gaidos}, Eric and {Luque}, Rafael and {Kasper}, David and {Bean}, Jacob L.},
        title = "{TOI-1685 b Is a Hot Rocky Super-Earth: Updates to the Stellar and Planet Parameters of a Popular JWST Cycle 2 Target}",
      journal = {\apjl},
         year = 2024,
        month = aug,
       volume = {971},
       number = {1},
          eid = {L12},
        pages = {L12},
archivePrefix = {arXiv},
 primaryClass = {astro-ph.EP},
       adsurl = {https://ui.adsabs.harvard.edu/abs/2024ApJ...971L..12B}
}

@ARTICLE{toi1685_2,
       author = {{Bluhm}, P. and {Pall{\'e}}, E. and {Molaverdikhani}, K. and {Kemmer}, J. and {Hatzes}, A.~P. and {Kossakowski}, D. and {Stock}, S. and {Caballero}, J.~A. and {Lillo-Box}, J. and {B{\'e}jar}, V.~J.~S. and {Soto}, M.~G. and {Amado}, P.~J. and {Brown}, P. and {Cadieux}, C. and {Cloutier}, R. and {Collins}, K.~A. and {Collins}, K.~I. and {Cort{\'e}s-Contreras}, M. and {Doyon}, R. and {Dreizler}, S. and {Espinoza}, N. and {Fukui}, A. and {Gonz{\'a}lez-{\'A}lvarez}, E. and {Henning}, Th. and {Horne}, K. and {Jeffers}, S.~V. and {Jenkins}, J.~M. and {Jensen}, E.~L.~N. and {Kaminski}, A. and {Kielkopf}, J.~F. and {Kusakabe}, N. and {K{\"u}rster}, M. and {Lafreni{\`e}re}, D. and {Luque}, R. and {Murgas}, F. and {Montes}, D. and {Morales}, J.~C. and {Narita}, N. and {Passegger}, V.~M. and {Quirrenbach}, A. and {Sch{\"o}fer}, P. and {Reffert}, S. and {Reiners}, A. and {Ribas}, I. and {Ricker}, G.~R. and {Seager}, S. and {Schweitzer}, A. and {Schwarz}, R.~P. and {Tamura}, M. and {Trifonov}, T. and {Vanderspek}, R. and {Winn}, J. and {Zechmeister}, M. and {Zapatero Osorio}, M.~R.},
        title = "{An ultra-short-period transiting super-Earth orbiting the M3 dwarf TOI-1685}",
      journal = {\aap},
         year = 2021,
        month = jun,
       volume = {650},
          eid = {A78},
        pages = {A78},
archivePrefix = {arXiv},
 primaryClass = {astro-ph.EP},
       adsurl = {https://ui.adsabs.harvard.edu/abs/2021A&A...650A..78B}
}

@ARTICLE{toi1634,
       author = {{Cloutier}, Ryan and {Charbonneau}, David and {Stassun}, Keivan G. and {Murgas}, Felipe and {Mortier}, Annelies and {Massey}, Robert and {Lissauer}, Jack J. and {Latham}, David W. and {Irwin}, Jonathan and {Haywood}, Rapha{\"e}lle D. and {Guerra}, Pere and {Girardin}, Eric and {Giacalone}, Steven A. and {Bosch-Cabot}, Pau and {Bieryla}, Allyson and {Winn}, Joshua and {Watson}, Christopher A. and {Vanderspek}, Roland and {Udry}, St{\'e}phane and {Tamura}, Motohide and {Sozzetti}, Alessandro and {Shporer}, Avi and {S{\'e}gransan}, Damien and {Seager}, Sara and {Savel}, Arjun B. and {Sasselov}, Dimitar and {Rose}, Mark and {Ricker}, George and {Rice}, Ken and {Quintana}, Elisa V. and {Quinn}, Samuel N. and {Piotto}, Giampaolo and {Phillips}, David and {Pepe}, Francesco and {Pedani}, Marco and {Parviainen}, Hannu and {Palle}, Enric and {Narita}, Norio and {Molinari}, Emilio and {Micela}, Giuseppina and {McDermott}, Scott and {Mayor}, Michel and {Matson}, Rachel A. and {Martinez Fiorenzano}, Aldo F. and {Lovis}, Christophe and {L{\'o}pez-Morales}, Mercedes and {Kusakabe}, Nobuhiko and {Jensen}, Eric L.~N. and {Jenkins}, Jon M. and {Huang}, Chelsea X. and {Howell}, Steve B. and {Harutyunyan}, Avet and {F{\H{u}}r{\'e}sz}, G{\'a}bor and {Fukui}, Akihiko and {Esquerdo}, Gilbert A. and {Esparza-Borges}, Emma and {Dumusque}, Xavier and {Dressing}, Courtney D. and {Fabrizio}, Luca Di and {Collins}, Karen A. and {Cameron}, Andrew Collier and {Christiansen}, Jessie L. and {Cecconi}, Massimo and {Buchhave}, Lars A. and {Boschin}, Walter and {Andreuzzi}, Gloria},
        title = "{TOI-1634 b: An Ultra-short-period Keystone Planet Sitting inside the M-dwarf Radius Valley}",
      journal = {\aj},
         year = 2021,
        month = aug,
       volume = {162},
       number = {2},
          eid = {79},
        pages = {79},
archivePrefix = {arXiv},
 primaryClass = {astro-ph.EP},
       adsurl = {https://ui.adsabs.harvard.edu/abs/2021AJ....162...79C}
}

@ARTICLE{toi1685_3_toi1634,
       author = {{Hirano}, Teruyuki and {Livingston}, John H. and {Fukui}, Akihiko and {Narita}, Norio and {Harakawa}, Hiroki and {Ishikawa}, Hiroyuki Tako and {Miyakawa}, Kohei and {Kimura}, Tadahiro and {Nakayama}, Akifumi and {Fujita}, Naho and {Hori}, Yasunori and {Stassun}, Keivan G. and {Bieryla}, Allyson and {Cadieux}, Charles and {Ciardi}, David R. and {Collins}, Karen A. and {Ikoma}, Masahiro and {Vanderburg}, Andrew and {Barclay}, Thomas and {Brasseur}, C.~E. and {de Leon}, Jerome P. and {Doty}, John P. and {Doyon}, Ren{\'e} and {Esparza-Borges}, Emma and {Esquerdo}, Gilbert A. and {Furlan}, Elise and {Gaidos}, Eric and {Gonzales}, Erica J. and {Hodapp}, Klaus and {Howell}, Steve B. and {Isogai}, Keisuke and {Jacobson}, Shane and {Jenkins}, Jon M. and {Jensen}, Eric L.~N. and {Kawauchi}, Kiyoe and {Kotani}, Takayuki and {Kudo}, Tomoyuki and {Kurita}, Seiya and {Kurokawa}, Takashi and {Kusakabe}, Nobuhiko and {Kuzuhara}, Masayuki and {Lafreni{\`e}re}, David and {Latham}, David W. and {Massey}, Bob and {Mori}, Mayuko and {Murgas}, Felipe and {Nishikawa}, Jun and {Nishiumi}, Taku and {Omiya}, Masashi and {Paegert}, Martin and {Palle}, Enric and {Parviainen}, Hannu and {Quinn}, Samuel N. and {Ricker}, George R. and {Schwarz}, Richard P. and {Seager}, Sara and {Tamura}, Motohide and {Tenenbaum}, Peter and {Terada}, Yuka and {Vanderspek}, Roland K. and {Vievard}, S{\'e}bastien and {Watanabe}, Noriharu and {Winn}, Joshua N.},
        title = "{Two Bright M Dwarfs Hosting Ultra-Short-Period Super-Earths with Earth-like Compositions}",
      journal = {\aj},
         year = 2021,
        month = oct,
       volume = {162},
       number = {4},
          eid = {161},
        pages = {161},
archivePrefix = {arXiv},
 primaryClass = {astro-ph.EP},
       adsurl = {https://ui.adsabs.harvard.edu/abs/2021AJ....162..161H}
}

@ARTICLE{ltt3780,
       author = {{Cloutier}, Ryan and {Eastman}, Jason D. and {Rodriguez}, Joseph E. and {Astudillo-Defru}, Nicola and {Bonfils}, Xavier and {Mortier}, Annelies and {Watson}, Christopher A. and {Stalport}, Manu and {Pinamonti}, Matteo and {Lienhard}, Florian and {Harutyunyan}, Avet and {Damasso}, Mario and {Latham}, David W. and {Collins}, Karen A. and {Massey}, Robert and {Irwin}, Jonathan and {Winters}, Jennifer G. and {Charbonneau}, David and {Ziegler}, Carl and {Matthews}, Elisabeth and {Crossfield}, Ian J.~M. and {Kreidberg}, Laura and {Quinn}, Samuel N. and {Ricker}, George and {Vanderspek}, Roland and {Seager}, Sara and {Winn}, Joshua and {Jenkins}, Jon M. and {Vezie}, Michael and {Udry}, St{\'e}phane and {Twicken}, Joseph D. and {Tenenbaum}, Peter and {Sozzetti}, Alessandro and {S{\'e}gransan}, Damien and {Schlieder}, Joshua E. and {Sasselov}, Dimitar and {Santos}, Nuno C. and {Rice}, Ken and {Rackham}, Benjamin V. and {Poretti}, Ennio and {Piotto}, Giampaolo and {Phillips}, David and {Pepe}, Francesco and {Molinari}, Emilio and {Mignon}, Lucile and {Micela}, Giuseppina and {Melo}, Claudio and {de Medeiros}, Jos{\'e} R. and {Mayor}, Michel and {Matson}, Rachel A. and {Martinez Fiorenzano}, Aldo F. and {Mann}, Andrew W. and {Magazz{\'u}}, Antonio and {Lovis}, Christophe and {L{\'o}pez-Morales}, Mercedes and {Lopez}, Eric and {Lissauer}, Jack J. and {L{\'e}pine}, S{\'e}bastien and {Law}, Nicholas and {Kielkopf}, John F. and {Johnson}, John A. and {Jensen}, Eric L.~N. and {Howell}, Steve B. and {Gonzales}, Erica and {Ghedina}, Adriano and {Forveille}, Thierry and {Figueira}, Pedro and {Dumusque}, Xavier and {Dressing}, Courtney D. and {Doyon}, Ren{\'e} and {D{\'\i}az}, Rodrigo F. and {Fabrizio}, Luca Di and {Delfosse}, Xavier and {Cosentino}, Rosario and {Conti}, Dennis M. and {Collins}, Kevin I. and {Cameron}, Andrew Collier and {Ciardi}, David and {Caldwell}, Douglas A. and {Burke}, Christopher and {Buchhave}, Lars and {Brice{\~n}o}, C{\'e}sar and {Boyd}, Patricia and {Bouchy}, Fran{\c{c}}ois and {Beichman}, Charles and {Artigau}, {\'E}tienne and {Almenara}, Jose M.},
        title = "{A Pair of TESS Planets Spanning the Radius Valley around the Nearby Mid-M Dwarf LTT 3780}",
      journal = {\aj},
         year = 2020,
        month = jul,
       volume = {160},
       number = {1},
          eid = {3},
        pages = {3},
archivePrefix = {arXiv},
 primaryClass = {astro-ph.EP},
       adsurl = {https://ui.adsabs.harvard.edu/abs/2020AJ....160....3C}
}

@ARTICLE{gj806,
       author = {{Palle}, E. and {Orell-Miquel}, J. and {Brady}, M. and {Bean}, J. and {Hatzes}, A.~P. and {Morello}, G. and {Morales}, J.~C. and {Murgas}, F. and {Molaverdikhani}, K. and {Parviainen}, H. and {Sanz-Forcada}, J. and {B{\'e}jar}, V.~J.~S. and {Caballero}, J.~A. and {Sreenivas}, K.~R. and {Schlecker}, M. and {Ribas}, I. and {Perdelwitz}, V. and {Tal-Or}, L. and {P{\'e}rez-Torres}, M. and {Luque}, R. and {Dreizler}, S. and {Fuhrmeister}, B. and {Aceituno}, F. and {Amado}, P.~J. and {Anglada-Escud{\'e}}, G. and {Caldwell}, D.~A. and {Charbonneau}, D. and {Cifuentes}, C. and {de Leon}, J.~P. and {Collins}, K.~A. and {Dufoer}, S. and {Espinoza}, N. and {Essack}, Z. and {Fukui}, A. and {Chew}, Y. G{\'o}mez Maqueo and {G{\'o}mez-Mu{\~n}oz}, M.~A. and {Henning}, Th. and {Herrero}, E. and {Jeffers}, S.~V. and {Jenkins}, J. and {Kaminski}, A. and {Kasper}, J. and {Kunimoto}, M. and {Latham}, D. and {Lillo-Box}, J. and {L{\'o}pez-Gonz{\'a}lez}, M.~J. and {Montes}, D. and {Mori}, M. and {Narita}, N. and {Quirrenbach}, A. and {Pedraz}, S. and {Reiners}, A. and {Rodr{\'\i}guez}, E. and {Rodr{\'\i}guez-L{\'o}pez}, C. and {Sabin}, L. and {Schanche}, N. and {Schwarz}, R. -P. and {Schweitzer}, A. and {Seifahrt}, A. and {Stefansson}, G. and {Sturmer}, J. and {Trifonov}, T. and {Vanaverbeke}, S. and {Wells}, R.~D. and {Zapatero-Osorio}, M.~R. and {Zechmeister}, M.},
        title = "{GJ 806 (TOI-4481): A bright nearby multi-planetary system with a transiting hot low-density super-Earth}",
      journal = {\aap},
         year = 2023,
        month = oct,
       volume = {678},
          eid = {A80},
        pages = {A80},
archivePrefix = {arXiv},
 primaryClass = {astro-ph.EP},
       adsurl = {https://ui.adsabs.harvard.edu/abs/2023A&A...678A..80P}
}

@ARTICLE{Zeng2019,
       author = {{Zeng}, Li and {Jacobsen}, Stein B. and {Sasselov}, Dimitar D. and {Petaev}, Michail I. and {Vanderburg}, Andrew and {Lopez-Morales}, Mercedes and {Perez-Mercader}, Juan and {Mattsson}, Thomas R. and {Li}, Gongjie and {Heising}, Matthew Z. and {Bonomo}, Aldo S. and {Damasso}, Mario and {Berger}, Travis A. and {Cao}, Hao and {Levi}, Amit and {Wordsworth}, Robin D.},
        title = "{Growth model interpretation of planet size distribution}",
      journal = {Proceedings of the National Academy of Science},
         year = 2019,
        month = may,
       volume = {116},
       number = {20},
        pages = {9723-9728},
archivePrefix = {arXiv},
 primaryClass = {astro-ph.EP},
       adsurl = {https://ui.adsabs.harvard.edu/abs/2019PNAS..116.9723Z}
}

@ARTICLE{nasa_exoplanet_archive,
       author = {{Christiansen}, Jessie L. and {McElroy}, Douglas L. and {Harbut}, Marcy and {Ciardi}, David R. and {Crane}, Megan and {Good}, John and {Hardegree-Ullman}, Kevin K. and {Kesseli}, Aurora Y. and {Lund}, Michael B. and {Lynn}, Meca and {Muthiar}, Ananda and {Nilsson}, Ricky and {Oluyide}, Toba and {Papin}, Michael and {Rivera}, Amalia and {Swain}, Melanie and {Susemiehl}, Nicholas D. and {Tam}, Raymond and {van Eyken}, Julian and {Beichman}, Charles},
        title = "{The NASA Exoplanet Archive and Exoplanet Follow-up Observing Program: Data, Tools, and Usage}",
      journal = {psj},
         year = 2025,
        month = aug,
       volume = {6},
       number = {8},
          eid = {186},
        pages = {186},
archivePrefix = {arXiv},
 primaryClass = {astro-ph.EP},
       adsurl = {https://ui.adsabs.harvard.edu/abs/2025PSJ.....6..186C}
}

@ARTICLE{Lin2025,
       author = {{Lin}, Zifan and {Cambioni}, Saverio and {Seager}, Sara},
        title = "{Most High-density Exoplanets Are Unlikely to Be Remnant Giant Planet's Cores}",
      journal = {\apjl},
         year = 2025,
        month = jan,
       volume = {978},
       number = {2},
          eid = {L41},
        pages = {L41},
archivePrefix = {arXiv},
 primaryClass = {astro-ph.EP},
       adsurl = {https://ui.adsabs.harvard.edu/abs/2025ApJ...978L..41L}
}

@ARTICLE{Bonomo2019,
       author = {{Bonomo}, Aldo S. and {Zeng}, Li and {Damasso}, Mario and {Leinhardt}, Zo{\"e} M. and {Justesen}, Anders B. and {Lopez}, Eric and {Lund}, Mikkel N. and {Malavolta}, Luca and {Silva Aguirre}, Victor and {Buchhave}, Lars A. and {Corsaro}, Enrico and {Denman}, Thomas and {Lopez-Morales}, Mercedes and {Mills}, Sean M. and {Mortier}, Annelies and {Rice}, Ken and {Sozzetti}, Alessandro and {Vanderburg}, Andrew and {Affer}, Laura and {Arentoft}, Torben and {Benbakoura}, Mansour and {Bouchy}, Fran{\c{c}}ois and {Christensen-Dalsgaard}, J{\o}rgen and {Collier Cameron}, Andrew and {Cosentino}, Rosario and {Dressing}, Courtney D. and {Dumusque}, Xavier and {Figueira}, Pedro and {Fiorenzano}, Aldo F.~M. and {Garc{\'\i}a}, Rafael A. and {Handberg}, Rasmus and {Harutyunyan}, Avet and {Johnson}, John A. and {Kjeldsen}, Hans and {Latham}, David W. and {Lovis}, Christophe and {Lundkvist}, Mia S. and {Mathur}, Savita and {Mayor}, Michel and {Micela}, Giusi and {Molinari}, Emilio and {Motalebi}, Fatemeh and {Nascimbeni}, Valerio and {Nava}, Chantanelle and {Pepe}, Francesco and {Phillips}, David F. and {Piotto}, Giampaolo and {Poretti}, Ennio and {Sasselov}, Dimitar and {S{\'e}gransan}, Damien and {Udry}, St{\'e}phane and {Watson}, Chris},
        title = "{A giant impact as the likely origin of different twins in the Kepler-107 exoplanet system}",
      journal = {Nature Astronomy},
         year = 2019,
        month = feb,
       volume = {3},
        pages = {416-423},
archivePrefix = {arXiv},
 primaryClass = {astro-ph.EP},
       adsurl = {https://ui.adsabs.harvard.edu/abs/2019NatAs...3..416B}
}

@ARTICLE{Plotnykov2024,
       author = {{Plotnykov}, Mykhaylo and {Valencia}, Diana},
        title = "{Observation uncertainty effects on the precision of interior planetary parameters}",
      journal = {\mnras},
         year = 2024,
        month = may,
       volume = {530},
       number = {3},
        pages = {3488-3499},
archivePrefix = {arXiv},
 primaryClass = {astro-ph.EP},
       adsurl = {https://ui.adsabs.harvard.edu/abs/2024MNRAS.530.3488P}
}

@ARTICLE{Plotnykov2020,
       author = {{Plotnykov}, Mykhaylo and {Valencia}, Diana},
        title = "{Chemical fingerprints of formation in rocky super-Earths' data}",
      journal = {\mnras},
         year = 2020,
        month = nov,
       volume = {499},
       number = {1},
        pages = {932-947},
archivePrefix = {arXiv},
 primaryClass = {astro-ph.EP},
       adsurl = {https://ui.adsabs.harvard.edu/abs/2020MNRAS.499..932P}
}

@ARTICLE{Valencia2006,
       author = {{Valencia}, Diana and {O'Connell}, Richard J. and {Sasselov}, Dimitar},
        title = "{Internal structure of massive terrestrial planets}",
      journal = {\icarus},
         year = 2006,
        month = apr,
       volume = {181},
       number = {2},
        pages = {545-554},
archivePrefix = {arXiv},
 primaryClass = {astro-ph},
       adsurl = {https://ui.adsabs.harvard.edu/abs/2006Icar..181..545V}
}

@INPROCEEDINGS{Szurgot2015,
       author = {{Szurgot}, M.},
        title = "{Core Mass Fraction and Mean Atomic Weight of Terrestrial Planets, Moon, and Protoplanet Vesta}",
    booktitle = {Comparative Tectonic and Geodynamics of Venus, Earth and Rocky Exoplanets},
         year = 2015,
       editor = {{LPI Editorial Board}},
       series = {LPI Contributions},
       volume = {1839},
        month = may,
        pages = {5001},
       adsurl = {https://ui.adsabs.harvard.edu/abs/2015LPICo1839.5001S}
}

@ARTICLE{Dorn2015,
       author = {{Dorn}, Caroline and {Khan}, Amir and {Heng}, Kevin and {Connolly}, James A.~D. and {Alibert}, Yann and {Benz}, Willy and {Tackley}, Paul},
        title = "{Can we constrain the interior structure of rocky exoplanets from mass and radius measurements?}",
      journal = {\aap},
         year = 2015,
        month = may,
       volume = {577},
          eid = {A83},
        pages = {A83},
archivePrefix = {arXiv},
 primaryClass = {astro-ph.EP},
       adsurl = {https://ui.adsabs.harvard.edu/abs/2015A&A...577A..83D}
}

@ARTICLE{Brinkman2024,
       author = {{Brinkman}, Casey L. and {Polanski}, Alex S. and {Huber}, Daniel and {Weiss}, Lauren M. and {Valencia}, Diana and {Plotnykov}, Mykhaylo},
        title = "{Revisiting the Relationship Between Rocky Exoplanet and Stellar Compositions: Reduced Evidence for a Super-Mercury Population}",
      journal = {\aj},
         year = 2024,
        month = dec,
       volume = {168},
       number = {6},
          eid = {281},
        pages = {281},
archivePrefix = {arXiv},
 primaryClass = {astro-ph.EP},
       adsurl = {https://ui.adsabs.harvard.edu/abs/2024AJ....168..281B}
}
\bibliographystyle{aasjournal.bst}

\onecolumn
\appendix

\counterwithin{table}{section}
\counterwithin{figure}{section}

\section{Supplementary data}

\begin{figure*}[h]
    \centering
    \begin{subfigure}{0.32\textwidth}
        \centering
        \includegraphics[width=\linewidth]{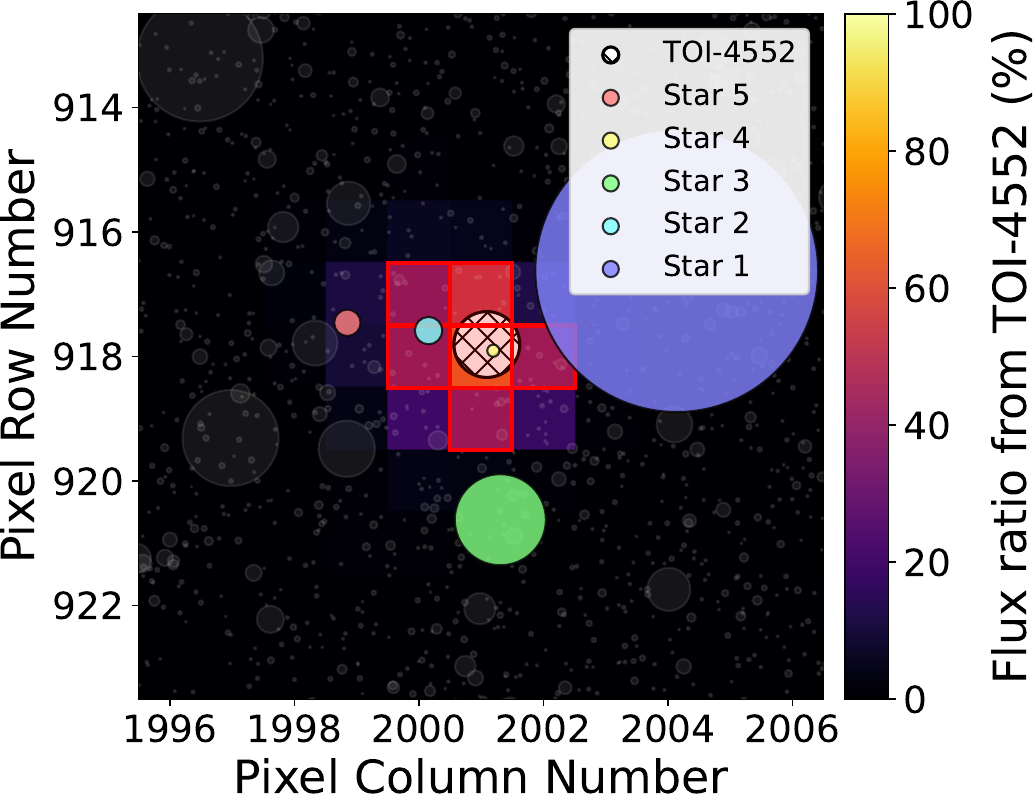}
    \end{subfigure}
    \hfill
    \begin{subfigure}{0.32\textwidth}
        \centering
        \includegraphics[width=\linewidth]{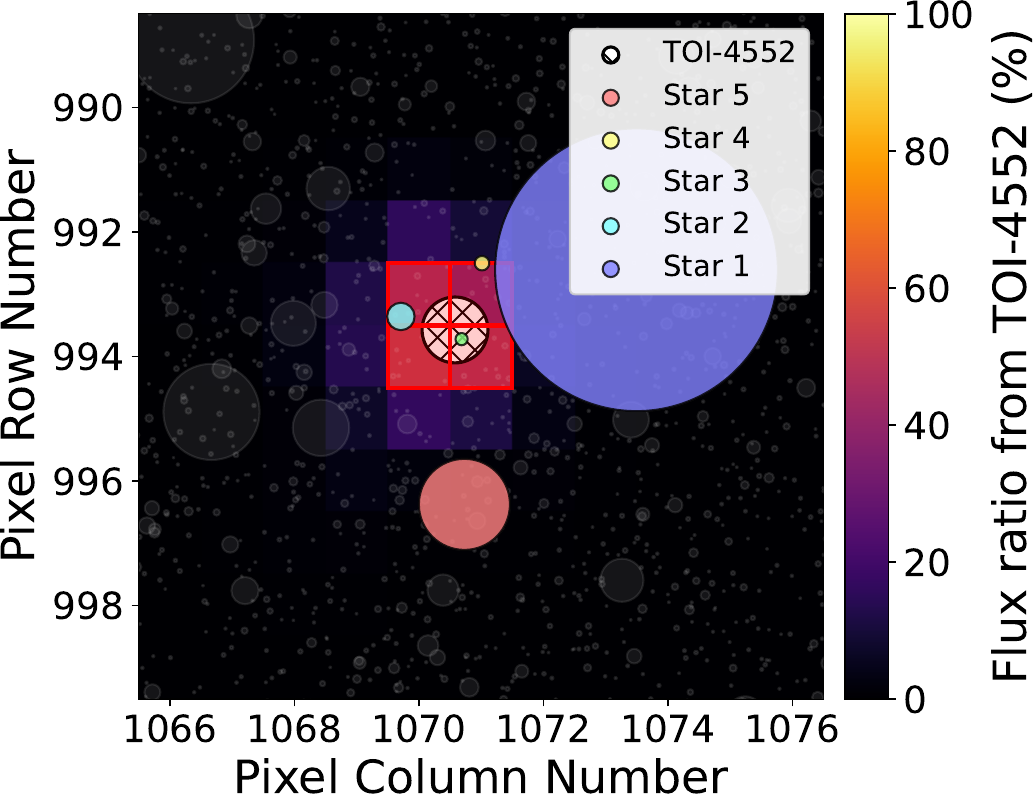}
    \end{subfigure}
    \hfill
    \begin{subfigure}{0.32\textwidth}
        \centering
        \includegraphics[width=\linewidth]{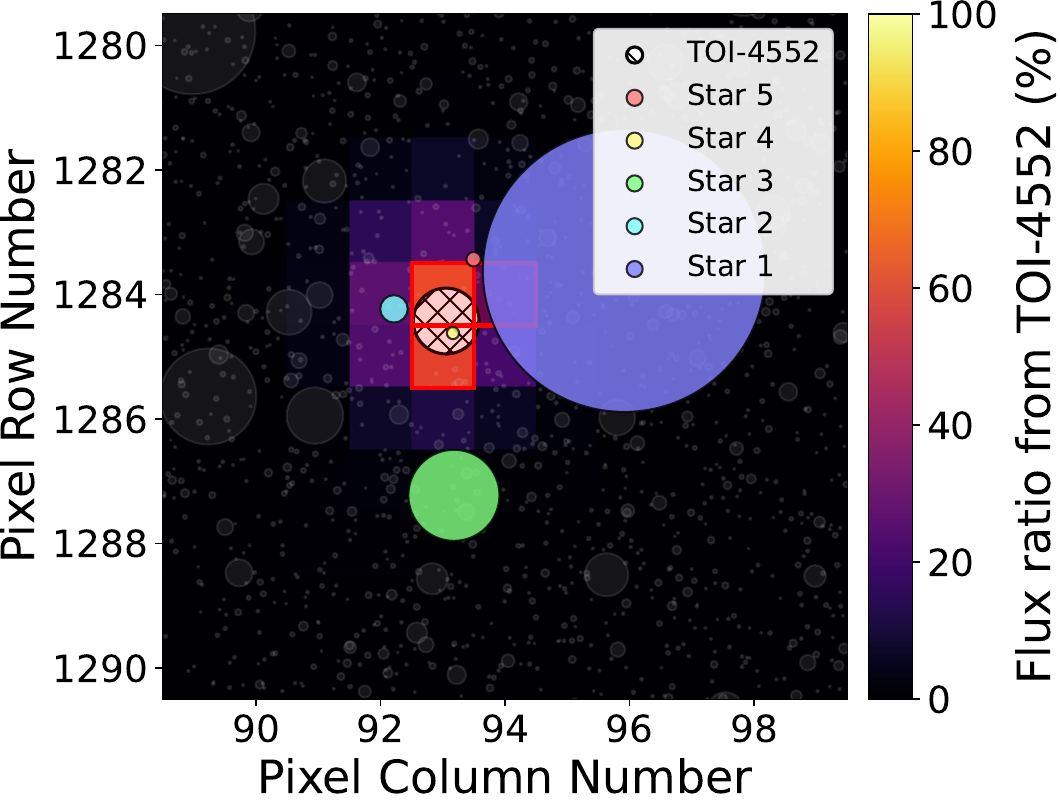}
    \end{subfigure}
    \caption{TESS target pixel file (TPF) of TOI-4552 from sectors 12 (left), 39 (centre) and 66 (right), created with \texttt{TESS-cont} \citep{Castro2024}. The aperture mask is highlighted in red and the pixel size is 21\arcsec. TOI-4552 is highlighted in white with the mesh pattern. The heatmap depicts the per-pixel relative flux contributed by TOI-4552, used to optimise the aperture mask. The five stars that contaminate the lightcurve of TOI-4552 the most are highlighted, with the size of the source being a relative measure of the contamination. The TESS SPOC lightcurves are already corrected for the dilution effect from these five sources.}
    \label{tesstpf_dil}
\end{figure*}

\begin{figure}[h]
    \centering
    \includegraphics[width=.5\linewidth]{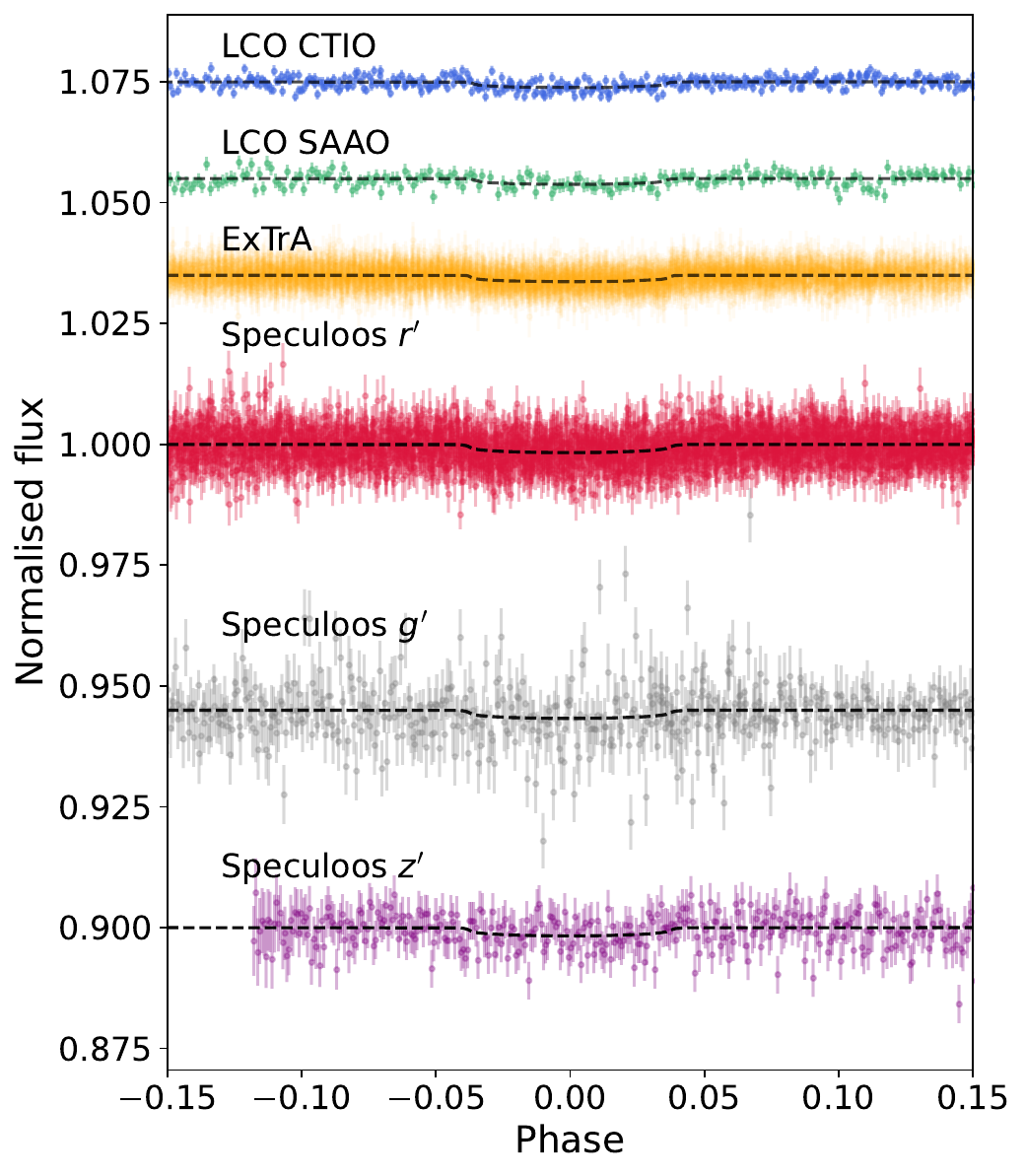}
    \caption{Ground based photometry of TOI-4552 separated by instruments and photometric bands. All lightcurves are phasefolded to refect the transit feature. Any offsets in the noralised flux axis are purely to represent all datasets within the same plot and are not representative of instrument-relative flux.}
    \label{transit-ground}
\end{figure}

The TESS target pixel files with the major contamination sources are depicted in Fig.~\ref{tesstpf_dil}. Ground based photometric phasefolded lightcurves are shown in Fig.~\ref{transit-ground}. Table~\ref{tab:RV_timeseries} provides the HAPRS and NIRPS timeseries corrected for tellurics used in this analysis.

\vspace{-1em}

\setlength{\tabcolsep}{12pt}

\begin{table}[H] 
\centering
\begin{threeparttable}
\caption{Radial velocity timeseries of NIRPS and HARPS used for the analysis (corrected for BERV crossing systematics).}
\label{tab:RV_timeseries}

\begin{tabular}{l c c c c c c c}

\toprule
\textbf{Time (BJD)} &
\textbf{$\Delta$RV (m/s)} &
\textbf{$\sigma_{RV}$ (m/s)} &
\textbf{$\Delta T_{3000K}$ (K)} &
\textbf{$\sigma_{\Delta T}$ (K)} & 
\textbf{D2V (m$^2$/s$^2$)} &
\textbf{$\sigma_{D2V}$ (m$^2$/s$^2$)} &
\textbf{Instrument} \\
\midrule

2460042.8854 & 7.90 & 5.42 &  25.43 & 1.0 & 292288.88 & 8872.02 & NIRPS \\
... & ... & ... &  ... & ... & ... & ... & ...\\

\bottomrule
\end{tabular}

\begin{tablenotes}
\footnotesize
\item \textbf{Notes:}
$\Delta$RV is the median subtracted relative radial velocity. The median systemic velocity for NIRPS is -25490.83\,m/s and for HARPS is -25247.14\,m/s. 
\end{tablenotes}

\end{threeparttable}
\end{table}

\pagebreak
\section{Stellar activity and joint fit table}

\begin{figure*}[h!]
    \centering
    \includegraphics[width=1\linewidth]{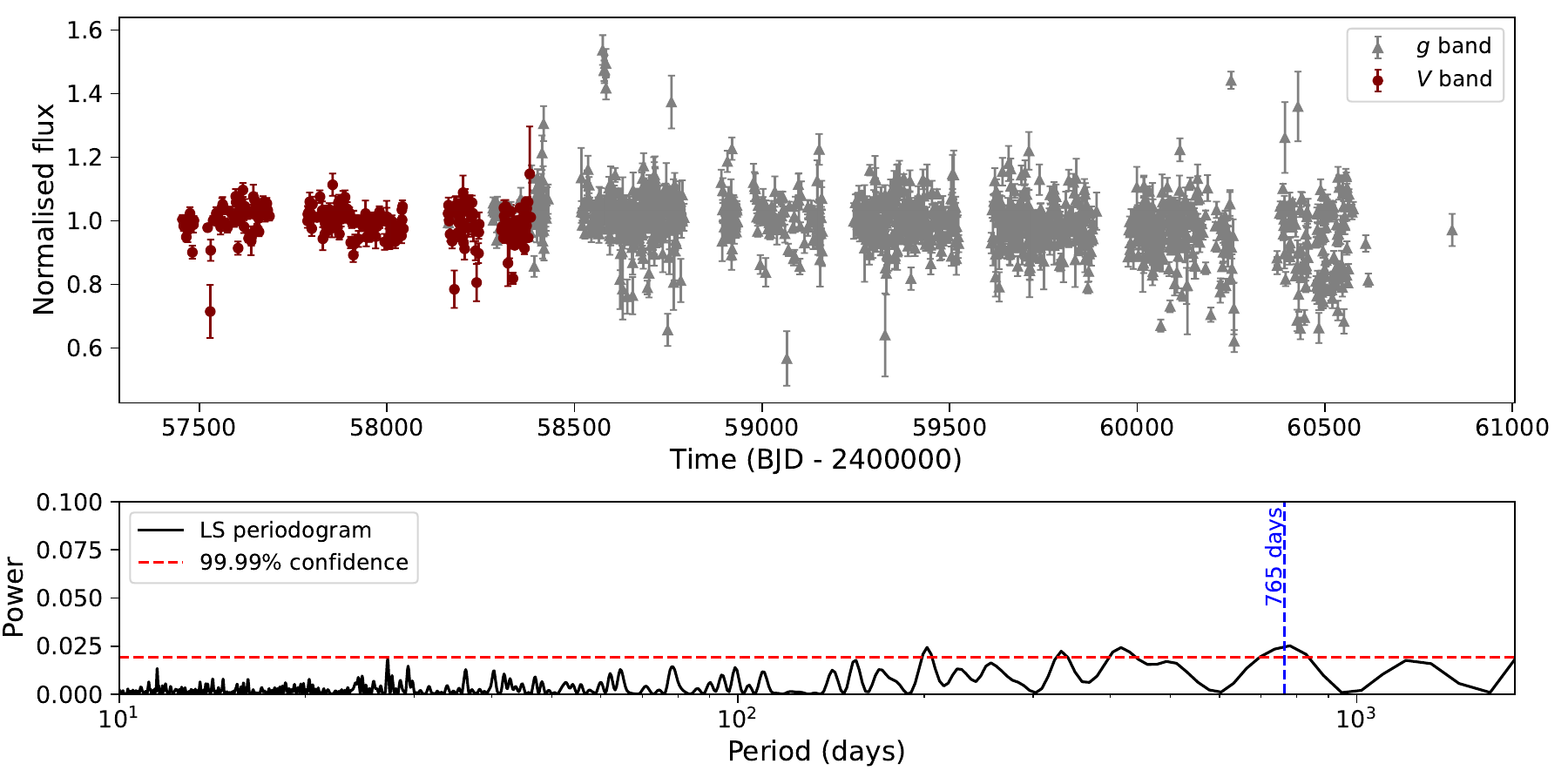}
    \caption{Long-term photometric monitoring of TOI-4552 done as part of the ASAS-SN survey \citep{Kochanek2017}. The data was recorded in the $V$ and $g'$ photometric bands. The periodogram peaks at a signal of 765 days, very close to the two year harmonic of Earth's orbital period, and likely an artifact of the window function. No other siginificant peaks are present that can be attributed to stellar rotation or magnetic cycle, thus TOI-4552 is a quiet star.}
    \label{fig:asassn_phot}
\end{figure*}

\begin{figure*}[h]
    \centering
    \includegraphics[width=1\linewidth]{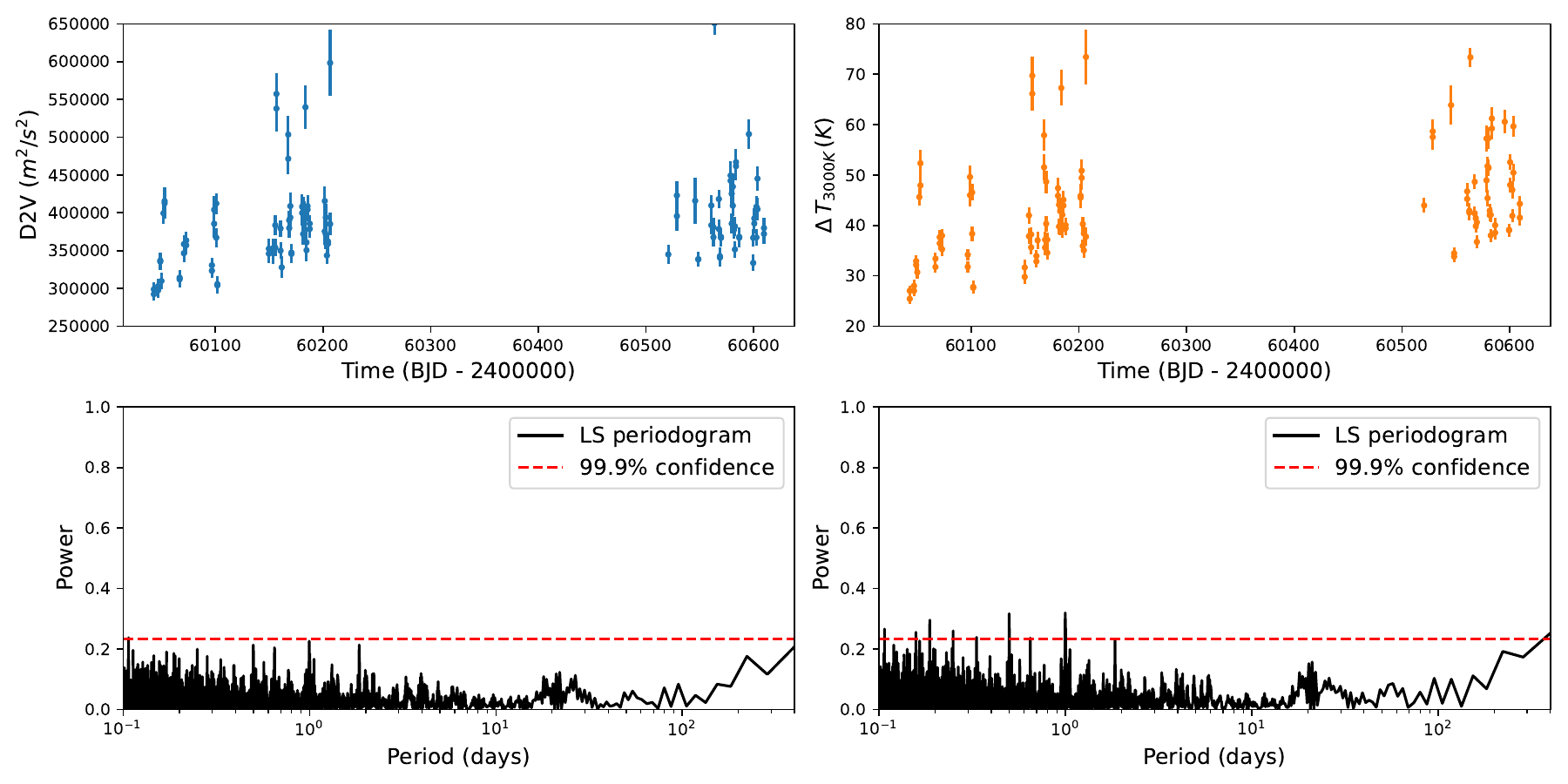}
    \caption{\textit{Left:} Second-order derivative of the velocity (top) and its respective periodogram (bottom). \textit{Right:} Differential temperature metric (top) and its periodogram (bottom). Both quantities are stellar activity indicators reported as part of the standard LBL RV extraction process \citep{Artigau2022, Artigau2024} for NIRPS and show no clear peaks in the periodogram that might correspond to any activity signal.}
    \label{fig:actvityNIRPS}
\end{figure*}

\onecolumn
\renewcommand{\arraystretch}{1.1}
\setlength{\tabcolsep}{28pt}
\begin{longtable}[t]{l c c}
\caption{Median values and 68$\%$ confidence intervals of the posterior distributions of the joint fit.\label{tab:posteriors_jointfit}}\\
\hline\hline
\textbf{Parameter} & \textbf{Prior} & \textbf{Posterior} \\
\hline
\\
\endfirsthead
\caption{continued.}\\
\hline\hline
\textbf{Parameter} & \textbf{Prior} & \textbf{Posterior} \\
\hline
\\
\endhead

\\
\hline
\endfoot

\\
\hline\hline
\endlastfoot

\textbf{Common transit parameters} & & \\

$r_1$ \dotfill & $\mathcal{U}(0,1)$ & $0.438^{+0.051} _{-0.052}$ \\
Scaled planetary radius, $R_{\mathrm{p}}$/$R_{*} = r_2$\dotfill & $\mathcal{U}(0, 1)$ & $0.0355^{+0.0008} _{-0.0008}$ \\
Eccentricity, $e$\dotfill & Fixed & $0.0$ \\
Argument of periastron, $\omega$ (deg)\dotfill & Fixed & $90.0$ \\
Stellar density, $\rho_*$ ($\mathrm{g/cm^{3}}$)\dotfill  & $\mathcal{N}(15.630, 1.513)$ & $16.059^{+0.700} _{-0.661}$ \\
Orbital period, $P$ (days)\dotfill  & $\mathcal{N}(0.3011, 0.001)$ & $0.30110023\pm0.00000012$\\
Transit epoch, $t_c$ (BJD)\dotfill & $\mathcal{N}(2459361.8822, 0.1)$ & $2459361.88588\pm0.00022$\\

\\
\textbf{TESS transit parameters} & & \\

$q_{1,\text{TESS-S12}}$\dotfill & $\mathcal{U}(0, 1)$ & $0.536^{+0.222} _{-0.219}$ \\
$q_{2,\text{TESS-S12}}$\dotfill & $\mathcal{U}(0, 1)$ & $0.625^{+0.231} _{-0.258}$\\
Dilution factor, $D_{\text{TESS-S12}}$\dotfill & $\mathcal{U}(0.5, 1)$ & $0.861^{+0.076} _{-0.077}$\\
Offset relative flux, $M_{\text{TESS-S12}}$\dotfill &$\mathcal{N}(0,0.1)$ & $-0.0012\pm0.0003$\\
Jitter, $\sigma_{w,\,\text{TESS-S12}}$ (ppm)\dotfill & $ \mathcal{L}(10^{-5},1000)$ & $3.323 ^{+55.673} _{-3.257}$ \\ 
GP amplitude, $\alpha_{\text{TESS-S12}}$ (m/s)\dotfill & $\mathcal{L}(10^{-5}, 1)$ & $0.0081^{+0.0069}_{-0.0018}$\\
GP length scale, $\beta_{\text{TESS-S12}}$ (days)\dotfill & $\mathcal{U}(1, 30)$ & $1.205^{+0.795}_{-0.163}$\\

$q_{1,\text{TESS-S39}}$\dotfill & $\mathcal{U}(0, 1)$ & $0.569^{+0.207} _{-0.207}$ \\
$q_{2,\text{TESS-S39}}$\dotfill & $\mathcal{U}(0, 1)$ & $0.608^{+0.239} _{-0.275}$\\
Dilution factor, $D_{\text{TESS-S39}}$\dotfill & $\mathcal{U}(0.5, 1)$ & $0.867^{+0.069} _{-0.075}$\\
Offset relative flux, $M_{\text{TESS-S39}}$\dotfill &$\mathcal{N}(0,0.1)$ & $-0.00031\pm0.0025$\\
Jitter, $\sigma_{w,\,\text{TESS-S39}}$ (ppm)\dotfill & $\mathcal{L}(10^{-5},1000)$ & $0.638 ^{+25.210} _{-0.633}$ \\  
GP amplitude, $\alpha_{\text{TESS-S39}}$ (m/s)\dotfill & $\mathcal{L}(10^{-5}, 1)$ & $0.0015^{+0.0031}_{-0.0009}$\\
GP length scale, $\beta_{\text{TESS-S39}}$ (days)\dotfill & $\mathcal{U}(1, 30)$ & $5.359^{+7.691}_{-3.875}$\\

$q_{1,\text{TESS-S66}}$\dotfill & $\mathcal{U}(0, 1)$ & $0.496^{+0.251} _{-0.249}$ \\
$q_{2,\text{TESS-S66}}$\dotfill & $\mathcal{U}(0, 1)$ & $0.398^{+0.269} _{-0.228}$\\
Dilution factor, $D_{\text{TESS-S66}}$\dotfill & $\mathcal{U}(0.5, 1)$ & $0.839^{+0.079} _{-0.078}$\\
Offset relative flux, $M_{\text{TESS-66}}$\dotfill &$\mathcal{N}(0,0.1)$ & $-0.0015\pm0.0014$\\
Jitter, $\sigma_{w,\,\text{TESS-S66}}$ (ppm)\dotfill & $\mathcal{L}(0.1,100000)$ & $0.157 ^{+212.789} _{-0.156}$ \\  
GP amplitude, $\alpha_{\text{TESS-S66}}$ (m/s)\dotfill & $\mathcal{L}(10^{-5}, 1)$ & $0.0043^{+0.0007}_{-0.0005}$\\
GP length scale, $\beta_{\text{TESS-S66}}$ (days)\dotfill & $\mathcal{U}(1, 30)$ & $1.089^{+0.115}_{-0.065}$\\

\\
\textbf{ExTrA transit parameters} & & \\

$q_{1,\text{ExTrA-T1}}$\dotfill & $\mathcal{U}(0, 1)$ & $0.299^{+0.272} _{-0.190}$ \\
$q_{2,\text{ExTrA-T1}}$\dotfill & $\mathcal{U}(0, 1)$ & $0.496^{+0.270} _{-0.281}$\\
Dilution factor, $D_{\text{ExTrA-T1}}$\dotfill & Fixed & $1.0$\\
Offset relative flux, $M_{\text{ExTrA-T1}}$\dotfill &$\mathcal{N}(0,0.1)$ & $-0.00001\pm0.00006$\\
Jitter, $\sigma_{w,\,\text{ExTrA-T1}}$ (ppm)\dotfill & $\mathcal{L}(10^{-5},1000)$ & $0.618 ^{+17.711} _{-0.609}$ \\  

$q_{1,\text{ExTrA-T2}}$\dotfill & $\mathcal{U}(0, 1)$ & $0.457^{+0.286} _{-0.293}$ \\
$q_{2,\text{ExTrA-T2}}$\dotfill & $\mathcal{U}(0, 1)$ & $0.250^{+0.295} _{-0.179}$\\
Dilution factor, $D_{\text{ExTrA-T2}}$\dotfill & Fixed & $1.0$\\
Offset relative flux, $M_{\text{ExTrA-T2}}$\dotfill &$\mathcal{N}(0,0.1)$ & $0.00000\pm0.00004$\\
Jitter, $\sigma_{w,\,\text{ExTrA-T2}}$ (ppm)\dotfill & $\mathcal{L}(10^{-5},1000)$ & $0.051 ^{+2.992} _{-0.050}$ \\  

$q_{1,\text{ExTrA-T3}}$\dotfill & $\mathcal{U}(0, 1)$ & $0.421^{+0.308} _{-0.268}$ \\
$q_{2,\text{ExTrA-T3}}$\dotfill & $\mathcal{U}(0, 1)$ & $0.300^{+0.252} _{-0.195}$\\
Dilution factor, $D_{\text{ExTrA-T3}}$\dotfill & Fixed & $1.0$\\
Offset relative flux, $M_{\text{ExTrA-T3}}$\dotfill &$\mathcal{N}(0,0.1)$ & $-0.00005\pm0.00003$\\
Jitter, $\sigma_{w,\,\text{ExTrA-T3}}$ (ppm)\dotfill & $\mathcal{L}(10^{-5},1000)$ & $0.001 ^{+0.045} _{-0.001}$ \\  

\\
\textbf{LCO transit parameters} & & \\

$q_{1,\text{LCO-CTIO}}$\dotfill & $\mathcal{U}(0, 1)$ & $0.526^{+0.251} _{-0.255}$ \\
$q_{2,\text{LCO-CTIO}}$\dotfill & $\mathcal{U}(0, 1)$ & $0.411^{+0.259} _{-0.241}$\\
Dilution factor, $D_{\text{LCO-CTIO}}$\dotfill & $\mathcal{U}(0.5, 1)$ & $0.749^{+0.078} _{-0.076}$\\
Offset relative flux, $M_{\text{LCO-CTIO}}$\dotfill &$\mathcal{N}(0,0.1)$ & $0.00001\pm0.00005$\\
Jitter, $\sigma_{w,\,\text{LCO-CTIO}}$ (ppm)\dotfill & $\mathcal{L}(10^{-5},1000)$ & $0.116 ^{+14.791} _{-0.115}$ \\  

$q_{1,\text{LCO-SAAO}}$\dotfill & $\mathcal{U}(0, 1)$ & $0.489^{+0.248} _{-0.242}$ \\
$q_{2,\text{LCO-SAAO}}$\dotfill & $\mathcal{U}(0, 1)$ & $0.555^{+0.229} _{-0.248}$\\
Dilution factor, $D_{\text{LCO-SAAO}}$\dotfill & $\mathcal{U}(0.5, 1)$ & $0.775^{+0.117} _{-0.118}$\\
Offset relative flux, $M_{\text{LCO-SAAO}}$\dotfill &$\mathcal{N}(0,0.1)$ & $0.00003\pm0.00010$\\
Jitter, $\sigma_{w,\,\text{LCO-SAAO}}$ (ppm)\dotfill & $\mathcal{L}(10^{-5},1000)$ & $0.137 ^{+15.483} _{-0.136}$ \\

\\
\textbf{SPECULOOS transit parameters} & & \\

$q_{1,\text{SPECULOOS}-E_{g'}}$\dotfill & $\mathcal{U}(0, 1)$ & $0.626^{+0.223} _{-0.265}$ \\
$q_{2,\text{SPECULOOS}-E_{g'}}$\dotfill & $\mathcal{U}(0, 1)$ & $0.561^{+0.266} _{-0.298}$\\
Dilution factor, $D_{\text{SPECULOOS}-E_{g'}}$\dotfill & $\mathcal{U}(0.5, 1)$ & $0.702^{+0.112} _{-0.110}$\\
Offset relative flux, $M_{\text{SPECULOOS}-E_{g'}}$\dotfill &$\mathcal{N}(0,0.1)$ & $0.00017\pm0.00025$\\
Jitter, $\sigma_{w,\,\text{SPECULOOS}-E_{g'}}$ (ppm)\dotfill & $\mathcal{L}(10^{-5},1000)$ & $0.005 ^{+0.623} _{-0.005}$ \\ 

$q_{1,\text{SPECULOOS}-E_{r'}}$\dotfill & $\mathcal{U}(0, 1)$ & $0.682^{+0.206} _{-0.268}$ \\
$q_{2,\text{SPECULOOS}-E_{r'}}$\dotfill & $\mathcal{U}(0, 1)$ & $0.497^{+0.253} _{-0.251}$\\
Dilution factor, $D_{\text{SPECULOOS}-E_{r'}}$\dotfill & $\mathcal{U}(0.5, 1)$ & $0.896^{+0.065} _{-0.080}$\\
Offset relative flux, $M_{\text{SPECULOOS}-E_{r'}}$\dotfill &$\mathcal{N}(0,0.1)$ & $0.00008\pm0.00005$\\
Jitter, $\sigma_{w,\,\text{SPECULOOS}-E_{r'}}$ (ppm)\dotfill & $\mathcal{L}(10^{-5},1000)$ & $0.055 ^{+9.581} _{-0.054}$ \\ 

$q_{1,\text{SPECULOOS}-G_{r'}}$\dotfill & $\mathcal{U}(0, 1)$ & $0.480^{+0.281} _{-0.273}$ \\
$q_{2,\text{SPECULOOS}-G_{r'}}$\dotfill & $\mathcal{U}(0, 1)$ & $0.438^{+0.278} _{-0.274}$\\
Dilution factor, $D_{\text{SPECULOOS}-G_{r'}}$\dotfill & $\mathcal{U}(0.5, 1)$ & $0.667^{+0.143} _{-0.107}$\\
Offset relative flux, $M_{\text{SPECULOOS}-G_{r'}}$\dotfill &$\mathcal{N}(0,0.1)$ & $0.00023\pm0.00019$\\
Jitter, $\sigma_{w,\,\text{SPECULOOS}-G_{r'}}$ (ppm)\dotfill & $\mathcal{L}(10^{-5},1000)$ & $0.095 ^{+17.365} _{-0.095}$ \\  

$q_{1,\text{SPECULOOS}-I_{g'}}$\dotfill & $\mathcal{U}(0, 1)$ & $0.416^{+0.268} _{-0.243}$ \\
$q_{2,\text{SPECULOOS}-I_{g'}}$\dotfill & $\mathcal{U}(0, 1)$ & $0.428^{+0.281} _{-0.262}$\\
Dilution factor, $D_{\text{SPECULOOS}-I_{g'}}$\dotfill & $\mathcal{U}(0.5, 1)$ & $0.813^{+0.117} _{-0.142}$\\
Offset relative flux, $M_{\text{SPECULOOS}-I_{g'}}$\dotfill &$\mathcal{N}(0,0.1)$ & $-0.00005\pm0.00012$\\
Jitter, $\sigma_{w,\,\text{SPECULOOS}-I_{g'}}$ (ppm)\dotfill & $\mathcal{L}(10^{-5},1000)$ & $0.057 ^{+5.620} _{-0.056}$ \\  

$q_{1,\text{SPECULOOS}-I_{r'}}$\dotfill & $\mathcal{U}(0, 1)$ & $0.594^{+0.227} _{-0.245}$ \\
$q_{2,\text{SPECULOOS}-I_{r'}}$\dotfill & $\mathcal{U}(0, 1)$ & $0.545^{+0.243} _{-0.254}$\\
Dilution factor, $D_{\text{SPECULOOS}-I_{r'}}$\dotfill & $\mathcal{U}(0.5, 1)$ & $0.839^{+0.102} _{-0.115}$\\
Offset relative flux, $M_{\text{SPECULOOS}-I_{r'}}$\dotfill &$\mathcal{N}(0,0.1)$ & $0.00009\pm0.00009$\\
Jitter, $\sigma_{w,\,\text{SPECULOOS}-I_{r'}}$ (ppm)\dotfill & $\mathcal{L}(10^{-5},1000)$ & $0.563 ^{+36.628} _{-0.558}$ \\  

$q_{1,\text{SPECULOOS}-I_{z'}}$\dotfill & $\mathcal{U}(0, 1)$ & $0.587^{+0.239} _{-0.270}$ \\
$q_{2,\text{SPECULOOS}-I_{z'}}$\dotfill & $\mathcal{U}(0, 1)$ & $0.520^{+0.246} _{-0.263}$\\
Dilution factor, $D_{\text{SPECULOOS}-I_{z'}}$\dotfill & $\mathcal{U}(0.5, 1)$ & $0.745^{+0.132} _{-0.129}$\\
Offset relative flux, $M_{\text{SPECULOOS}-I_{z'}}$\dotfill &$\mathcal{N}(0,0.1)$ & $0.00011\pm0.00023$\\
Jitter, $\sigma_{w,\,\text{SPECULOOS}-I_{z'}}$ (ppm)\dotfill & $\mathcal{L}(10^{-5},1000)$ & $0.319 ^{+6.179} _{-0.0317}$ \\  

\\
\textbf{NIRPS and HARPS RV parameters} & & \\

RV semi-amplitude, $K_\mathrm{p}$ (m/s) \dotfill & $\mathcal{U}(0, 10)$ & $4.324^{+1.036}_{-1.080}$ \\
Relative systemic RV offset, $\mu_\mathrm{NIRPS}$ (m/s) \dotfill& $\mathcal{U}(0,100)$ & $1.672^{+1.712}_{-1.163}$\\
Jitter, $\sigma_{w,\mathrm{NIRPS}}$ (m/s) \dotfill & $\mathcal{U}(0,10)$ & $1.317^{+1.256}_{-0.975}$ \\ 

GP amplitude, $\alpha_\mathrm{NIRPS}$ (m/s) \dotfill& $\mathcal{U}(0,10)$ & $3.568^{+0.529}_{-0.728}$\\
GP length scale, $\beta_{\mathrm{NIRPS}}$ (m/s) \dotfill & $\mathcal{U}(1,200)$ & $89.935^{+61.742}_{-44.931}$ \\ 

Relative systemic RV offset, $\mu_\mathrm{HARPS}$ (m/s) \dotfill& $\mathcal{U}(0,100)$ & $3.337^{+2.793}_{-1.921}$\\
Jitter, $\sigma_{w,\mathrm{HARPS}}$ (m/s) \dotfill & $\mathcal{U}(0,100)$ & $27.146^{+4.226}_{-3.523}$ \\ 

\end{longtable}

\begin{tablenotes}
\item
\textbf{Notes:} $\mathcal{N}(\mu, \sigma)$ indicates a normal distribution with mean $\mu$ and variance $\sigma^{2}$, $\mathcal{U}(a, b)$ a uniform distribution between $a$ and $b$ and $\mathcal{L}(a, b)$ a log-uniform distribution between $a$ and $b$. The TESS sectors 12 (TESS-S12), 39 (TESS-S39) and 66 (TESS-S66) were modelled separate from each other. ExTrA-T1, ExTrA-T2, ExTrA-T3 refer to the three ExTrA telescopes. LCO-CTIO and LCO-SAAO are the two telecope part of the LCO network. SPECULOOS-south observed TOI-4552 using three of the four telescopes: Europa (SPECULOOS-E), Ganymede (SPECULOOS-G) and Io (SPECULOOS-I), with $g',\,r'\,$ and $z'$ referring to the various Sloan photometric filters. NIRPS and HARPS are the high-resolution spectrographs used for the velocimetry. The systemic RV offset for NIRPS is -25490.828 m/s and for HARPS is -25247.139 m/s. $\mu_{NIRPS}$ and $\mu_{HARPS}$ are relative to these values
\end{tablenotes}

\section{TOI-4552\,b interior structure modelling}

\begin{figure*}[h]
    \centering
    \includegraphics[width=1\linewidth]{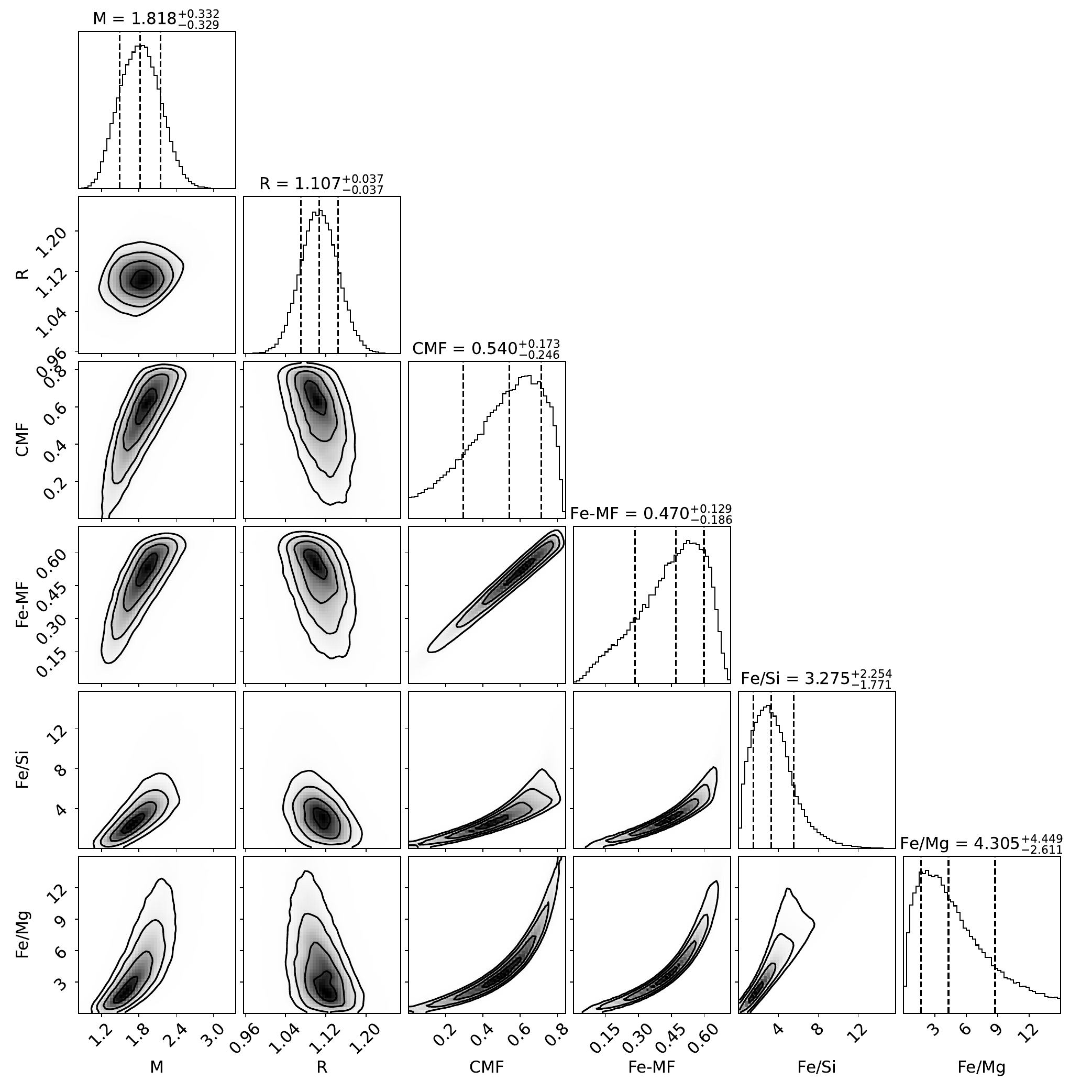}
    \caption{Corner plot depicting the interior structure model of TOI-4552\,b using \texttt{exopie} \citep{Plotnykov2024} with the planetary mass and radius as priors. The abundance of iron in the mantle and silicon in the core is left to vary freely between 0 and 20\%. The resulting CMF of 0.54 is partway between that of Mercury (=\,0.7; \citealt{Szurgot2015}) and Earth (=\,0.33), suggesting a marginal over-abundance of iron (also evident from the high iron mass fraction, Fe-MF).}
    \label{fig:cmf_no_prior}
\end{figure*}

\begin{figure}
    \centering
    \includegraphics[width=1\linewidth]{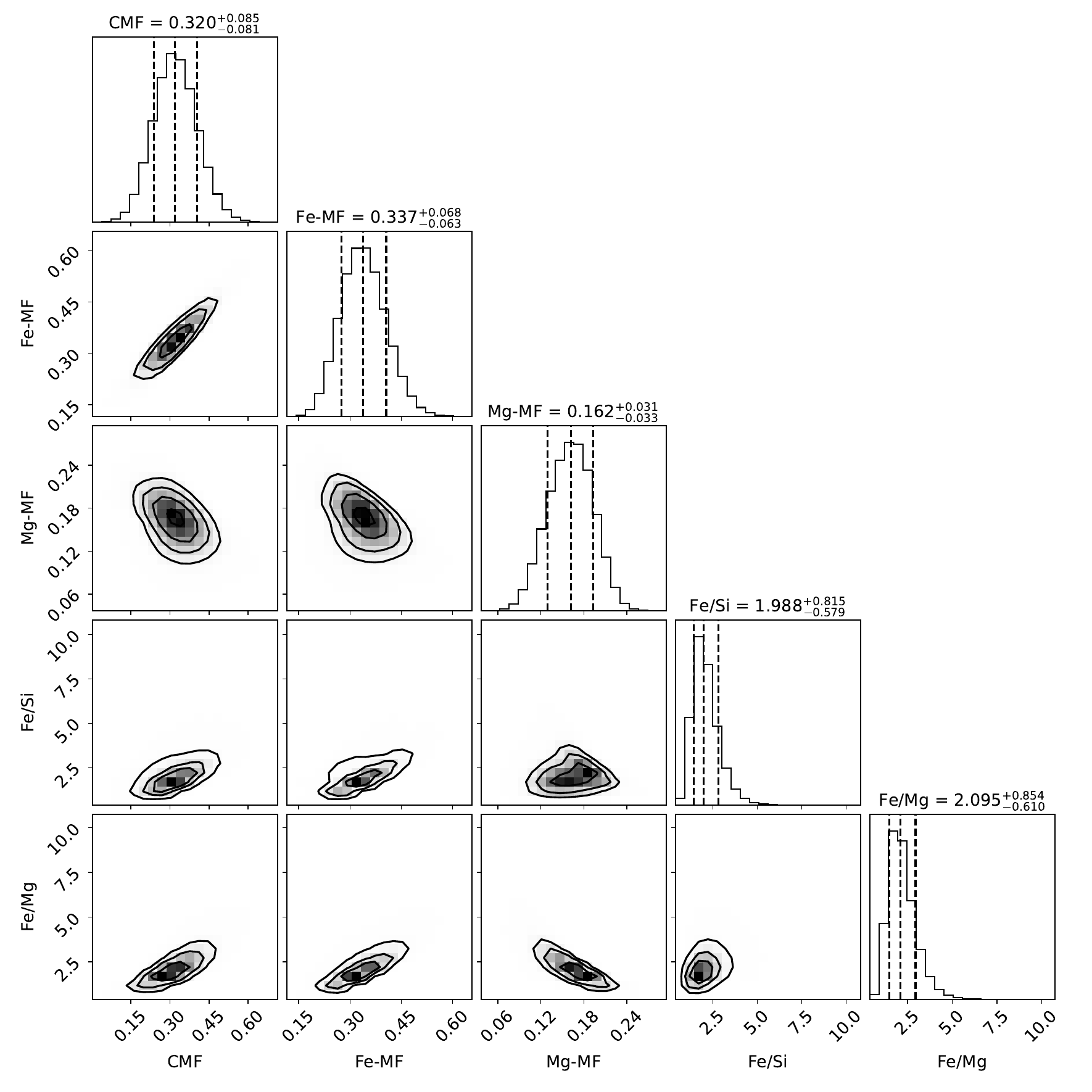}
    \caption{Corner plot of an alternative interior structure model illustrating the expected planetary properties under the assumption that the refractory element abundances of the planet match those of the host star. Therefore, only the stellar element abundances ([Fe/H], [Mg/H] and [Si/H]) were input as parameters, with [$\alpha/H$] used as a proxy for [Mg/H] and [Si/H] due to very few lines used in the abundance measurements for those elements. No constraints were put on planetary mass and radius. The expected CMF for a rocky planet around TOI-4552 is 0.32, consistent with that of the Earth. This is just within the lower uncertainty bound predicted by the interior model constrained by the mass and radius measurements (CMF\,=\,0.54$^{0.17}_{0.25}$; Section~\ref{sec:composition}; Fig~\ref{fig:cmf_no_prior}).}
    \label{fig:cmf_stellar_prior}
\end{figure}

\end{document}